\documentclass[floatfix,nofootinbib,longbibliography,notitlepage]{revtex4-1}

\usepackage{xcolor}
\usepackage[T1]{fontenc}
\usepackage{hyperref}

\usepackage{amsthm}
\usepackage{amsmath}
\usepackage[mathscr]{euscript}
\usepackage{bbm}
\usepackage{times}
\usepackage{amsfonts}
\usepackage{amssymb}
\usepackage{thmtools}
\usepackage{physics}

\definecolor{Fabulous}{RGB}{220,0,100}
\definecolor{David}{RGB}{100,80,150}

\def\css{\textup{CSS}}
\def\ad{\mathrm{Ad}}
\def\pos{\mathcal{X}}

\def\h{\mathrm{H}}
\def\p{\mathrm{P}}
\def\cadd{\mathrm{CADD}}
\def\cnot{\mathrm{CNOT}}
\def\cphase{\mathrm{CPHASE}}

\def\stabs{\mathrm{STABS}}
\def\hilb{\mathcal{H}}

\def\center{\mathcal{Z}}

\def\grass{\mathcal{G}} 
\def\orb{\mathcal{O}}
\def\rk{\mathrm{rk}}

\def\u{\mathrm{U}} 
\def\cliff{\mathrm{Cl}}
\def\rcliff{\mathrm{RCl}}
\def\pcliff{\mathrm{PCl}}
\def\dcliff{\mathrm{D}} 
\def\rdcliff{\mathrm{RD}} 
\def\pauli{\mathscr{P}}
\def\rpauli{\mathrm{R}\mathscr{P}}

\def\sp{\mathrm{Sp}}
\def\orth{\mathrm{O}}   
\def\stoch{\mathrm{St}} 
\def\gl{\mathrm{Gl}}
\def\borel{\mathcal{N}} 
\newcommand*{\com}[1]{\mathrm{A}_{#1}}      
\def\semigroup{\mathcal{A}}  
\def\css{\mathrm{CSS}}

\def\normal{\triangleleft}        
\def\normaleq{\trianglelefteq}    

\def\sym{\mathrm{Sym}}  
\def\alt{\mathrm{Alt}}  
\def\bil{\mathrm{Bil}}
\def\q{\mathrm{Q}}      
\def\endom{\mathrm{End}}
\def\hom{\mathrm{Hom}}
\def\ones{\mathbf{1}}

\def\arf{\mathrm{Arf}}
\def\garf{\tilde{\mathrm{Arf}}}
\def\discr{\mathrm{dis}}
\def\iso{\mathrm{Iso}}
\def\levi{\mathrm{Levi}}

\def\range{\mathrm{range}\,}
\def\vspan{\mathrm{span}\,}
\def\supp{\mathrm{supp}\,}
\def\rank{\mathrm{rank}\,}
\def\ann{\mathrm{Ann}}
\def\dim{\mathrm{dim}\,}
\def\irr{\mathrm{Irr}\,}

\def\diag{\mathrm{diag}}
\def\tr{\mathrm{tr}\,}
\def\rad{\mathrm{rad}\,}
\def\div{\,|\,}
\def\DIV{\,\Big{|}\,}
\def\darg{\,\cdot\,}
\def\t{\tilde}

\def\ff{\mathbb{F}}
\def\cc{\mathbb{C}}
\def\rr{\mathbb{R}}
\def\zz{\mathbb{Z}}

\def\nn{\mathbb{N}}
\def\hh{\mathbb{H}}

\def\ii{\mathbbm{1}}
\def\id{\mathrm{id}}

\newcommand*{\ind}[2]{\mathrm{Ind}_{#1}^{#2}\,}
\newcommand*{\res}[1]{\mathrm{Res}_{#1}\,}
\newcommand*{\vect}[1]{\mathrm{vec}(#1)}

\newcommand*{\cm}[1]{\mathcal{C}_{#1}}
\newcommand*{\dm}[1]{\mathcal{D}_{#1}}
\def\cones{C_{\mathbf{1}_t}}
\def\pones{P(\mathbf{1}_t)}
\newcommand*{\spinful}[1]{\mathcal{K}_{#1}}
\newcommand*{\spinless}[1]{\mathcal{L}_{#1}}
\newcommand*{\egal}[1]{\mathcal{M}_{#1}}

\newcommand*{\mcal}{\mathcal}

\newcommand*{\mrm}{\mathrm}
\newcommand*{\mbf}{\mathbf}
\newcommand*{\mfrak}{\mathfrak}

\newtheorem{theorem}{Theorem}[section]

\newtheorem{lemma}[theorem]{Lemma}
\newtheorem{proposition}[theorem]{Proposition}
\newtheorem{conjecture}[theorem]{Conjecture}
\newtheorem{corollary}[theorem]{Corollary}
\newtheorem{remark}[theorem]{Remark}
\newtheorem{example}[theorem]{Example}
\newtheorem{statement}[theorem]{Statement}

\begin{document}

\title{Duality theory for Clifford tensor powers}

\author{Felipe Montealegre-Mora}
\email{fmonteal@thp.uni-koeln.de}

\author{David Gross}
\affiliation{\
Institute for Theoretical Physics, University of Cologne, 50937 Cologne, Germany}


\date{\today}

\begin{abstract}
	The representation theory of the Clifford group is playing an increasingly prominent role in quantum information theory, including in such diverse use cases as the construction of protocols for quantum system certification, quantum simulation, and quantum cryptography.
	In these applications, the tensor powers of the defining representation seem particularly important.
  The representation theory of these tensor powers is understood in two regimes.
	1.\ For odd qudits in the case where the power $t$ is not larger than the number of systems $n$:
	Here, a duality theory between the Clifford group and certain discrete orthogonal groups can be used to make fairly explicit statements about the occurring irreps (this theory is related to \emph{Howe duality} and the \emph{$\eta$-correspondence}).
	2.\ For qubits: Tensor powers up to $t=4$ have been analyzed on a case-by-case basis.
	In this paper, we provide a unified framework for the duality approach that also covers qubit systems.
	To this end, we translate the notion of \emph{rank} of symplectic representations to representations of the qubit Clifford group, and generalize the $\eta$ correspondence between symplectic and orthogonal groups to a correspondence between the Clifford and certain orthogonal-stochastic groups.
  As a sample application, we provide a protocol to efficiently implement the complex conjugate of a black-box Clifford unitary evolution.
\end{abstract}

\maketitle


\section{Introduction and summary of results}

Tensor power representations of the Clifford group $\cliff$,
\begin{align*}
  \cliff \ni U \mapsto U^{\otimes t},
\end{align*}
have played an increasingly prominent role in quantum information theory in the past years. 
Studying these has, for example, lead to efficient protocols for quantum device characterization~\cite{rothrecovering,RBfewsamples,processtomography,kimmel2017phase}, 
quantum state distinction~\cite{kueng2016distinguishing}, 
compressed sensing~\cite{kueng2016low}, 
stabilizerness testing~\cite{gnw}, bounding the stabilizer rank of magic states~\cite{lowrankstab}, 
as well as the construction of complex projective~\cite{gracefully,gnw} and unitary~\cite{homeopathy} $t$-designs.

Tensor power representations with $t\leq4$ are now well understood~\cite{zhu3design,webb2016clifford,kueng2015qubit,gracefully}.
An important characteristic here is that the Clifford group is a \emph{unitary 3-design}, ie.\ the uniform distribution on the Clifford group has the same third order moments as the Haar distribution on the full unitary group.
Moreover, the Clifford group is singled out among unitary designs as the only infinite family of finite groups forming 3-designs~\cite{bannai_t}.
Generalising these results to larger $t$, Ref.~\cite{gnw} gave an explicit description of the commutant of general $t$-th tensor power representations. 
However, that description does not directly give information about irreducible representations.

For qudits, the state of the art is better, though.
There, a duality theory between the Clifford group and certain discrete orthogonal groups can be used to make fairly explicit statements about the occurring irreps  \cite{ghsmall,rank-deficient}.
The results are valid whenever the dimension $d$ of an individual system is an \emph{odd} power of a prime, 
the power $t$ does not exceed the number of subsystems $n$,
and the characteristic $p$ of the finite field $\ff_d$ does not divide $t$.

In this paper, we generalize these results. 
Specifically, our contribution goes beyond previous works on various regards:
\begin{enumerate}
  \item 
		We generalize the main result of Ref.~\cite{rank-deficient} to the case of qubits ($d=2$), and to the case where $t$ is a multiple of $p$.
  \item
  While the main result of Ref.~\cite{rank-deficient} was inspired by Ref.~\cite{gnw}, the proof strategies used in the former were independent of the latter results.
  Thus, the question of whether the proofs of Ref.~\cite{rank-deficient} could be simplified by exploiting directly the main result of Ref.~\cite{gnw} arises.
  Here we answer this question in the affirmative.
  \item
		Refs.~\cite{ghsmall,ghtakagi} develop a \emph{rank theory} for representations of the symplectic group.
		This rank theory is crucial in the generalization of the so-called \emph{Theta correspondence}~\cite{howe89,kashiwara1978segal} between the real orthogonal and symplectic groups, to the scenario over finite fields~\cite{ghsmall,ghtakagi}.
  Here, we extend this theory to a rank theory for Clifford representations.
	Using this, we generalize the results of~\cite{ghsmall} to the Clifford group.
  \item
  Refs.~\cite{nebe_invariants,bannai_invariants,nebe-book,runge96} study the related problem of classifying polynomial invariants of the Clifford group---equivalently, they study the trivial $\cliff$-subrepresentation of $\sym_t(\cc^{d^n})$.
  There, certain self-dual codes span the space of these invariants.
  Here, we show that these invariants correspond to rank-0 subrepresentations of Clifford tensor powers.
  We furthermore show that all subrepresentations with rank $<t$ are contained in the span of certain self-orthogonal codes, of which a subset are the aforementioned self-dual codes.
  In this sense, our result extends Refs.~\cite{runge96,nebe_invariants,bannai_invariants} to non-zero rank representations.
  \item
  In Ref.~\cite{gracefully}, the code space with stabilizer group $\{P^{\otimes 4}\div P \text{ an $n$-qubit Pauli }\}$ was studied in detail.
  In particular, it was found that the fourth tensor power of the \emph{real} Clifford group acts as a permutation representation on this space.
  Here we generalize this result to $t=4k$, $k\geq1$, and show that this representation gives rise to a duality between the real Clifford group and a symplectic group.
  We conjecture that, in analogy to~\cite{ghsmall}, on a given maximal-rank subspace, this duality gives a correspondence between irreps of the real Clifford group and the symplectic group. 
  \item
  As a sample application of our results, we study the problem of using a Clifford black box $U$ to implement $\bar U$ (\emph{Clifford black box conjugation}).
  We provide an optimal parallel inversion protocol for Clifford black boxes and show that the number of black box uses needed for this is $\sim d$.
  This is in stark contrast to the problem of inverting arbitrary unitaries, which requires $d^n-1$ black box uses.
\end{enumerate}

\subsection{Technical relations to prior works}

As should by now be apparent, many results in this paper generalize previous ones.
Here, we describe these interrelations  in a more technical way.

Ref.~\cite{ghsmall} develops a theory of rank for $\sp(2n,d)$ representations.
They show that certain ‘‘maximal rank’’ subrepresentations of tensor powers of the \emph{oscillator representation}\footnote{
	In odd characteristic, the Clifford group decomposes as a semi-direct product between the Weyl-Heisenberg group (``the Paulis'') and a representation of the symplectic group $\sp(2n,d)$.
	The restriction to the symplectic group is known as the \emph{oscillator representation} of $\sp(2n,d)$.
}
set up a correspondence between $\irr\sp(2n,d)$ and $\irr \orth^{\pm}(t,d)$.
The rank theory developed in that reference is analogous to the rank theory in Sec.~\ref{sec:rank}.
In fact, Thm.~\ref{thm:eta duality} is a generalization of the main theorem in~\cite{ghsmall} and their proofs are conceptually very similar.
The analogy between both rank theories is broken in two prominent places.
First, it is broken in a mild way in the qubit case. 
Here, we have not been able to show that the only rank zero representation is the trivial one (compare Lem.~\ref{lem:rank 0} to~\cite[Lem.~1.3.1]{ghsmall}). 
Instead, our results seem to only imply that rank zero representations of the qubit Clifford group are in general $\pm1$ valued.
Second, while~\cite{ghsmall} requires $t\leq n$, we are able to also describe certain low-rank sectors even when $t>n$ (see the relevant condition in Thm.~\ref{thm:eta duality}).

Ref.~\cite{gnw} provides a description for the commutant of Clifford tensor powers. 
This amounts to computing the trivial component of ‘‘balanced’’ tensor powers $U\mapsto U^{\otimes t}\otimes \bar{U}^{\otimes t}$.
Here we build on this result to find a description of the full representation theory of \emph{arbitrary} tensor powers $U\mapsto U^{\otimes r}\otimes \bar{U}^{\otimes s}$.
In this way, while the methods of~\cite{ghsmall} (appropriately generalized) describe the ‘‘maximal rank’’ component of tensor powers, and the methods of~\cite{gnw} describe the ‘‘rank zero’’ component, this paper describes all other rank components.

Our main result, Thm.~\ref{thm:main} is closely analogous to~\cite[Thm.~1.2]{rank-deficient}. 
Three points are important in this regard.
First, that~\cite{rank-deficient} holds over finite fields of characteristic not two, while our results hold for arbitrary prime fields (in particular, our results hold for the qubit case).
We believe that our results can be straightforwardly generalized to arbitrary finite fields but we will not attempt to show this.
Second, Ref.~\cite[Prop.~4.2]{rank-deficient} describes the representation theory of Clifford tensor powers indirectly, heavily relying on its symplectic counter-part~\cite[Thm.~1.2]{rank-deficient}.
Here, our proof of Thm.~\ref{thm:main} directly describes the Clifford representation theory.
This allows us to generalize the aforementioned result to the qubit case and the case where $t=0\mod d$.
Third, even if the statements Thm.~\ref{thm:main} and~\cite[Thm.~1.2]{rank-deficient} are analogous, our proof strategies are considerably different.
Importantly, while in~\cite{rank-deficient} we resorted to probabilistic proof strategies~\cite{alon2004probabilistic}, the proofs presented here are much more constructive.

As mentioned, Sec.~\ref{sec:rcliff} looks at a subrepresentation which generalizes the one studied in~\cite[Sec.~3.4 and App.~B]{gracefully}.
Our emphasis here is different, however.
That reference focuses on singling out the full Clifford action on this space, where they find that it is equivalent to an analog of the oscillator representation of $\sp(2n,2)$.
Here, we focus on the action of the real Clifford group on it, which gives rise to a new correspondence between $\orth^+(2n,2)$ and $\sp(2t',2)$, where $t=2t'$ is even.

Finally,~\cite[Thm.~3]{nwstabnet} provides a list of Clifford tensor powers which lead to an exact correspondence between the Clifford group and an orthogonal group $\orth(\zz_d^t)$.
In Sec.~\ref{sec:exact} we show that this list is essentially complete.

\subsection{Guided tour through the paper}

The main body of this paper is fairly technical. 
To make it more accessible, we sketch the main concepts and results in this self-contained expository section.

We will focus on tensor power representations of the $n$ \emph{qubit} Clifford group,
\begin{align*}
  \Delta_t(U)=U^{\otimes t}, \qquad U\in\cliff, t \leq n.
\end{align*}

The representation $\Delta_t$ acts on the Hilbert space $\hilb_{n,t}:=((\cc^2)^{\otimes n})^{\otimes t}$.
A handy visualization of the computational basis states for this space is as a $t\times n$ grid of qubit basis states,
\begin{align*}
  \begin{matrix}
    \ket{F_{11}} & \otimes & \cdots & \otimes & \ket{F_{n1}}\\
		\otimes 			&	       &        &         & \otimes \\
    \vdots       &         &        &         & \vdots\\
		\otimes 			&	       &        &         & \otimes \\
    \ket{F_{1t}} & \otimes & \cdots & \otimes & \ket{F_{nt}}
  \end{matrix}\,,
  \qquad
  F_{ij}\in\{0,1\}.
\end{align*}
Equivalently, we may gather the coefficients $F_{ij}$ into an $t\times n$ matrix $F\in\zz_2^{t\times n}$ and write the computational basis state as $\ket{F}$.

In our proof methods, the discrete vector space $\zz_2^t$ plays a key role.
A subspace $N\subset\zz_2^t$ is called \emph{isotropic} if the following two conditions hold:
\begin{enumerate}
  \item for every $v\in N$, it holds that $v\cdot v = 0\mod 4$,
  \item for every $v,v'\in N$, it holds that $v\cdot v' = 0\mod 2$.
\end{enumerate}
For such an $N$, we may define a $t$-qubit \emph{Calderbank-Shor-Steane (CSS)} code $\css(N)\subseteq(\cc^2)^{\otimes t}$ as the space spanned by the \emph{coset states} $\ket{[u]_N}$:
\begin{align*}
  \css(N)
  :=
  \vspan\left\{
    \ket{[u]_N}
    :=
    \frac{1}{\sqrt{|N|}}
    \sum_{v\in N} \ket{u+v}
    \DIV
    u\in N^\perp
  \right\},
\end{align*}
where $N^\perp=\{u \div u\cdot v=0\mod 2,\ \forall\ v\in N\}$ is the \emph{orthocomplement} of $N$.
Notice that $N\subseteq N^\perp$ so that all vectors in $\css(N)$ are linear combinations of computational basis states $\ket{u}$ with $u\in N^\perp$.
These CSS codes turn out to be a useful tool towards our goal:

\begin{statement}
  \label{sta:css subrep}
  The tensor power CSS code 
  \begin{align*}
    C_N := \css(N)^{\otimes n} \subset 
    ((\cc^2)^{\otimes t})^{\otimes n}
    \simeq
    \hilb_{n,t}
  \end{align*}
  is invariant under $\Delta_t$.
\end{statement}

By Statement~\ref{sta:css subrep}, the projector $P_N$ onto $C_N$ commutes with the Clifford action,
\begin{align*}
  P_N U^{\otimes t} = U^{\otimes t} P_N.
\end{align*}
Similarly, one may define the action of a certain discrete orthogonal group which commutes with $\Delta_t$.
Namely, consider the group
\begin{align*}
  \stoch(\zz_2^t) 
  =
  \{
  O\in\zz_2^{t\times t}
  \div
  (Ov)\cdot(Ov)=v\cdot v \mod 4,\ \forall\ v\in \zz_2^t
  \}.
\end{align*}
(The choice of notation ‘‘$\stoch$’’ will become clear in the main text.)
This group acts on $\hilb_{n,t}$ through the following representation $R$,
\begin{align*}
  R(O)\ket{F}=\ket{OF},
\end{align*}
i.e., it permutes the standard basis.
This action of $\stoch(\zz_2^t)$ commutes with the Clifford group.
Moreover, Ref.~\cite{gnw} shows that the \emph{commutant} $\com{t}$ of $\Delta_t$ (i.e.\ the space of matrices that commute with the Clifford tensor power action) is generated by taking products and linear combinations of the projectors $P_N$ and orthogonal operators $R(O)$.

The basis of $C_N$ consists of $n$-fold tensor products of coset states which we denote as
\begin{align*}
  \ket{[F]_N}:=\ket{[f_1]_N}\otimes\cdots\otimes\ket{[f_n]_N}, 
  \qquad
  f_i\in N^\perp,
\end{align*}
where $F\in\zz_2^{t\times  n}$ has as columns the vectors $f_i$ (i.e. $F_{ij}=(f_j)_i$).
We will refer to the states $\ket{[F]_N}$ simply as the \emph{coset basis of $C_N$}.
One may show that, for any $\ket{[F]_N}\in C_N$ and any $O\in\stoch(\zz_2^t)$,
\begin{align}
  \label{eq:stoch permutes}
    R(O)\ket{[F]_N}
    =
    \ket{[OF]_{ON}},
\end{align}
so the action of $\stoch(\zz_2^t)$ permutes the codes as $R(O)C_N = C_{ON}$.

By Eq.~\eqref{eq:stoch permutes}, the subgroup which preserves $N$,
\begin{align*}
  \stoch(\zz_2^t)^N :=
  \{
   O\in\stoch(\zz_2^t) \div ON=N
  \},
\end{align*}
acts on $C_N$.
It turns out that, for our purposes, it is useful to consider the restriction of this action to a subgroup $G_N\subset\stoch(\zz_2^t)^N$.
In the main text this group arises in a rather canonical way from the action  of  $\stoch(\zz_2^t)^N$  on $C_N$.
However, here we introduce it in a non-canonical manner to simplify the dicussion.
Let $K_N$ be a subspace of $\zz_2^t$ which complements $N^\perp$, i.e.\ a subspace for which $\zz_2^t= K_N\oplus N^\perp$.
Similarly, let $T_N$ be such that $N^\perp = N\oplus T_N$.
In total, we have the following decomposition:
\begin{align*}
  \zz_2^t = K_N \oplus  \overbrace{N \oplus T_N}^{N^\perp}.
\end{align*}
Then $G_N$ is the subgroup of $\stoch(\zz_2^t)^N$ of matrices which act trivially on $K_N\oplus  N$, that is, matrices of the form
\begin{align*}
  O
  =
  \begin{pmatrix}
    \ii_{K_N} &       &\\
              & \ii_N &\\
              &       &  O|_{T_N}
  \end{pmatrix}
  \in
  \stoch(\zz_2^t)^N.
\end{align*}
As one would expect, this is the symmetry group of the mod 4 dot product restricted to $T_N$.
Namely, letting $O_T:=O|_{T_N}$ for $O\in G_N$, it holds that
\begin{align*}
  (O_Tv_T)\cdot(O_Tv_T)=v_T\cdot v_T \mod 4,\qquad \forall\ v_T\in T_N.
\end{align*}

The code spaces $C_N$ obey certain inclusion relations: if $N\subset N'$, then their corresponding codes obey the opposite inclusion $C_{N'}\subset C_N$.
This inclusion is strict: the state
\begin{align}
  \label{eq:ket F}
  \ket{[F]_N}
  \in C_N,
  \quad
  \text{s.t. }\rank  F = \dim N^\perp,
\end{align}
is orthogonal to any $C_{N'}$ with $N\subset N'$.

Our proof relies -- more generally -- on singling out the subspace of $C_N$ which is orthogonal to all $C_{N'}$ with $\dim  N'  >  \dim N$. 
To this end, we use a  notion of \emph{rank}.

The coset states $\ket{[F]_N}$ given in Eq.~\eqref{eq:ket F} are of ‘‘maximal rank’’ within $C_N$ in the sense that for any $\ket{[F']_N}\in C_N$, it holds that $\rank F'\leq\dim N^\perp=\rank F$.
This follows because, by definition, the columns of $F'$ are contained in $N^\perp$.

\begin{example}
  \label{ex:max rank}
	In the case of the trivial isotropic subspace $N=\{0\}$ (henceforth written as $N=0$), the corresponding code is all of Hilbert space, $C_{N=0}=\hilb_{n,t}$.
  Moreover, the coset states are the computational basis, $\ket{[F]_{N=0}}=\ket{F}$.
  Consider an  $F$ with rank $t$.
  Then, by  the arguments above, $\ket{F}$ is of maximal rank within $\hilb_{n,t}$ and, thus, is orthogonal to \emph{all} non-trivial codes $C_{N'}$ (i.e., such that $N'\neq0$).
\end{example}

Let us consider the subrepresentation that is generated by acting with the Clifford group on the set of maximal rank coset states:
\begin{align}
  \label{eq:code max}
  C_N^{\max}
  :=
  \vspan\big\{
    U^{\otimes  t}\ket{[F]_N}
    \;\big|\;
    U\in\cliff,\
		\ket{[F]_N}\in  C_N\
    \text{s.t. column space of } F =  N^\perp
  \big\} \subseteq C_N.
\end{align}

In the core of our proof, Sec.~\ref{sec:main theorem}, we show that the spaces $C_N^{\max}$ are linearly independent and span all of $\hilb_{n,t}$.
This reduces the problem of decomposing $\hilb_{n,t}$ to the problem of decomposing each of the subrepresentations $C_N^{\max}$.

\begin{statement}
  \label{sta:cmax span}
  One may decompose
  \begin{align*}
    \hilb_{n,t}
    =
    \bigoplus_N C_N^{\max}
  \end{align*}
  as a Clifford representation.
  Here, the sum ranges over all isotropic subspaces $N\subset\zz_2^t$.
\end{statement}

As previously  mentioned, the space $C_N^{\max}$ is invariant  under  $\stoch(\zz_2^t)^N$  and, in particular,  also under its subgroup  $G_N\subset\stoch(\zz_2^t)^N$.
In Sec.~\ref{sec:rank and duality} we decompose $C_N^{\max}$ by obtaining a duality between $G_N$ and $\cliff$ arising from their action on $C_N^{\max}$.
In particular, the Clifford characters in $C_N^{\max}$ are labeled by irreps of $G_N$.
We summarize this result as:

\begin{statement}
  \label{sta:code duality}
  Let $N$ be an isotropic subspace and $C_N^{\max}$ be as above.
  Then:
  \begin{enumerate}
    \item
    \emph{(Compression of the commutant to $C_N^{\max}$.)}
    Let $P(C_N^{\max})$ be the projector onto $C_N^{\max}$.
    Then, it holds that the compression of the commutant $\com{t}$ of $\Delta_t$ to this space, i.e.\ $P(C_N^{\max}) \com{t} P(C_N^{\max})$, is spanned by 
    \begin{align*}
      \{P(C_N^{\max})R(O) \div O \in G_N\}.
    \end{align*}
    That is, the action of $G_N$ on $C_N^{\max}$ spans the commutant of the action  of  $\cliff$ on $C_N^{\max}$.
    \item
    \emph{(Decomposition of the representation.)}
    The decomposition of $C_N^{\max}$ is, as a representation of $G_N\times\cliff$, given by
    \begin{align*}
      C_N^{\max}\simeq
      \bigoplus_{\tau \in \irr G_N} \tau \otimes \eta_\tau,
    \end{align*}
    where the sum is over all irreps $\tau$ of $G_N$, and where $\{\eta_\tau\}_\tau$ is a set of pairwise inequivalent irreps of the Clifford group.
  \end{enumerate}
\end{statement}

Finally, some of the spaces $\{C_N^{\max}\}$ are equivalent as Clifford representations.
By Eq.~\eqref{eq:stoch permutes}, it follows that for any $O\in\stoch(\zz_2^t)$, the action  of $R(O)$,
\begin{align}
  \label{eq:rep isomorphism}
  R(O)C_N^{\max} = C_{ON}^{\max},
\end{align}
permutes these representation spaces.
Since $R(O)$ commutes with  the  Clifford action, Eq.~\eqref{eq:rep isomorphism} is an isomorphism of Clifford representations by Schur's lemma.
That is, $C_N^{\max}$ and $C_{ON}^{\max}$ are equivalent as Clifford representations.
Moreover, in Sec.~\ref{sec:main theorem} we show the converse.
Namely, that if $N'$ is not of the form $ON$ for some $O\in\stoch(\zz_2^t)$, then  $C_{N'}^{\max} \not\simeq C_N^{\max}$.
In fact we show a  stronger result: that in the latter case, $C_N^{\max}$ and $C_{N'}^{\max}$ have no Clifford irreps in common.
In  other words, if $\chi_N$ is the character of $C_N^{\max}$ as a Clifford representation, then
\begin{align*}
  \langle \chi_N,\ \chi_{N'}\rangle_\cliff = 0.
\end{align*}

To understand these equivalences better, in Sec.~\ref{sec:quad} we characterize the orbits of $\stoch(\zz_2^t)$ on the set of isotropic subspaces.
These orbits are classified by two invariants:
\emph{1.} the dimension $m=\dim N$, and \emph{2.} whether the all-ones vector ($\ones_t:=(1,\dots,1)^T$) is an element of $N$ or not.
(The second invariant can only be non-trivial when $t$ is a multiple of 4, as only in that case does it hold that $\ones_t\cdot\ones_t=0\mod 4$.) 
For example, the following pair of isotropic subspaces of $\zz_2^8$ are in the same orbit:
\begin{align*}
  N_1 &= \{(00000000),  (11110000)\},\\
  N_2 &= \{(00000000),  (00001111)\},
\end{align*}
since they're  both  one-dimensional and neither contains the  all-ones vector.
On the  other hand, the following pair are  \emph{not} in the same orbit:
\begin{align*}
  N_1 &= \{(00000000),  (11110000)\},\\
  N_2 &= \{(00000000),  (11111111)\}.
\end{align*}
The reason for  the  appearance of a seemingly arbitrary vector as an invariant of these orbits is that it is in fact \emph{not} arbitrary  at all---in fact it is preserved by  $\stoch(\zz_2^t)$,
\begin{align*}
  O\ones_t = \ones_t, \qquad \forall\  O\in\stoch(\zz_2^t).
\end{align*}

Let $\grass_{m,i}$, with $i\in\{0,1\}$, be the orbit of isotropic subspaces of dimension $m$ which contain the all-ones vector if $i=0$:
\begin{align*}
  \grass_{m,0} 
  &:= 
  \{ N\subset\zz_2^t \text{ isotr. subsp.} 
  \div 
  \dim N=m,\ \ones_t\in N
  \},\\
  \grass_{m,1} 
  &:= 
  \{ N\subset\zz_2^t \text{ isotr. subsp.} 
  \div 
  \dim N=m,\ \ones_t\notin N
  \}.
\end{align*}
Then, by the discussion  above,
\begin{align*}
  \egal{m,i}
  :=
  \bigoplus_{ N\in\grass_{m,i} } C_{N}^{\max}
  \simeq
  \bigoplus_{\tau \in \irr G_N} \tau\otimes\eta_\tau \otimes \cc^{|\grass_{m,i}|}.
\end{align*}
Our main result may be summarized as the previous equation together with
\begin{align*}
  \hilb_{n,t}=\bigoplus_{m,i}\egal{m,i}.
\end{align*}

\subsection{Explicit decomposition for $t=5$}
\label{sec:example}

To showcase our result, we provide an explicit decomposition for the fifth tensor power representation of the qubit Clifford group.

The space $\zz_2^5$ only has five non-trivial isotropic spaces, all of them one-dimensional: $N_i = \langle \bar e_i:=\ones_5+e_i\rangle$, where $\{e_i\}_i$ is the standard basis.
Moreover, all of these isotropic subspaces  are in a single  $\stoch(\zz_2^5)$ orbit.
By the arguments in Sec.~\ref{sec:main theorem}, it holds that $C_{N_i}=C_{N_i}^{\max}$.
By Statement~\ref{sta:cmax span}, then,
\begin{align*}
  \hilb_{n,5}
  \simeq
  C_{N=0}^{\max}
  \oplus
  (C_{N_1}\otimes\cc^5).
\end{align*}

Now, $\stoch(\zz_2^5)\simeq S_5$ so that
\begin{align*}
  C_{N=0}^{\max}
  \simeq
  \bigoplus_{\lambda\vdash 5} \lambda\otimes\eta_\lambda,
\end{align*}
where the sum runs over Young diagrams of size 5 and $\eta_\lambda$ are rank 5 Clifford representations (where ‘‘rank’’ is defined as in Sec.~\ref{sec:rank}).
To obtain an explicit form of the projector onto the isotypic components, we need to first obtain the projector onto $C_{N=0}^{\max}=\vspan\{C_{N_1},\dots,C_{N_5}\}^\perp$.
While these code spaces are linearly independent (by Statement~\ref{sta:cmax span}), they are not orthogonal to one another.
Specifically, the set of coset states, 
\begin{align*}
  \{\ket{[F_i]_{N_i}} \div i=1,\dots,5,\ \ket{[F_i]_{N_i}}\in C_{N_i}\},
\end{align*}
forms a basis but not an \emph{orthogonal} basis for $\vspan\{C_{N_1},\dots,C_{N_5}\}$.
(Note, however, that the set of coset states \emph{on any given code} $C_{N_i}$ are an orthonormal basis for that code).
One may painstakingly -- but straightforwardly -- obtain an orthonormal basis for this space using the Gram matrix for the set of coset states.
This allows one to construct a projector $P(\langle C_{N_1},\dots,C_{N_5}\rangle)$ onto the span of the CSS codespaces.
Thus the projector onto the isotypic component of $\eta_\lambda$ is proportional to
\begin{align*}
  (\ii-P(\langle C_{N_1},\dots,C_{N_5}\rangle)) \sum_{O\in S_5} \chi_\lambda^*(O)R(O).
\end{align*}

\begin{remark}
  \label{rem:gram matrix}
  For completeness, we include the overlap of coset states $\ket{[F_i]_{N_i}}\in C_{N_i}$, $\ket{[F_j]_{N_j}}\in C_{N_j}$ with $i\neq j$.
  With this information one may construct the Gram matrix of the set of CSS coset states and thus an orthonormal basis of $\langle C_{N_1},\dots,C_{N_5}\rangle$.
  
  Let $E_{ij}:=\ii+\bar e_i\bar e_j^T$.
  Then, the columns of $E_{ij}F_i$ are in $N_j^\perp$ so that $\ket{[E_{ij}F_i]_{N_j}}\in C_{N_j}$.
  Moreover,
  \begin{align*}
    \braket{[F_i]_{N_i}}{[F_j]_{N_j}}
    =
		2^{-n}
    \delta_{[F_j]_{N_j},[E_{ij}F_i]_{N_j}}.
  \end{align*}
\end{remark}
\begin{remark}
	\label{rem:approx  projector}
	Rem.~\ref{rem:gram matrix} implies that for large $n$, the operator $Q:=\sum_i P_{N_i}$
	gives a good approximation of $P(\langle C_{N_1},\dots,C_{N_5}\rangle)$ in operator norm.
	
	Specifically, for $i,j,k\in\{1,\dots,5\}$ all distinct and $\ket{\psi_i}\in C_{N_i}$ with norm $c_i:=\braket{\psi_i}$, it holds that 
	\begin{align*}
		\bra{\psi_i}P_{N_j}\ket{\psi_i} &\leq 4^{-n}c_i,\\
		|\bra{\psi_i}P_{N_j}\ket{\psi_j}| &\leq 2^{-n}c_i,\\
		|\bra{\psi_i}P_{N_j}\ket{\psi_k}| &\leq 4^{-n}\max\{c_i,c_k\}.
	\end{align*}
	It follows that for any normalized vector $\ket{\psi}=\sum_i\ket{\psi_i}\in\langle C_{N_1},\dots,C_{N_5}\rangle$,
	\begin{align*}
		\bra{\psi}Q\ket{\psi} 
		&= 
		\sum_{i,j,k} \bra{\psi_i}P_{N_j}\ket{\psi_k}\\
		&\leq
		1 +
		5\times (4^{-n}+2^{-n})
		+
		2\times{5 \choose 2} 4^{-n}
		=: 1+\epsilon(n).
	\end{align*}
	This implies  that
	\begin{align*}
		P(\langle C_{N_1},\dots,C_{N_5}\rangle) 
		\leq
		Q
		\leq
		P(\langle C_{N_1},\dots,C_{N_5}\rangle)+\epsilon(n)\ii
	\end{align*}
	and thus, asymptotically, that
	\begin{align*}
		\|Q-P(\langle C_{N_1},\dots,C_{N_5}\rangle)\|_\infty \in o(\exp(-n)).
	\end{align*}
\end{remark}

Moving on to the lower rank sector, let us focus on the code spaces $C_{N_i}$.
By Lem.~\ref{lem:restriction to css},
\begin{align*}
  \Delta_5|_{C_{N_i}}\simeq\bar\Delta_3.
\end{align*} 
In particular,
\begin{align*}
  C_{N_i}
  \simeq
  \bigoplus_{\lambda \vdash 3} \lambda\otimes\eta_\lambda.
\end{align*}
In particular, the action of $\Delta_5$ on  $C_{N_i}$ is a unitary 3-design relative to $\u(C_{N_i})$.
Moerover: 
\begin{enumerate}
  \item
  $\eta_{\lambda=(3)}$ is isomorphic to the action of $\bar\Delta_3$ on symmetric vectors of $\hilb_{n,3}$,
  \item
  $\eta_{\lambda=(1,1,1)}$ is isomorphic to the action of $\bar\Delta_3$ on skew-symmetric vectors of $\hilb_{n,3}$,
  \item
  $\eta_{\lambda=(2,1)}$ is isomorphic to the action of $\bar\Delta_3$ on the space $\ker(\ii+\pi_{\text{cycl.}}+\pi_{\text{cycl.}}^2)$, where
  \begin{align*}
    \endom(\hilb_{n,3})\ni \pi_{\text{cycl.}}: 
    \ket{a}\otimes\ket{b}\otimes\ket{c}
    =
    \ket{b}\otimes\ket{c}\otimes\ket{a}.
  \end{align*}
\end{enumerate}

In order to obtain the projector onto  each component, we will to characterize more explicitly the subgroup $G_{N_i}\simeq S_3$.
For concreteness, consider $i=1$. 
In this case,  we may choose
\begin{align*}
  T_{N_1} = \vspan\{00011,\ 00101,\ 10000\}.
\end{align*}
Then, $G_{N_1}$ fixes $10000$ as this is the only vector with support of  size 1 modulo 4 in $T_{N_1}$.
Moreover, one may see that $G_{N_1}$ acts as an arbitrary invertible matrix on the space $\vspan\{00011,\ 00101\}$ and therefore $G_{N_1}\simeq\gl(\zz_2^2)\simeq S_3$.
That is,  $G_{N_1}$ is the set of matrices of the following form
\begin{align*}
  \begin{pmatrix}
    1 &   & \\
      & 1 & \\
      &   & \pi
  \end{pmatrix}
  \in \stoch(\zz_2^5),
  \qquad
  \pi\in S_3\subset\zz_2^{3\times 3}.
\end{align*}
Then, the $\lambda\otimes\eta_\lambda$ component in $C_{N_1}$ has a projector proportional to
\begin{align*}
  \sum_{\pi\in S_3} \lambda^*(\pi)
  R\begin{pmatrix}
    1 & & \\ & 1& \\ & &  \pi
  \end{pmatrix}
  P(N_1).
\end{align*}

All in all, the following decomposition holds:
\begin{align*}
  \hilb_{n,5}
  \simeq
  \left( \bigoplus_{\lambda\vdash 5} \lambda\otimes\eta_\lambda \right)
  \oplus
  \left( \bigoplus_{\lambda\vdash 3} \lambda\otimes\eta_\lambda \right)\otimes\cc^5.
\end{align*}

\section{Preliminaries}
\label{sec:prelim}

Here we summarize some relevant results found in the literature.
Being encyclopedic, we defer the proofs of claims in this section to App.~\ref{app:proofs prelim}. 

\subsection{Representation Theory}
\label{sec:rep theory}

A representation of a finite group $G$ is a matrix-valued function $\rho:G\to\ff^{k\times k}$ for which
\begin{align*}
  \rho(g_1)\rho(g_2)=\rho(g_1g_2),
\end{align*}
where $\ff$ is an arbitrary field.
The space of functions $f:\ff^k\to S$, where $S$ is a set, is canonically a $G$-representation through
\begin{align}
  \label{eq:inherited rep}
  gf(\darg) := f(\rho(g^{-1})\darg).
\end{align}

The representations with which we deal with here are usually complex, $\ff=\cc$, and unitary, $\range\rho\subset \mrm{U}(\cc^k)$.
The \emph{representation space} of $\rho$ is the space on which it acts, namely $\cc^k$.
A subspace $\mcal{V}\subseteq\cc^k$ is called a \emph{$G$-invariant space}, or simply an invariant space, if $\rho(G)\mcal{V}=\mcal{V}$.
A representation with no non-trivial invariant subspaces is \emph{irreducible}, the set of all such representations is $\irr G$.
If $\rho$ is \emph{reducible}, there is a $U\in\mrm{U}(\cc^k)$ for which
\begin{align*}
  U\rho(g)U^\dagger = \rho_1(g)\oplus\cdots\oplus\rho_m(g),
  \quad
  \forall\ g\in G,
\end{align*}
where $\rho_i$ are irreducible representations.
Equivalently, we may decompose the representation space into invariant subspaces
\begin{align*}
  \cc^k = \bigoplus_i \mcal{V}_i,
\end{align*}
where $\rho(g)\mcal{V}_i=\rho_i(g)\mcal{V}_i$.
We summarize this into the equation $\rho\simeq\oplus_i\rho_i$, where $\simeq$ denotes isomorphism between representations.

Let $H\normaleq G$ and $G'=G/H$. 
Any representation $\rho$ of $G'$ can be extended to a representation of $G$ which has $H$ in its kernel.
For simplicity, we will also call this representation $\rho$.
It follows that there is a canonical embedding $\irr G'\subseteq\irr G$.

The \emph{regular representation} of $G$ is the space $\cc[G]$ of formal linear combinations of group elements, with $G$ acting from the left as
\begin{align*}
  g\ket{g'}=\ket{gg'}.
\end{align*}
The regular representation decomposes as
\begin{align*}
  \cc[G] \simeq \bigoplus_{\tau \in \irr G} \tau\otimes\cc^{\dim\tau},
\end{align*}
where the right factors are multiplicity spaces.
The commutant of the regular representation is spanned by the right-action of $G$,
\begin{align*}
  g:\ket{g'}\mapsto\ket{g'g^{-1}}.
\end{align*}

Suppose $G$ acts freely and transitively on a set $S$ (that is, no non-trivial element acts trivially and the action has a single orbit).
Then $|S|=|G|$ and the permutation action obtained on the space $\cc[S]$ is isomorphic to the regular representation.

\subsection{The Pauli and Clifford groups}
\label{sec:pauli and clifford}

Consider $n$ qudits, with Hilbert space $\hilb_n:=\left(\cc^d\right)^{\otimes n}$ and standard basis states $\ket{x}$ where $x \in \zz_d^n =: \pos$. 
The main results of this paper hold only in the case where $d$ is a prime, however the presentation within this subsection holds for arbitrary $d$.

Let the \emph{shift} and \emph{clock} operators be
\begin{align*}
  X(q)\ket{x} = \ket{x+q} \qquad Z(p) \ket{x} = \omega^{p\cdot x}\ket{x}, \qquad
  p,q\in\pos,
\end{align*}
where $\omega = \exp(2\pi i/d)$.
Because the construction of the Pauli and Clifford groups differs slightly depending on whether $d$ is even or odd, it is convenient to introduce $\tau = (-1)^d\exp(i\pi/d)$, and let $D$ be the order of $\tau$. 
Notice that $\tau^2 =\omega$, so $d=D$ if $d$ is odd, and $D=2d$ if $d$ is even.
Then, the \emph{Weyl operators} (also known as \emph{displacement operators}) are
\begin{align*}
  W_v := \tau^{-(v_z\cdot v_x\ \mrm{mod}\ d)} Z(v_z)X(v_x),
\end{align*}
where $v=(v_z,v_x)\in\zz_d^{2n}:= V$.
The $n$-qudit Pauli group $\pauli$ is generated by the displacement operators and $\tau\ii$.
The Pauli group is also known as the finite Heisenberg group.
When $d=p$ is a prime, it moreover is one of the two extra-special $p$-groups.
Its defining representation, known as the \emph{Weyl representation}, is denoted $W$.

Each $m\in\zz_d^\times:=\zz_d\setminus\{0\}$ defines an inequivalent representation $W^{(m)}$ of $\pauli$ acting on $\hilb_n$, through
\begin{align*}
  W^{(m)}(\tau\ii) = \tau^m\ii, \qquad
  W^{(m)}_v = W_{(mv_z,v_x)}.
\end{align*}
All non-Abelian irreps arise this way, a fact sometimes known as the \emph{Stone-von Neumann Theorem}.
One may verify that
\begin{align*}
  W^{(m)}_vW^{(m)}_u = \tau^{m[v,u]}W^{(m)}_{v+u},
\end{align*}
where $[v,u] = v_z\cdot u_x - v_x\cdot u_z$ is a symplectic form on $V$, which directly implies
\begin{align*}
  W^{(m)}_vW^{(m)}_u = \omega^{m[v,u]}W^{(m)}_uW^{(m)}_v.
\end{align*}

The Clifford group~\cite{farinholt2014ideal,gnw,hostens2005stabilizer,nielsen2002universal}, denoted $\cliff$, is the subgroup of $\u(\hilb_n)$ generated by: 
the discrete Fourier transform (also known as the \emph{Schur matrix} or,  when $d=2$ the \emph{Hadamard gate}) on any qudit,
\begin{align}  
	\h = \frac{1}{\sqrt{d}} \sum_{x,y \in \zz_d}\omega^{xy} \ket{x}\bra{y},
\end{align}
the phase gate acting on any qudit,
\begin{align}
  \p = \sum_{x\in \zz_d} \tau^{x^2}\ket{x}\bra{x}\quad(d \text{ even}),
  \quad
  \p &= \sum_{x\in \zz_d} \tau^{x(x-1)}\ket{x}\bra{x}\quad(d \text{ odd}),
\end{align}
and the controlled addition acting on any pair of qudits,
\begin{align}
  \cadd = \sum_{x,y}\ket{x,x+y}\bra{x,y}.
\end{align}
When $d=2$ we use, alternatively, the standard notation $\cnot:=\cadd$.

The Pauli group is a normal subgroup of the Clifford group group, $\pauli\normal\cliff$.
Furthermore every character-preserving automorphism of the Pauli group can be realized by conjugating with some Clifford matrix, and in particular 
\begin{align*}
  \cliff/(\center(\cliff)\pauli)\simeq\sp(V),
\end{align*}
where $\center(\cliff)$ is the \emph{center} of the Clifford group.
Each automorphism corresponds to a coset $\center(\cliff)U\subset\cliff$, where $U\in\cliff$.
In the odd $d$ case, $\center(\cliff)=\langle\omega\ii\rangle\simeq\zz_p$, while in the qubit case $\center(\cliff)=\langle\omega_8\ii\rangle\simeq\zz_8$ whenever $n\geq2$~\cite{nebe_invariants}, where $\omega_8=\exp(2\pi i/8)$.%
\footnote{For $n=1$ the center is $\langle i\ii\rangle\simeq\zz_4$.
Furthermore, a careful choice of the phase in front of the $\h$ generator can make the center be $\zz_4$ for larger $n$ as well~\cite{gracefully}.}
If $d$ is \emph{odd} it further holds that $\cliff\simeq \pauli\rtimes\sp(V)$ and any $U\in\cliff$ can be expressed~\cite{grosshudson} as
\begin{align*}
  U = e^{i\varphi} W(v)\mu(S),
\end{align*}
where $\varphi\in\rr$ and $\mu$ is a unitary representation of the symplectic group $\sp(V)$ known as the \emph{oscillator representation} or the \emph{Weil representation}.
The \emph{projective Clifford group} is $\pcliff:=\cliff/\center(\cliff)$. 
If $d$ is odd, one can see that $\pcliff\simeq V\rtimes\sp(V)$ is the affine symplectic group.
In the qubit case, $d=2$, this separation of $\cliff$ and $\pcliff$ into a semi-direct product ceases to hold~\cite{zhuWigner}.

The oscillator representation satisfies
\begin{align*}
  \mu(S)W(v)\mu(S)^\dagger = W(Sv).
\end{align*}
One may similarly define the oscillator representation with mass $m$, denoted by $\mu^{(m)}$, as the unique representation of $\sp(V)$ satisfying
\begin{align}
  \label{eq:consistent}
  \mu^{(m)}(S)W^{(m)}(v)\mu^{(m)}(S)^\dagger = W^{(m)}(Sv).
\end{align}
It turns out that $\mu^{(m)}$ only depends on whether $m$ is a square or not in $\zz_d$~\cite{ghsmall}.
In the former case, $\mu^{(m)}=\mu$, while if $m$ is a non-square then $\mu^{(m)}=\bar\mu$.
Finally, by~\eqref{eq:consistent}, it follows that $W^{(m)}$ and $\mu^{(m)}$ generate a representation of $\cliff$ which we will denote by $\cliff_{(m)}$.

In previous studies of the oscillator representation~\cite{ghsmall,ghtakagi,rank-deficient} the following subgroup of $\sp(V)$ has figured prominently,
\begin{align} 
  \label{eq:borel}
  \borel := 
  \Big\{
    N_B=
    \begin{pmatrix}\ii_n & B\\ 0 & \ii_n\end{pmatrix} \,\Big|\, B=B^T \in \zz_d^{n\times n}
  \Big\}.
\end{align} 
One may verify~\cite{ghsmall} that
\begin{align*}
  \mu(N_B)
  &=
  \sum_{x\in\pos} \tau^{2^{-1}x^T B x} \ketbra{x}.\\
\end{align*}

\subsection{Quadratic and bilinear forms}
\label{sec:quad}

Here we review some general results on quadratic and bilinear forms over finite fields.
There are several equivalent ways of thinking of quadratic and bilinear forms, here we will use three: as functions, as tensors, and as matrices.
In the odd $d$ case, as in the case of real-valued forms, these three points of view are straightforwardly connected to each other.
The $d=2$ case, however, requires some care.

A bilinear form on a $k$-dimensional space $K\simeq\zz_d^k$ is a bilinear function $\beta:K\times K\to \zz_d$. 
The space of such forms is denoted by $\bil(K)$.
The subspace of symmetric forms, i.e.\ forms satisfying $\beta(u,v)=\beta(v,u)$ for all $u,v\in K$, is denoted by $\sym(K)$.
Similarly, the subspace of alternating forms, i.e.\ those satisfying $\beta(u,u)=0$ for all $u$, is denoted by $\alt(K)$.

There is a canonical isomorphism $\bil(K)\simeq K^*\otimes K^*$ 
associating a bilinear form $\beta$ to the tensor $\phi$ satisfying
\begin{align*}
  \beta(u,v) = \phi(u\otimes v),\quad \forall\ u,v \in K.
\end{align*}
Here, we will refer to $\phi$ as the \emph{tensor representation} of $\beta$, and conversely to $\beta$ as the \emph{functional representation} of $\phi$.

We may equivalently find a \emph{matrix representation} $M_\beta$ of $\beta$ with respect to a basis $\{e_i\}\subset K$ given by
\begin{align*}
  (M_\beta)_{ij} = \beta(e_i,e_j),
\end{align*}
which determines $\beta$ uniquely.
A matrix $M$ is \emph{symmetric} if $M_{ij}=M_{ji}$,
\emph{anti-symmetric} if $M_{ij}=-M_{ji}$,
and \emph{alternating} if it is anti-symmetric and $M_{ii}=0$.
In odd characteristic, any anti-symmetric matrix is alternating, but in characteristic $2$, the latter condition is strictly stronger.
The matrix $M_\beta$ is symmetric if and only if $\beta\in\sym(K)$
and alternating matrix if and only if $\beta\in\alt(K)$.

The following proposition translates these statements to tensor language.
For a subset $S \subset K$, let 
\begin{align*}
  \ann(S)=\{\varphi\in K^* \div \varphi(s) =0,\ \forall\ s\in S\}
\end{align*}
be the associated \emph{annihilator}, i.e.\ the space of functions vanishing on $S$.

\begin{proposition}
  \label{prop:sym alt tensors}
  Let $\pi$ be the map which permutes the two tensor factors in $K\otimes K$, moreover let $(K\otimes K)^\pi$ and $(K^*\otimes K^*)^\pi$ be the set of tensors invariant under this action.
  Then, the canonical isomorphism $\bil(K)\simeq K^*\otimes K^*$ restricts to
  \begin{align*}
    \sym(K) \simeq (K^*\otimes K^*)^\pi,
    \quad
    \alt(K) \simeq \ann((K\otimes K)^\pi).
  \end{align*}
\end{proposition}

A \emph{quadratic form} is a function $q:K\to\zz_d$ which satisfies $q(au)=a^2 q(u)$ for all $a\in\ff$, and that the function $\Xi(q):K\times K\to\zz_d$ defined through
\begin{align*}
  \Xi(q)(u,v):=
  q(u+v)-q(u)-q(v)
\end{align*}
is a symmetric bilinear form.
The space of quadratic forms over $K$ is denoted by $\q(K)$.
The form $\Xi(q)$ is the \emph{polarization of $q$}, and conversely, $q$ is a \emph{quadratic refinement of $\Xi(q)$}.

The space of quadratic forms is canonically isomorphic to $((K\otimes K)^\pi)^*$, where the isomorphism associates to $q\in\q(K)$ the functional $\phi\in((K\otimes K)^\pi)^*$ satisfying
\begin{align*}
  \phi(u\otimes u) = q(u), \quad \forall\ x\in K.
\end{align*}
We say that $\phi$ is the \emph{tensor polarization of $q$}.

The spaces $\bil(K)$ and $\t\q(K)$ are naturally $\gl(K)$ representations through
\begin{align}
  gq(\darg) &= q(g^{-1}\darg),\label{eq:q gl}\\
  g\beta(\darg,\darg) &= \beta(g^{-1}\darg,g^{-1}\darg).\label{eq:sym gl}
\end{align}
By Prop.~\ref{prop:sym alt tensors}, we conclude:

\begin{corollary}
  \label{cor:sym q duality}
  There are canonic isomorphisms $\sym(K^*)\simeq(K\otimes K)^\pi\simeq\q(K)^*$ as $\gl(K)$ spaces.
\end{corollary}

A matrix representation of $q$ with respect to $\{e_i\}$ is a matrix $M$ satisfying
\begin{align*}
  (u_1,\dots,u_k) M(u_1,\dots,u_k)^T = q(u),\quad\forall\ u\in K,
\end{align*}
where $u=\sum_i u_ie_i$.
Matrix representations of quadratic forms are not uniquely determined by the basis.
More precisely, two matrices $M, M'$ represent $q$ exactly if they difference $A=M-M'$ is alternating.

When $d$ is odd, we may however find a unique \emph{symmetric} matrix representation of $q$, namely $2^{-1}(M+M^T)$ where $M$ is an arbitrary matrix representation of $q$.
This works, because in this case, the space of matrices is the direct sum of the symmetric and the anti-symmetric ones.
If $d=2$, on the other hand, alternating matrices form a subset of symmetric matrices.
This has two implications:
First, some quadratic forms have no symmetric matrix representation.
An example of such a form is $q(u)=u_1u_2$, with representations
\begin{align*}
  \begin{pmatrix}
    0 & 1 \\ 0 & 0
  \end{pmatrix},
  \qquad
  \begin{pmatrix}
    0 & 0 \\ 1 & 0
  \end{pmatrix}.
\end{align*}
Second, when quadratic forms have a symmetrix matrix representation, it is not unique.
For example the form $q(u)=u_1^2+u_2^2$ has the two following symmetric representations
\begin{align*}
  \begin{pmatrix}
    1 & 0 \\ 0 & 1
  \end{pmatrix},
  \qquad
  \begin{pmatrix}
    1 & 1 \\ 1 & 1
  \end{pmatrix}.
\end{align*}

Cor.~\ref{cor:sym q duality} allows us to define a canonical inner product between $\sym(K^*)$ and $\q(K)$. 
This can be most easily expressed using matrices,
\begin{align*}
  (\beta,q) = \tr(M_qM_\beta),
\end{align*}
where $M_q$ is a matrix representation of $q$ with respect to $\{e_i\}$ and $M_\beta$ is the matrix representation of $\beta$ with respect to the dual basis $\{e_i^*\}$.
To see that the inner product is in fact basis-independent, notice that Cor.~\ref{cor:sym q duality} associates to $\beta$ a $p_\beta\in\q(K)^*$ such that
\begin{align}
  \label{eq:sym q pairing}
  (\beta,q):=p_\beta(q) = \phi_q(\phi_\beta).
\end{align}
It may be checked that $\tr(M_qM_\beta) = \phi_q(\phi_\beta)$ for any matrix representation of $q$.

In the following, we will draw the connection between the polarization and the tensor polarization of quadratic forms.
We begin by asking: which symmetric forms arise as the polarization of a quadratic form?
In the odd $d$ case this can be easily seen to be, as in the real-valued case, all symmetric forms, i.e.\ $\range\Xi=\sym(K)$. 
This is because in this case $\Xi$ is invertible, with 
\begin{align*}
  \Xi^{-1}(\beta):u\mapsto \frac12 \beta(u,u).
\end{align*}
If $d=2$, $\Xi$ is no longer invertible:
its kernel is exactly the set of linear forms
\begin{align*}
  q(u)=\sum_{i}q_i u_i^2 = \sum_i q_i u_i,
  \qquad
  q_i\in\zz_2,
\end{align*}
equivalently, the set of quadratic forms with a diagonal matrix representation, $\diag(q_1,\dots,q_k)$.
Indeed, for such forms $q(u+v)=q(u)+q(v)$.

\begin{proposition}
  \label{prop:xi}
  Let $d=2$. Then it holds that
  \begin{align*}
    K^*\subset\q(K),
    \qquad
    \ker\Xi = K^*,
    \qquad
    \range\Xi=\alt(K).
  \end{align*}
\end{proposition}
\begin{proof}
  One may immediately verify that $K^*\subset\q(K)$ and that $K^*\subseteq\ker\Xi$.
  Furthermore, any $q$ satisfying $q(v+u)=q(u)+q(v)$ is linear and so part of $K^*$.
  Thus $\ker\Xi = K^*$.
  
  We now show the last statement. 
  If a form $\beta\in\sym(K)$ has a refinement $q$, then
  \begin{align*}
    \beta(u,u)=q(u+u)+q(u)+q(u)=q(0)+0 = 0.
  \end{align*} 
  Converseley, consider the basis $b_{ij}\in\alt(K)$, $i\neq j$, given by $b_{ij}(u,v)=u_iv_j+u_jv_i$.
  It is sufficient to show that each basis element has a refinement, and, indeed,
  the refinement of $b_{ij}$ is $q_{b_{ij}}(u)=u_iu_j$.
\end{proof}

Translating this statement to tensor language, we may simply state that if $d$ is odd, then $(K^*\otimes K^*)^\pi = ((K\otimes K)^\pi)^*$, while if $d=2$, this equation does not hold.
Indeed, the subspace of tensors associated with $\alt(K)\subset\sym(K)$ act as the trivial functional on  $(K\otimes K)^\pi$.
In matrix language: 
The trace inner product restricts to a non-degenerate form on symmetric matrices if and only if $d$ is odd. 
In the $d=2$ case, the subspace of alternating matrices -- recall that $\alt(K)\subset\sym(K)$ here -- is the radical of this restricted form.

In App.~\ref{app:proofs prelim} we will show that this reflects a representation-theoretical fact---that the action of $\{\ii,\pi\}\simeq S_2$ on $K\otimes K$ is semi-simple if and only if $d$ is odd.

Finally, we show the relationship between the polarization and the tensor polarization of a quadratic form.

\begin{proposition}
  \label{prop:refinable bilinear}
  Let $\beta=\Xi(q)$, and $\phi$ be the tensor polarization of $q$.
  Then the tensor representation of $\beta$ is given by $\phi\circ(\ii+\pi)$.
\end{proposition}

\subsubsection{Generalized quadratic forms}

When $d=2$ one may define a non-trivial generalization of quadratic refinements, leading to the concept of generalized quadratic forms.
Here, we will introduce these generalized forms in a way which allows us to discuss the $d=2$ and odd $d$ cases uniformly.
In the following, given a $\zz_2$-valued function $f$, we let $2f$ denote a $\zz_4$-valued function defined by
\begin{align*}
  2f(x)=
  \begin{cases}
    2 \quad &f(x)=1,\\
    0 \quad &f(x)=0.
  \end{cases}
\end{align*}

Recall from Sec.~\ref{sec:pauli and clifford} that $D=4$ if $d=2$ and $D=d$ else.
A \emph{generalized quadratic form} on $K$ is a function $q:K\to\zz_D$ satisfying
\begin{align*}
  q(\alpha u) &= \alpha^2 q(u),\\
  q(u+v) - q(u) - q(v) &= 2\beta(u,v),
\end{align*}
for some $\beta\in\sym(K)$.
The form $q$ is a \emph{generalized} quadratic refinement of $\beta$.
Notice that, in the odd $d$ case, the generalized quadratic refinement and the quadratic refinement of a symmetric form differ by a factor of 2.

The space of generalized forms is denoted by $\t\q(K)$.
As mentioned, if $d$ is odd then $\q(K)=\t\q(K)$.
In the case $d=2$, there is a canonical injection $\q(K)\subset\t\q(K)$ given by $\q(K)\ni q \mapsto 2q \in\t\q(K)$.

One may define the tensor polarization and a matrix representation of a generalized quadratic form.
While we will not use this formulation in the main text of the paper, we believe that this less well-known formulation might provide additional intuition on generalized quadratic forms.
Because of this, we have included some discussion on it in App.~\ref{app:gen quad}. 

\subsubsection{Classification of forms}

The classification of forms is widely known over the real numbers and over $\zz_d$ when $d$ is odd (see e.g.~\cite[Chap.~VI]{omeara}).
On the other hand, while such a classification exists for $d=2$, it is much less well-known. 
Here we review the results classify symmetric, quadratic, and generalized quadratic forms for $d=2$ and for odd $d$.

We denote the $\gl(K)$ equivalence of forms by $\sim$, e.g. $q\sim q' \in\q(K)$ if there is some $g\in\gl(K)$ such that $q' = gq$.

\begin{proposition}[See, e.g., Statement~62:1a in \cite{omeara}.]
  \label{prop:odd symmetric}
  Let $d$ be odd and $\beta\in\sym(K)$. 
	Let $\beta_0$ be the induced form on $K/\rad(\beta)$.
	Define the \emph{rank} of $\beta$ to be the dimension of the quotient space $K/\rad(\beta)$.
  The \emph{discriminant} of $\beta$ is $\discr(\beta):=\ell(\det M_{\beta_0}))$, where $\ell(\darg)=\pm1$ is the \emph{Legendre symbol} and the determinant is computed with respect to  an arbitrary basis of $K/\rad(\beta)$.  
	
	Then the $\gl(K)$ orbits on $\sym(K)$ are classified by \emph{rank} and \emph{discriminant}, i.e.\ $\rank\beta'=\rank\beta$ and $\discr(\beta')=\discr(\beta)$ if and only if $\beta\sim\beta'$.
  
  In particular, any $\beta\in\sym(K)$ has a diagonal matrix representation of the form
  \begin{align*}
    M_\beta=\diag(\
    \overbrace{1,\dots,1,a}^{\rank\beta}\ ,0,\dots,0),
  \end{align*} 
  where $a$ is such that $\ell(a)=\discr(\beta)$.
\end{proposition}

Because in the odd characteristic case $\Xi^{-1}$ exists and preserves $\gl(K)$ orbits, Prop.~\ref{prop:odd symmetric} also gives a criterion for equivalence in $\q(K)$.
Following the exposition of Ref.~\cite[App.~A]{klausthesis}, 
we now proceed to summarize the classification results for quadratic and bilinear forms in the $d=2$ case.

First a definition, which will be used repeatedly below.
For any $d$, the \emph{hyperbolic plane} is the space $\hh = \zz_d^2$ with basis $\{e,f\}$, equipped with a form $\beta_{\hh}$ given by
\begin{align*}
	\beta_{\hh}(e,e)=\beta_{\hh}(f,f) = 0, \qquad \beta_{\hh}(e,f) = 1.
\end{align*} 

\begin{proposition}
  \label{prop:even symmetric}
  Let $d=2$ and $\beta\in\sym(K)$.
	Let $\beta_0$ be the induced form on $K/\rad(\beta)$.
  Then exactly one of the following holds:
  \begin{enumerate}
    \item There exists an orthonormal basis $\{e_i\}$ of $K/\rad(\beta)$,
    \begin{align*}
      \beta_0(e_i,e_j)=\delta_{ij}.
    \end{align*}
		In this case, we say that the \emph{type} of $\beta$ is \emph{odd}.
    \item The rank of $\beta$ is even, $\dim(K/\rad(\beta))=2m$, and $\beta_0\sim\beta_{\hh}^{\oplus m}$.
    We say that the type of $\beta$ is even in this case.
  \end{enumerate}
  In particular, the $\gl(K)$ orbits on $\sym(K)$ are labeled by rank and type.
\end{proposition}

This statement is typically only shown as a classification of non-degenerate forms. 
For completeness, in App.~\ref{app:proofs prelim} we show how Prop.~\ref{prop:even symmetric} follows from such a classification.

Because when $d=2$ the canonical isomorphism $\Xi^{-1}$ between $\sym(K)$ and $\q(K)$ no longer exists, Prop.~\ref{prop:even symmetric} does not directly imply a classification of quadratic forms.
To see how the two can differ, we first look at a simple example, which will turn out to contain all ingredients required for the general result.

\begin{example}
  \label{ex:hyperbolic}
  In the case $d=2$, the following two quadratic forms are compatible with the hyperbolic plane $\beta_{\hh}$,
  \begin{align*}
    q_{\hh}^0(e)=q_{\hh}^0(f)=0 \qquad q_{\hh}^0(e+f)=1,\\
    q_{\hh}^1(e)=q_{\hh}^1(f)=1 \qquad q_{\hh}^1(e+f)=1.
  \end{align*}
	The two forms are not $\gl(\hh)$-equivalent, because
	\begin{align*}
		3 = 
		\big(q_{\hh}^0\big)^{-1}(\{ 0 \}) \neq
		\big(q_{\hh}^1\big)^{-1}(\{ 0 \}) = 1.
	\end{align*}
\end{example}

We now turn to general non-degenerate quadratic forms.

\begin{proposition}[Prop.~A.2.8 in \cite{klausthesis}]
  \label{prop:even quadratic}
  Let $d=2$ and $q\in\q(K)$ be a quadratic refinement of the non-degenerate form $\beta$.
  Then $\beta$ is even and $\dim K = 2m$.
  Furthermore, consider the \emph{Arf invariant} of $q$, $\arf:\q(K)\to\zz_2$, defined by
  \begin{align*}
    (-1)^{\arf(q)} = \mathrm{sgn}\,\sum_{u\in K} (-1)^{q(u)}.
  \end{align*}
  Then exactly one of the following holds:
  \begin{enumerate}
    \item $\arf(q)=0$ and $q \sim (q_{\hh}^0)^{\oplus m}$,
    \item $\arf(q)=1$ and $q \sim q_{\hh}^1\oplus(q_{\hh}^0)^{\oplus (m-1)}.$
  \end{enumerate}
\end{proposition}

Finally, Prop.~\ref{prop:generalized even quadratic} provides a classification of generalized quadratic forms which refine non-degenerate symmetric forms.
These were studied in~\cite{brown1972generalizations} in the context of algebraic topology.
An accessible and self-contained proof is found in~\cite[App.~A.3]{klausthesis}.

\begin{proposition}[\cite{brown1972generalizations}]
  \label{prop:generalized even quadratic}
  Let $d=2$ and $q\in\t\q(K)$ be a generalized quadratic refinement of the non-degenerate form $\beta$.

	Consider the \emph{generalized Arf invariant} of $q$, $\garf:\t\q(K)$ defined by
  \begin{align*}
    \exp(\frac{2i\pi}{8}\garf(q)) = 
    \textup{phase of }
    \sum_{u\in K} i^{q(u)}.
  \end{align*}

	Then $\garf$ takes values in $\zz_8$.

  If $\beta$ is of even type, $\dim K= 2m$, and exactly one of the following holds,
  \begin{enumerate}
    \item $\garf(q)=0$ and $q\sim(2q_{\hh}^0)^{m}$,
    \item $\garf(q)=4$ and $q\sim (2q_{\hh}^1)\oplus(2q_{\hh}^0)^{\oplus (m-1)}.$ 
  \end{enumerate}
  Otherwise, $\beta$ is of odd type and exactly one of the following holds,
  \begin{enumerate}
    \item 
    $\garf(q)= k \mod 8$ and $q\sim q_{k,0}$,
    \item 
    $\garf(q)= k-2 \mod 8$ and $q\sim q_{k-1,1}$,
    \item 
    $\garf(q)= k-4 \mod 8$ and $q\sim q_{k-2,2}$,
    \item 
    $\garf(q)= k-6 \mod 8$ and $q\sim q_{k-3,3}$, 
  \end{enumerate}
  where $k:=\dim K$.
\end{proposition}

\subsubsection{Model quadratic space}

Throughout the remainder of the paper we take $T:=\zz_d^{r+s}$, with $r+s=t$ and let $\beta_{r,s}:T\times T \to \zz_d$ be the symmetric form defined on the standard basis 
\begin{align*}
  \beta_{r,s}(e_i,e_j) = s_i\delta_{ij},
\end{align*}
where $s_i = 1$ if $i\leq r$, and $s_i = -1$ if $i>r$. 
Equivalently, its matrix and tensor representations are
\begin{align*}
  M_{r,s}
  =
  \diag(\
  \overbrace{+1, \dots, +1}^{r}\ ,
  \overbrace{-1, \dots, -1}^{s} \
  ),
  \qquad
  \phi_{r,s}=\sum_i s_i e_i^*\otimes e_i^*,
\end{align*}
where $\{e_i^*\}$ is the dual basis of $\{e_i\}$.
In the case $d=2$, $\beta_{r,s}$ is the standard dot product, however we keep the notation above for uniformity.

Let $q_{r,s}:T\to\zz_D$ be the 
generalized quadratic refinement of $\beta_{r,s}$ that is given in matrix form as%
\footnote{
For a discussion on matrix representations of generalized quadratic forms over $\zz_2$, see App.~\ref{app:gen quad}. 
} 
\begin{align*}
  \diag(\
  \overbrace{+1, \dots, +1}^{r}\ ,
  \overbrace{-1, \dots, -1}^{s} \
  ).
\end{align*}
Here, the matrix entries are elements in $\zz_D$, whereas the matrix entries of $M_{r,s}$ are in $\zz_d$.

Equivalently, on the standard basis $q_{r,s}$ evaluates to
\begin{align*}
  q_{r,s}(e_i)  = 
  \begin{cases}
    +1\mod D, \qquad i\leq r,\\
    -1\mod D, \qquad i>r,
  \end{cases}
\end{align*}
and 
\begin{align}
  \label{eq:qrs}
  q(u)=\sum_{e_i\in\supp u}q(e_i).
\end{align}

A subspace $N\subset T$ is \emph{isotropic} if $q_{r,s}|_N = 0$. 
It is \emph{stochastic} 
if $\ones_t\in N^\perp$, where
\begin{align*}
	\ones_t &:= (1,1,\dots,1)\in T,
\end{align*}
and
\begin{align*}
	N^\perp &:= \{ u\in T \div \beta(u,v)=0,\, \forall\ v\in N\}
\end{align*}
is the \emph{orthocomplement} of $N$. 
Notice that for an isotropic subspace $N$, it holds that $N\subseteq N^\perp$.

\begin{remark}
  The all-ones vector, $\ones_t$, will appear rather naturally when describing the commutant of tensor power representations of the Clifford group (Sec.~\ref{sec:commutant}).
  In the qubit case, $\ones_t$ additionally acquires a \emph{geometrical} meaning.
  Namely, in this case $\beta_{r,s}(u,u)=\beta_{r,s}(\ones_t,u)$ for all $u\in T$, and thus $\ones_t$ is known as the \emph{Wu class} of $\beta$.
\end{remark}

We use the following notation:
\begin{align*}
  \grass_m :&=\{ N\subset T \div N \text{ stoch. isotr.}, \ones_t\notin N, \dim N = m\}\\
  \grass :&= \bigcup_m\grass_m,\\
  \grass_m^0 :&=\{ N\subset T \div N \text{ stoch. isotr.}, \ones_t\in N,  \dim N = m\}\\
  \grass^0 :&= \bigcup_m\grass_m^0,\\
  T_N :&= N^\perp/N.
\end{align*}
The maximal $m$ for which $\grass_m$ is non-empty is denoted by $m(T)$. 
From~\eqref{eq:qrs}, it follows that $\ones_t$ is isotropic if and only if $r-s=0 \mod D$.
In this case, the largest $m$ for which $\grass_m^0$ is non-empty is $m(T)+1$.
Otherwise, the spaces $\grass_m^0$ are empty.

Because
\begin{align*}
  \rad\left(q_{r,s}|_{N^\perp}\right)
  =
  N,
\end{align*}
the form $q_{r,s}$ is well defined on $T_N$.
We write $q_N\in\t\q(T_N)$ for the non-degenerate form inherited. 

The orthogonal group, $\orth(T)$, is the subgroup of $\gl(T)$ leaving $q_{r,s}$ invariant, 
\begin{align*}
  q_{r,s}(O\darg)=q_{r,s}(\darg) \mod D, \qquad O\in\orth(T).
\end{align*}
The orthogonal stochastic group, $\stoch(T)$, is the subgroup of $\orth(T)$ leaving $\ones_t$ invariant, 
\begin{align*}
  O\ones_t=\ones_t \mod d, \qquad O\in\stoch(T).
\end{align*}

\begin{proposition}
  \label{prop:ortho stochastic}
  Let $d=2$. Then $\orth(T)=\stoch(T)$.
\end{proposition}
\begin{proof}
  Because $q_{r,s}$ is a generalized refinement of $\beta_{r,s}$, the following equation over $\zz_4$ holds for any $O\in\orth(T)$,
  \begin{align*}
    2\beta_{r,s}(Ou,Ov) 
    =
    q_{r,s}(Ou+Ov)-q_{r,s}(Ou) - q_{r,s}(Ov) 
    =
    q_{r,s}(u+v)-q_{r,s}(u) - q_{r,s}(v) 
    =
    2\beta_{r,s}(u,v),
  \end{align*}
  so that $\beta_{r,s}$ is $\orth(T)$-invariant.
  But, $\beta_{r,s}(u,u)=\beta_{r,s}(\ones_t,u)$, and so
  \begin{align*}
    \beta_{r,s}(O\ones_t,u)
    =
		\beta_{r,s}(\ones_t,O^{-1}u)
    =
		\beta_{r,s}(O^{-1}u,O^{-1}u)
    =
    \beta_{r,s}(u,u)
    =
    \beta_{r,s}(\ones_t,u).
  \end{align*} 
  The claim follows from non-degeneracy of $\beta_{r,s}$.
\end{proof}

Model spaces of the same dimension are sometimes isometric, the following proposition classifies all such isometries.

\begin{proposition}
  \label{prop:equivalence qrs}
  Let $r,r',s,s,'\in\nn$ with $r+s=r'+s'$. 
  Then $q_{r,s}\sim q_{r',s'}$ if and only if one of the following conditions holds,
  \begin{enumerate}
    \item $d=1\mod 4$,
    \item $d=3\mod 4$, $s=s'\mod 2$,
    \item $d=2$, $r-s=r'-s'\mod 8$.
  \end{enumerate}
\end{proposition}
\begin{proof}
  Point 2. follows from $\discr(\beta_{r,s})=\text{ sq. class of }(-1)^{s}$.
  Point 1. uses this identity together with the fact that $-1$ is a square over $\zz_d$ in this case.
  
  For Point 3. we begin by noticing that both quadratic forms are of odd type.
  Therefore they are equivalent if and only if their generalized Arf invariant is equal.
  \begin{align}
    \exp(\frac{2i\pi}{8}\garf(q_{r,s}))
    &=
    \text{ phase of }
    \sum_{u_1\in \zz_d^r}\sum_{u_2\in \zz_d^s}
    i^{q_{r,0}(u_1)-q_{s,0}(u_2)} \nonumber\\
    &=
    \exp(\frac{2i\pi}{8}\big(\garf(q_{r,0})-\garf(q_{s,0}) \big)) \nonumber\\
    &=
    \exp(\frac{2i\pi}{8}(r-s)),\label{eq:garf qrs}
  \end{align}
  where the last line follows from Prop.~\ref{prop:generalized even quadratic}. 
\end{proof}

It is convenient to define $q_{r,s}$ for possibly negative $r,s\in\zz$.
Namely, we let $q_{r,s}$ be any generalized quadratic form equivalent to some $q_{r',s'}$ with $r',s'\geq0$ such that one of points 1.-3. of Prop.~\ref{prop:equivalence qrs} holds. 

\begin{proposition}
  \label{prop:form restriction}
  Let $N\in\grass_m$, then $q_N\simeq q_{r-m,s-m}$.
\end{proposition}

This proposition is proven in App.~\ref{app:proofs prelim}. 

\begin{remark}
  Prop.~\ref{prop:form restriction} does not hold if $N\in\grass_m^0$, that is, if $\ones_t\in N$.
  This is because, in that case, $N^\perp/N$ mods out the Wu class of the bilinear form $\beta_{r,s}$.
  In particular, $q_N$ is in this case the generalized refinement of an alternating form $\beta_N:=\beta_{r,s}|_{N^\perp/N}$, and thus $q_N\in2\q(T)\subset\t\q(T)$ is $\{0,2\}$-valued.
  
  Morally speaking, Prop.~\ref{prop:form restriction} shows that if $N$ is the Lagrangian of the sum of $m$ hyperbolic planes in $T$, then as long as $N\in\grass_m$, the Witt cancellation theorem holds for the equation
  \begin{align*}
    (N^\perp/N)\oplus\hh^{m}\simeq T.
  \end{align*} 
  In general, the Witt cancellation theorem is known not to hold~\cite[A.3.4]{klausthesis} for generalized quadratic forms over $\zz_2$. 
\end{remark}

\begin{proposition}
  \label{prop:orthogonal to symplectic}
  Let $r-s=0\mod4$, $d=2$, and $N\in\grass_m^0$. 
  Then, $t=0\mod2$, and it holds that 
  \begin{align*}
    \beta_N\simeq \beta_{\hh}^{\oplus(t-2m)/2}.
  \end{align*}
\end{proposition}
\begin{proof}
  The first statement follows from $r=s\mod2$.

	Regarding the second statement:
	For any $u\in N^\perp$, let $[u]\in T_N=N^\perp/N$ be the corresponding element of the quotient space. 
	Then
  \begin{align*}
		\beta_N([u],[u])
		=\beta_{r,s}(u,u)
		=\beta_{r,s}(\ones_t,u)
		=0
  \end{align*}
  because $\ones_t\in N$.
  Thus, the inherited form does not contain an orthonormal basis and is thus of even type.
\end{proof}

Prop.~\ref{prop:orthogonal to symplectic} motivates us to define $\sp(T_N)$, the isometry group of $\beta_N$ for $N\in\grass_m^0$.
It is clear that $\sp(T_N)\simeq\sp(2,t-2m)$, and that $\stoch(T_N)\subseteq\sp(T_N)$ where this inclusion is strict as long as $t>2m$.
Finally, notice that because $\ones_t\in N$, $\stoch(T_N)=\orth(T_N)$ is the isometry group of $q_N$.


\section{Clifford tensor powers}
\label{sec:tensor powers}

In this paper we study the tensor power representations of $\cliff$,
\begin{align*}
  \Delta_{r,s}(U) = U^{\otimes r}\otimes \bar{U}^{s} =: U^{\otimes(r,s)},
  \quad U\in\cliff.
\end{align*}
The representation space corresponding to $\Delta_{r,s}$ is $\hilb_{n,t}:=\hilb_n^{\otimes t}$ where $t:=r+s$.
It is useful to think of $\hilb_{n,t}$ as an $n$ by $t$ ``grid'' of qudit Hilbert spaces,
\begin{align}
  \label{eq:grid}
  \hilb_{n,t} =\,
  \begin{matrix}
    \cc^d & \otimes & \cdots & \otimes & \cc^d\\
    \vdots &   \vdots     & &       \vdots  & \vdots\\
    \cc^d & \otimes & \cdots & \otimes & \cc^d
  \end{matrix}
	\qquad\qquad
	\text{($n$ horizontal factors, $t$ vertical ones).}
\end{align}
We label elements 
of the standard basis of $\hilb_{n,t}$ using $t\times n$ matrices $F$ over $\zz_d$.
More precisely, we define
\begin{align*}
  \ket{F} :=\,
  \begin{matrix}
    \ket{F_{11}} & \otimes & \cdots & \otimes & \ket{F_{n1}}\\
		\otimes 			&	       &        &         & \otimes \\
    \vdots       &         &        &         & \vdots\\
		\otimes 			&	       &        &         & \otimes \\
    \ket{F_{1t}} & \otimes & \cdots & \otimes & \ket{F_{nt}}
  \end{matrix}\,,
  \qquad
  F_{ij}\in\zz_d.
\end{align*}
We furthermore identify columns of $F$ with vectors in $T$, and its rows with vectors in $\pos$.
This way, we obtain several equivalent representations of $\hilb_{n,t}$:
\begin{align*}
  \hilb_{n,t}
  \simeq
  \cc[\pos]^{\otimes t}
  \simeq
  \cc[T]^{\otimes n}
  \simeq
  \cc[\hom(\pos\to T)]
  \simeq
  \cc[\zz_d^{t\times n}],
\end{align*}
where $\hom(\pos\to T)$ is the space of linear functions $F:\pos\to T$.
We will use these representations of $\hilb_{n,t}$ interchangeably throughout.

The Hilbert space $\hilb_{n,t}$ hosts its own “global'' Pauli group $\pauli_{nt}$, generated by $\tau\ii$ and tensor products of displacement operators acting on any $\cc^d$ factor.
We can see that the phase space associated to this Pauli group is
\begin{align}
  \label{eq:global pauli}
  \pauli_{nt}/\center(\pauli_{nt}) 
  \simeq 
  \zz_d^{2n\times t} 
  \simeq
  T\otimes V
  \simeq
  \hom(V\to T).
\end{align}

Notice that $\Delta_{1,1}$ is isomorphic to the action of $\cliff$ on $\endom(\hilb_n)$ by conjugation.
This space decomposes into two irreducible subrepresentations: the trivial component spanned by the identity, and the space of traceless matrices.
The latter irrep will be referred to as the \emph{adjoint} representation of $\cliff$, denoted $\ad(\cliff)$.

The following lemma establishes equivalences between different tensor power representations. 
The proof, presented in App.~\ref{app:proofs tensor}, builds on the geometric equivalences of Prop.~\ref{prop:equivalence qrs}.

\begin{restatable}[Equivalent tensor powers]{lemma}{deltars}
  \label{lem:rs equivalence}
  Let $d,r,s,r',s'\in\nn$ be such that $r+s=r'+s'$, and let $\Delta_{r,s}$ be as above.
  Furthermore, if $d$ is odd, let $r-s=r'-s'\mod d$. 
  Then for all the following cases we have that $\Delta_{r,s}\simeq\Delta_{r',s'}$:
  \begin{enumerate}
    \item
    If $d=1\mod4$
    \item
    If $d=3\mod4$, and $s=s'\mod 2$,
    \item
    If $d=2$, $r-s = r'-s'\mod 8$.
  \end{enumerate}
\end{restatable}

\begin{remark}
Compare the condition $r-s=r'-s'\mod d$ to~\cite[Lem.~4.2]{rank-deficient}.
In the notation of that reference, consider two orthogonal bases $\{e_i\}$ and $\{f_i\}$ of the space $U$, satisfying $[\beta(e_i,e_j)]_{ij} = \ii_r\oplus(-\ii_s)$, $[\beta(f_i,f_j)]_{ij} = \ii_{r'}\oplus(-\ii_{s'})$.
Then our $\Delta_{r,s}$ corresponds to $\mathcal{C}l^{(e)}_{U\otimes V}$ from the reference, and $\Delta_{r',s'}$ to $\mathcal{C}l^{(f)}_{U\otimes V}$, where $e=\sum_ie_i$ and $f=\sum_i f_i$.
These two representations of the Clifford group are non-equivalent whenever $\beta(e,e)\neq\beta(f,f)$.
To see this, one must simply compute the action of the central matrices of these representations, denoted in the proof of~\cite[Lem.~4.2]{rank-deficient} as $W_{U\otimes V}(\iota_e(\lambda,0))$ and $W_{U\otimes V}(\iota_f(\lambda,0))$.
\end{remark}

As we did with $q_{r,s}$, we will extend the definition of $\Delta_{r,s}$ to possibly negative values of $r, s$.
Namely, take some $r',s'\in\nn$.
Then if $d$ is odd, for any $r,s\in \zz$ with
\begin{align*}
  r+s=r'+s',\qquad r-s=r'-s'\mod d,\qquad s=s'\mod 2,
\end{align*}
we let $\Delta_{r,s}$ be a representaiton equivalent to $\Delta_{r',s'}$.
On the other hand, if $d=2$, then for any $r,s\in \zz$ with
\begin{align*}
  r+s=r'+s',\qquad r-s=r'-s'\mod 8,
\end{align*}
we let $\Delta_{r,s}$ be equivalent to $\Delta_{r',s'}$.

\subsection{The commutant algebra}
\label{sec:commutant}

In~\cite{gnw} the \emph{commutant} $\com{r,s}$ of $\Delta_{r,s}$ was studied, i.e.\ the subalgebra of $\endom(\hilb_{n,t})$ which commutes with all images $\Delta_{r,s}(C)$, where $C\in\cliff$.
Important to this construction where two classes of operators.
With respect to eq.~\eqref{eq:grid}, these will be $n$-th tensor power operators---each tensor factor acting on a $t$-qubit column of the grid.

The first is a class of \emph{tensor power CSS code} projectors associated with an isotropic subspace $N\subset T$
\begin{align*}
  P(N)
  =
  d^{-2n\dim N}\Big(
    \sum_{u,v\in N} X(u)Z(v)
  \Big)^{\otimes n}.
\end{align*}
As in~\eqref{eq:global pauli} , we identify the operator
\begin{align*}
  X(u_1)Z(v_1)\otimes\cdots\otimes X(u_n)Z(v_n)
\end{align*} 
with the global Pauli operator $W(M)$, where $M\in T\otimes V \simeq \hom(V\to T)$ maps the standard canonical basis of $V$ to $\{u_1,v_1,\dots,u_n,v_n\}$.
Notice that for any such $M$, $\range M \subseteq N$.
This leads to the following alternative formula which will be handy for calculations later on: 
\begin{align}
  \label{eq:css definition}
  P(N) = d^{-2n\dim N}\sum_{M\in \hom(V\to N)} W(M).
\end{align}

Letting $C_N := \range P(N)$, these code spaces have the following \emph{coset state basis}:
\begin{align}
  \label{eq:factorization coset}
  \ket{[f_1]_N}\otimes \cdots \otimes\ket{[f_n]_N}
  \in ((\cc^d)^{\otimes t})^{\otimes n}
  \simeq\hilb_{n,t}
\end{align}
where $f_1,\dots,f_n\in N^\perp$ and 
\begin{align*}
  \ket{[f_i]_N}=d^{-\dim N/2}\sum_{u\in N}\ket{f_i + u}\in(\cc^d)^{\otimes t},
  \qquad
  [f_i]_N\in T_N.
\end{align*}
In particular, the tensor power code $C_N$ encodes $\dim T_N=t-2\dim N$ qudits into each $t$-qudit column of eq.~\eqref{eq:grid}. 
Equivalently, identifying $\hilb_{n,t}\simeq\cc[\hom(\pos\to T)]$ and letting $F$ be the $t\times n$ matrix $F=[f_1\dots f_n]$, the coset basis of the code is
\begin{align*}
    \ket{[F]_N} := d^{-\dim N/2} \sum_{F'\in \hom(\pos\to N)} \ket{F+F'}\in\cc[\hom(\pos\to T)],
    \qquad
    F \in \hom(\pos\to N^\perp).
\end{align*}
That is $\ket{[F]_N}\simeq\ket{[f_1]_N}\otimes \cdots \otimes\ket{[f_n]_N}$.

In summary, we have the natural identifications
\begin{align*}
  C_N\simeq
  \cc[T_N]^{\otimes n}
  \simeq
  ((\cc^d)^{\otimes (\dim T_N)})^{\otimes n}
  \simeq
  \cc[\hom(\pos\to T_N)],
\end{align*}
where the last isomorphism arises from the fact that $[F]_N\in\hom(\pos\to T_N)$.

The second is the following representation of $\orth(T)$,
\begin{align}
  \label{eq:R def}
  R(O) = \sum_{F\in\hom(\pos\to T)} \ketbra{OF}{F}.
\end{align}
On a tensor product standard basis state $\ket{F}=\ket{f_1,\dots,\ f_n}$, where $f_i\in T$, this representation acts as
\begin{align*}
  R(O)\ket{f_1,\dots,\ f_n} = \ket{Of_1,\dots,\ Of_n}.
\end{align*}

We will use the following slight generalization of the main result of~\cite{gnw}, which dealt with the representation $\Delta_{t,0}$.

\begin{proposition}
  \label{prop:gnw}
  Let $P(N)$ be as in~\eqref{eq:css definition}, where $N$ is isotropic, and $R$ be as in~\eqref{eq:R def}.
  
  Then, if $t\leq n-1$, the commutant $\com{r,s}$ of $\Delta_{r,s}$ has the following basis
  \begin{align*}
    \semigroup_{r,s}=\{ R(O)P(N) \div O\in\stoch(T), \, N\text{ isotropic stochastic}\}.
  \end{align*}
  Furthermore, if $d$ is odd, the commutant of $\res{Sp(V)}\Delta_{r,s}$ is generated as an algebra by
  \begin{align*}
    \{ R(O), P(N) \div O\in\orth(T), N\text{ isotropic}\}.
  \end{align*}
\end{proposition}
\begin{proof}[Proof sketch.]
  One can show (e.g. as in~\cite[Lem.~4.5]{gnw}) that the operators $R(O)$ and $P(N)$ commute with $\Delta_{r,s}$.
  Following the same procedure as in~\cite[Lem.~4.7]{gnw} one can show that $\semigroup_{r,s}$ is a set of linearly independent operators (because $t-1\leq n$).
  Finally, by the fact that the dimension of the commutant of $\Delta_{r,s}$ is the same as the dimension of the commutant of $\Delta_{t,0}$, the proof of~\cite[Thm.~4.9]{gnw} shows our first claim.
  The second claim follows similarly.
\end{proof}

It can be easily checked that 
\begin{align}
  \label{eq:o on code}
  R(O)P(N)R(O)^\dagger = P(ON).
\end{align}

A variant of Witt's lemma establishes a converse:
\begin{lemma}
  \label{lem:stoch single orbit}
  Let $N$, $N'\subset T$ be stochastic isotropic subspaces of the same dimension.
  Furthermore, let it be the case that $\ones_t$ is either an element of both of $N, N'$ or is not an element of either.
  Then, there exists an $O\in\stoch(T)$ for which $N'=ON$.
\end{lemma}
\begin{proof}
  Consider the two spaces $M=\vspan\{N,\ones_t\}$, $M'=\vspan\{N',\ones_t\}$.
	There is a linear isomorphism between $M$ and $M'$ that also maps $\ones_t$ to $\ones_t$. 
	Because $M$ and $M'$ are isometric, this isomorphism extends to an orthogonal map $O\in O(T)$, by Witt's lemma (cf.\ e.g.~\cite[Thm.~3.3]{finitesimple} for the odd $d$ case, and~\cite{wittmod4} for the case of generalized quadratic forms when $d=2$).
	But $O$ preserves $\ones_t$, so it is an element of $\stoch(T)$.
\end{proof}

Lem.~\ref{lem:stoch single orbit} motivates the definition of the following subsets of $\semigroup_{r,s}$,
\begin{align*}
	\semigroup_{r,s}^m     &:= \{ R(O) P(N) \div O\in\stoch(T),\ 
  \dim N\geq m\},\\
  \semigroup_{r,s}^{m,0} &:= \{ R(O) P(N) \div O\in\stoch(T),\ 
  \dim N\geq m,\ \ones_t\in N\},
\end{align*}
and, for each isotropic stochastic $N\subset T$,
\begin{align*}
  \semigroup_{r,s}^N 
  &:= 
	\{ R(O) P(N') \div O\in\stoch(T),\ N\subseteq N'
  \}.
\end{align*}

By~\cite[Eq.~(4.24)]{gnw}, $\semigroup_{r,s}$ forms a semigroup.
Namely, for any $N_1,\ N_2$, there exists 
an $O_{N_1,N_2}\in\stoch(T)$ and
a stochastic isotropic subspace $I_{N_1,N_2}$ for which
\begin{align}
  \label{eq:gnw}
	P({N_1})P({N_2}) = 
	R(O_{N_1,N_2}) P(I_{N_1,N_2}).
\end{align}

We will require the following properties of this semi-group, which follow directly from the arguments in \cite{gnw}:

\begin{proposition}
  \label{prop:semi group}
	The commutant semi-group has the following properties:
	\begin{enumerate}
		\item
			The set $\semigroup_{r,s}^N$ is 
			invariant under left-multiplication by $\stoch(T)$ and right-multiplication by $\stoch(T)^N$.
			What is more, its span is a left-ideal in the commutant algebra, i.e.\
		  $N \subseteq I_{N',N}$ for all $N'$.
		\item
			The sets $\semigroup_{r,s}^m, \semigroup_{r,s}^{m,0}$ are
			invariant under left- and right-multiplication by $\stoch(T)$.
			What is more, they span ideals in the commutant algebra, i.e.\
			$\dim I_{N_1, N_2}\geq \max(\dim N_1, \dim N_2)$ and $\ones_t\in I_{N_1, N_2}$ if $\ones_t \in N_1$ or $\ones_t \in N_2$.
	\end{enumerate}
\end{proposition}

We denote by $\com{r,s}^m,\ \com{r,s}^{m,0}\subset\com{r,s}$ the ideals spanned respectively by $\semigroup_{r,s}^m$ and $\semigroup_{r,s}^{m,0}$.

\begin{lemma}
  \label{lem:light cone}
  Let $N$ be an isotropic stochastic subspace of $T$.
  Then the code $C_N$ is a $\cliff$ subrepresentation of $\Delta_{r,s}$.
	The compression $P(N)\com{r,s} P(N)$ of the commutant to this subspace is generated by 
	$\semigroup_{r,s}^N$.
\end{lemma}

\begin{proof}
  The first claim follows from the fact that the projection $P(N)$ onto $C_N$ lies in the commutant.
	The second claim is a consequence of $\semigroup_{r,s}^N$ being a left-ideal (point 1. in Prop.~\ref{prop:semi group}), so that
  $P(N)\com{r,s} P(N)\subset \semigroup_{r,s}^N$.
\end{proof}

\begin{remark}
  \label{rem:compression basis}
  Lem.~\ref{lem:light cone} provides a set of operators, namely $\semigroup_{r,s}^N$, which contains a basis for the compression of the commutant $P(N)\com{r,s}P(N)$.
  For the purposes of our main theorem this is sufficient.
  It would be interesting, however, to go further and find a basis for this compression.
  Indeed, this would help characterize the action of $\cliff$ on the codes $C_N$ in more detail.
  
  Let $\stoch(T)^N$ be the subgroup of $\stoch(T)$ which preserves $N$.
  Lem.~\ref{lem:kernel on code action} points out that $\stoch(T)^N$ acts on $C_N$.
  Let this representation be $R_N$, it instantiates the homomorphism
  \begin{align*}
    \stoch(T)^N\mapsto \stoch(T)^N/\ker(R_N)\simeq \stoch(T_N),
    \quad
    T_N:=N^\perp/N.
  \end{align*}
  It is plausible to expect that the basis of this commutant is of the form
  $R(O)P(N')$,
  where one orthogonal stochastic matrix $O\in\stoch(T)^N$ is chosen as representative for each $\ker(R_N)$-coset, and where $N\subseteq N'$.
  Indeed, if $N\in\grass_m$ this follows from Lem.~\ref{lem:restriction to css} and Prop.~\ref{prop:gnw}.  
  
  We leave the question of whether this holds as well for $N\in\grass_m^0$ open.
  In particular, it is not clear whether such a commutant contains an operator of the form
  \begin{align*}
    R(O)P(N'),
  \end{align*}
  where $N\subset N'$ and $N\subset ON'$, but $O\notin \stoch(T)^N$.
\end{remark}

\subsection{Code representation spaces}

By Prop.~\ref{prop:gnw}, the code space $C_N$ corresponding to a stochastic isotropic $N$ is a $\cliff$ representation.
Furthermore, by Lem.~\ref{lem:stoch single orbit}, for any pair $N,\ N'\in\grass_m$ or $N,\ N'\in\grass_m^0$, the corresponding code spaces $C_N$ and $C_{N'}$ are isomorphic as $\cliff$ representations.
 
The following lemmas specify this representation and can be seen as a generalization of~\cite[Lem.~2.7]{rank-deficient}.
We prove them in App.~\ref{app:proofs tensor}. 
Lem.~\ref{lem:restriction to css} is essentially a consequence of Prop.~\ref{prop:form restriction}.
On the other hand, Lem~\ref{lem:odd cones} follows from~\cite[Lem.~2.7]{rank-deficient}.

\begin{restatable}[Code representations]{lemma}{codereps}
  \label{lem:restriction to css}
  Let $N\in \grass_m$ and $C_N\subset\hilb_{n,t}$ be the associated code.
  Then, $\Delta_{r,s}|_{C_N}\simeq \Delta_{r-m,s-m}$.
\end{restatable}

\begin{restatable}[Qudit $\cones$ representation]{lemma}{quditsubcode}
  \label{lem:odd cones}
  Let $d$ be odd and $r-s=0\mod d$. 
  As a $\cliff$-subrepresentation of $\Delta_{r,s}$, we have that $\ker(\Delta_{r,s}|_{\cones})=\pauli$ and
  \begin{align*}
    \Delta_{r,s}|_{\cones} \simeq
    \begin{cases}
      \mu^{\otimes(r-1,s-1)},\qquad &s>0,\\
      \mu^{\otimes(r-3,1)},  \qquad &s=0.
    \end{cases}
  \end{align*}
\end{restatable} 

\begin{remark}
  \label{rem:not cones}
  The proof of Lem.~\ref{lem:restriction to css} does \emph{not} hold when $N\in\grass_m^0$.
  To see this most easily, notice that in this case $\pauli\subseteq\ker(C_N)$, whereas $\pauli\not\subseteq\ker(\Delta_{r-m,s-m})$.
  By Lem.~\ref{lem:odd cones}, when $d$ is odd this isomorphism nevertheless continues to hold if one restricts attention to the action of $\sp(V)\subset\cliff$.
  
  While Lems.~\ref{lem:restriction to css},~\ref{lem:odd cones} were crucial for the proof of our previous result in~\cite{rank-deficient}, they are not needed for the main result in this paper.
  Because of this, we have also left open the question of what is a suitable generalization of Lem.~\ref{lem:odd cones} to the qubit case. 
\end{remark}

Now we work out the action of stochastic orthogonal matrices on the code spaces.

\begin{lemma}
  \label{lem:kernel on code action}
  Let $\stoch(T)^N$ be the subgroup of $\stoch(T)$ which preserves the stochastic isotropic subspace $N$.
  Then, $\stoch(T)^N$ acts on $C_N$. 
  Letting this action be $R_N$, as in Rem.~\ref{rem:compression basis}, it holds that $\stoch(T)^N/\ker(R_N)\simeq\stoch(T_N)$.
\end{lemma}
\begin{proof}
  It suffices to consider the $n=1$ case.
  Because $O\in\stoch(T)^N$ acts on $N$-cosets, there exists some $\t O\in\gl(T_N)$ for which
  \begin{align*}
    R(O)\ket{[f]_N}=\ket{[Of]_N}=\ket{\t O[f]_N},\qquad f\in N^\perp.
  \end{align*}
  Moreover, $\t O$ preserves $q_N$ and $[\ones_t]_N$, so $\t O\in\stoch(T_N)$.
  Finally, Lem.~\ref{lem:stoch single orbit} implies that every $\t O\in\stoch(T_N)$ can be extended to an $O\in\stoch(T)^N$ such that $[Of]_N=\t O[f]_N$ for all $f\in N^\perp$.
\end{proof}


\section{Rank theory of Clifford representations}
\label{sec:rank}

Several notions of ``rank'' have been introduced in the study of the representation theory of discrete symplectic and orthogonal groups~\cite{ghsmall,ghtakagi}. 
Here we extend a part of this formalism to the Clifford group.

Let $\dcliff\subset\cliff$ be the subgroup of diagonal Clifford matrices.
In App.~\ref{app:proofs rank} we show: 

\begin{restatable}{lemma}{diagonals}
  \label{lem:diagonals}
  Any $U\in\dcliff$ is, up to a phase, of the form
  \begin{align}
    \label{eq:diag quad}
    U = \sum_{x\in\pos} \tau^{q(x)+2x'\cdot x}\ketbra{x}
  \end{align}
  for some $q\in\t\q(\pos)$ and $x'\in\pos$.
  If $d=2$ one can furthermore set $x'=0$.
\end{restatable}

Consider the subgroup $\rdcliff\subset\dcliff$ of elements $U$, as in Lem.~\ref{lem:diagonals}, for which $x'=0$ and $q\in2\q(\pos)$.
For odd qudits, this corresponds to the complement of diagonal Paulis in the group $\rcliff$, that is,  $\rdcliff=\mu(\borel)$ where $\borel\subset\sp(V)$ is given as in Eq.~\eqref{eq:borel}. 
For the qubit case, this is simply the group of real diagonal Cliffords (hence the notation).
Notice  that for qubits, $\rdcliff$ contains Pauli $Z$-type matrices.
In both cases, by the proof of Lem.~\ref{lem:diagonals}, $\rdcliff\simeq\q(\pos)$ so label group elements with quadratic forms,
\begin{align*}
  \rdcliff = \{U_q\div q\in\q(\pos)\}.
\end{align*}
This group will be used to define a notion of \emph{rank} on representations of the Clifford group, in close analogy to how the subgroup $\borel\subset\sp(V)$ was used to define the rank of a symplectic representation in~\cite{ghsmall,ghtakagi}.%

Consider some $\cliff$ representation $\rho$.
Then, its restriction to $\rdcliff$ decomposes into one-dimensional representations because $\rdcliff$ is Abelian.
These one dimensional representations are characters of $\q(\pos)$ labeled by functionals in $\q(\pos)^*$. 
By Cor.~\ref{cor:sym q duality}, these characters are labeled by forms $B\in\sym(\pos^*)$.
In other words, there is an orthonormal basis $\{\ket{\psi_B}\}$ labeled by some set of matrices $B\in\sym(\pos^*)$ for which $U_q\in\rdcliff$ acts as
\begin{align*}
  \rho(U_q)\ket{\psi_B} = \omega^{p_B(q)}\ket{\psi_B}.
\end{align*}

We call the eigenvectors $\ket{\psi_B}$ \emph{weight vectors} of $\rho$, and the symmetric forms $B$ their corresponding \emph{weights.}
The span of all weight vectors corresponding to the same weight $B$ is called the \emph{weight space} of $B$.
The rank of a representation $\rho$ of $\cliff$ is
\begin{align}
  \rk(\rho) = \max_{B\text{ wght. of }\rho} \rank(B).
\end{align}
By Cor.~\ref{cor:sym q duality}, weight-spaces are permuted by the subgroup $\gl(\pos)\subset\cliff$ generated by the $\cadd$ gates.
That is, if $g\in\gl(\pos)$, then
\begin{align*}
  \rho(g)\ket{\psi_B}
\end{align*}
is a weight vector with weight $gBg^T$.
Because of this, weights which are equivalent in the sense of Sec.~\ref{sec:quad} appear with the same multiplicity in $\res{\rdcliff}\rho$.

Given that in $\Delta_{r,s}(\rdcliff)$ acts diagonally, all standard basis elements $\ket{F}\in\hilb_{n,t}$ are weight vectors.
The following lemma specifies their weights.

\begin{lemma}
  \label{lem:weights tensor power}
  The weight $w_F$ corresponding to the standard basis state $\ket{F}\in\hilb_{n,t}$, where $F$ has rows $f_1,\dots,f_t\in \pos$, has tensor representation
  \begin{align*}
    \phi_F := \sum_{i=1}^t s_i f_i\otimes f_i,
  \end{align*}
  where $s_i = +1$ if $i\leq r$ and $s_i=-1$ otherwise.
  Equivalently, the matrix representation of $w_F$ is
  \begin{align*}
    M_F := F M_{r,s} F^T.
  \end{align*}
  This way, $\rank(w_F)=\rank(M_F)=\rank \beta_{r,s}|_{\range(F)}$ and $\rk(\Delta_{r,s})=\min\{n,t\}$.
\end{lemma}

\begin{remark}
  \label{rem:ranks}
  In Lem.~\ref{lem:weights tensor power}, three distinct (but related) notions of rank are used:
  The rank of a matrix, say, $\rank M_F$, is the dimension of its column space.
  The rank of a form is the dimension of the space on which the form acts, minus the dimension of the radical (see Sec.~\ref{sec:quad}).
  For example, in the aforementioned lemma, $\rank(\beta_{r,s}|_{\range F})=\dim\range  F - \dim\rad(\beta_{r,s}|_{\range F})$.
  Finally, the rank of a Clifford representation (e.g., $\rk(\Delta_{r,s})$), is defined  in this section.
\end{remark}

\begin{proof}
  We may directly compute
  \begin{align*}
    U_q^{\otimes(r,s)}\ket{F} = \omega^{\sum_i s_i q(f_i)}\ket{F}
    =
    \omega^{\sum_i s_i (f_i\otimes f_i)(\phi_q)}\ket{F},
  \end{align*}
  where $\phi_q$ is the tensor polarization of $q$. 
  The first statement follows by Eq.~\eqref{eq:sym q pairing}.
  
  The second statement from the fact that, in the weight's functional representation, $w_F(\darg,\darg)=\beta_{r,s}(F\darg,F\darg)$ and so $\rank w_F =\rank \beta_{r,s}|_{\range F}$. 
\end{proof}

The following lemma is analogous to~\cite[Lem.~1.3.1.]{ghsmall} which states that the only irrep of the symplectic group with rank zero is the trivial one.
We prove it in App.~\ref{app:proofs rank}.

\begin{restatable}[Rank 0 irreps]{lemma}{rankzero}
  \label{lem:rank 0}
  If $d$ is odd, the unique rank zero $\cliff$ irrep is the trivial one.
  If $d=2$ and $n\geq3$, a rank zero representation is one dimensional, $\pm1$ valued, and uniquely specified by its restriction to $\center(\cliff)$.
  Namely, if $\rho$, $\rho'$ are rank zero repesentations with
  \begin{align*}
    \rho(\omega_8\ii)=\rho'(\omega_8\ii),
  \end{align*}
  then $\rho\simeq\rho'$.
\end{restatable}

As claimed at the beginning of the subsection, the formalism presented here extends the theory of rank studied in Refs.~\cite{ghsmall,ghtakagi}. 
There, each irrep of the symplectic group with odd $d$ is assigned a rank. 
Rank, as defined here, coincides with the rank introduced there when restricted to $\irr\sp(\zz_d^{2n})\subset\irr\cliff$ with $d$ odd (i.e.\ to irreps $\rho$ of the odd qudit Clifford group for which $\pauli\subseteq\ker\rho$).
Specifically it coincides with the ``U-rank'' $\rk_{\text{U}}$ defined in~\cite{ghtakagi}:
\begin{align*}
  \rk_{\text{U}}(\rho)=\rk(\rho),
  \qquad
  \text{with}
  \qquad
  d \text{ odd, }
  \rho\in\irr\sp(\zz_d^{2n}).
\end{align*}

\section{Classification of subrepresentations}
\label{sec:classification}

\subsection{Rank and duality}
\label{sec:rank and duality}

Here, we use the formalism developed in Sec.~\ref{sec:rank} to obtain a series of dualities.
Our proof techniques are an extension of those introduced in~\cite{ghsmall}.

Consider some, possibly zero, stochastic isotropic subspace $N\subset T$ with $\dim N = m$.
We will show that a duality very similar to the \emph{$\eta$ correspondence}, studied in~\cite{ghsmall}, holds on the span of all maximal-rank irreps in $C_N$.
This result is closely related to the main theorem of that reference:
There, tensor powers $\mu^{\otimes t}$ of the oscillator representation (with $d$ odd) are studied and the $\eta$ correspondence arises in the span of maximal rank irreps of $\hilb_{n,t}$.
Thm.~\ref{thm:eta duality} proves an equivalent result, valid for $\Delta_{r,s}$ with arbitrary $d$, and valid for arbitrary codes $C_N$ (as opposed to being valid only on the maximal code $C_{N=0}=\hilb_{n,t}$).

\begin{theorem}
  \label{thm:eta duality}
  Let $N\subset T$ be a stochastic isotropic subspace of dimension $m\geq\frac{t-n}{2}$, and let $\Delta^{(k)}_{r,s}$ be the subrepresentation of $\Delta_{r,s}$ spanned by all $\cliff$ irreps with rank $k$.
  Then, there exists some injective function $\eta_N:\irr\stoch(T_N)\to\irr\cliff$ for which, as a $\stoch(T)^N\times\cliff$ representation,
  \begin{align}
    \label{eq:shamgar spinful}
    C_N\cap\Delta^{(t-2m)}_{r,s} 
    \simeq
    \bigoplus_{\tau\in\irr\stoch(T_N)}\tau\otimes\eta_N(\tau).
  \end{align}
  It furthermore holds that $\dim\tau\leq\dim\eta_N(\tau)$.
\end{theorem}

To show this, we require an intermediate result whose proof we postpone to App.~\ref{app:proofs class}. 
Throughout this section we will use the identification 
\begin{align*}
  C_N\simeq\cc[\hom(\pos\to T_N)].
\end{align*}
As before, let $R_N$ be the representation of $\stoch(T)^N$ acting on $C_N$.

Any surjective $F: \pos \to T_N$ pulls back the form $q_N$ to $X$, as $q^F(\darg):= q_N(F\darg)$.
(Notice that $N$ is determined by $F$, which justifies our choice of notation.)

For any complement $\pos_F\subset\pos$ of $\ker F$, i.e.\ for any subspace which satisfies $\pos = \ker F \oplus \pos_F$, $F|_{\pos_F}$ is invertible.
Let $i_F$ be the inverse of $F|_{\pos_F}$ for some such choice of $\pos_F$, that is
\begin{align*}
	i_F: T_N \to \pos_F\subseteq \pos, \qquad \text{is such that} \qquad F i_F(v) = v.
\end{align*}
The space $\pos_F$ is not canonically associated to $F$---indeed any complement of $\ker F$ is a possible choice of $\pos_F$---, however, $i_F$ is determined by the choice of $\pos_F$.
The space $\pos_F$, equipped with $q^F|_{\pos_F}$, is isometric to $T_N$ through
\begin{align*}
  q^F(i_F(v))=q_N(v), \qquad \forall v\in T_N.
\end{align*}
This isometry pulls back the group action of $\gl(T_N)$ to $\pos_F$: for any $O\in \gl(T_N)$, we can define
\begin{align*}
	O_F := (i_F O i_F^{-1})\oplus \ii_{\ker F}\in \gl(\pos), 
\end{align*}
which fulfills
\begin{align}
	FO_F = OF.
\end{align}
The set of all such elements  of $\gl(\pos)$ forms a subgroup, $\gl_F\subseteq\gl(\pos)$,
\begin{align*}
  \gl(T_N)
  \simeq
  \gl_F:=
  \{ (i_F O i_F^{-1})\oplus\ii_{\ker F}  
  \div
  O\in\gl(T_N) \}
  \subseteq
  \gl(\pos)
  \subset 
  \cliff.
\end{align*}
At some points it will be convenient to use the following notation: for any $g\in \gl_F$, let $O_g\in\gl(T_N)$ be the unique element such that 
\begin{align}
  \label{eq:fof}
  O_g F = Fg.
\end{align}

\begin{remark}
  \label{rem:pull back of stoch}
  Because $i_F$ is an isometry, it pulls back $\orth(T_N)$ to the isometry group of $q^F|_{\pos_F}$.
  Moreover, it pulls back $\stoch(T_N)$ to the subgroup of all $q^F|_{\pos_F}$-isometries which preserve $i_F([\ones_t]_N)$.
\end{remark}

\begin{restatable}[]{lemma}{hfregular}
  \label{lem:hf regular}
  It holds that
  $J\in\hom(\pos\to T_N)$ is such that $q^F=q^J,\ F^{-1}([\ones_t]_N)=J^{-1}([\ones_t]_N)$ if and only if $J = OF$ for some $O \in \stoch(T_N)$.
  Moreover, $\stoch(T_N)\times\stoch(T_N)$ acts on the space
  \begin{align*}
    \hilb_F :&=
      \vspan\{
        \ket{OF} \div O\in\stoch(T_N)
      \}
  \end{align*}
  through $\stoch(T_N)\times\stoch(T_N)\ni(L,R):\ket{J}\mapsto\ket{LJ R_F^{-1}}$ as the regular representation.
\end{restatable}

The left action on the regular representation is given by $R_N$.
The right action, on the other hand, is given by $\Delta_{r,s}(G_F)$, where
\begin{align*}
  \stoch(T_N)
  \simeq
  G_F:=
  \{ (i_F O i_F^{-1})\oplus\ii_{\ker F}  
  \div
  O\in\stoch(T_N) \}
  \subseteq
  \gl(\pos)
  \subset 
  \cliff.
\end{align*}

With the preceding three lemmas, we may prove the main result of this section---Thm.~\ref{thm:eta duality}. 

\begin{proof}[Proof of Thm.~\ref{thm:eta duality}]
  The space $\hilb_F$ is a $\stoch(T_N)\times G_F$ representation.
  By the preceding discussion, we may decompose it as
  \begin{align*}
    \hilb_F
    \simeq
    \bigoplus_{\tau\in\irr\stoch(T_N)}\tau\otimes\tau^*,
  \end{align*}
  where the action on the right-hand side factors is obtained from the action of $\Delta_{r,s}(G_F)$ on $\hilb_F$.
  Acting with $\cliff$ on $\hilb_F$, we obtain some subrepresentation
  \begin{align}
    \label{eq:max rank CN}
    C_N \supseteq
    \cc[\Delta_{r,s}(\cliff)]\hilb_F\simeq
    \bigoplus_{\tau\in\irr\stoch(T_N)}\tau\otimes\eta_{F,N}(\tau),
  \end{align}
  where $\eta_{F,N}(\tau)]\in\irr\cliff$.
  
  The irrep $\eta_{F,N}$ is independent of $F$:
  For any $g\in\gl(\pos)\subset\cliff$ it holds that $\Delta_{r,s}(g)\hilb_F=\hilb_{Fg^T}$, and  since $n\geq t-2m$, any surjective $F'\in\hom(\pos\to T_N)$ is of this  form, $F'=Fg^T$ for some $g\in\gl(\pos)$.
  Therefore $\hilb_{F'}\subseteq\cc[\Delta_{r,s}(\cliff)]\hilb_F$ for all surjective $F$.
  
  This implies that every weight vector of $C_N$ with rank $t-2m$ is contained in the subspace of Eq.~\eqref{eq:max rank CN}.
  Conversely, every $\eta_N$ has rank $t-2m$.
  
  Finally, by Lem.~\ref{lem:hf regular}, if $\tau\not\simeq\tau'$ then $\res{G_F}\eta_N(\tau)\not\simeq\res{G_F}\eta_N(\tau')$ and therefore $\eta_N(\tau)\not\simeq\eta_N(\tau')$.
  The same lemma implies $\dim\tau=\dim\tau^*\leq\dim\eta_N(\tau)$. 
\end{proof}

\begin{remark}
  \label{rem:Gl}
  A similar procedure could be followed to obtain a duality between $\stoch(T)$ and any subgroup $G\subset\cliff$ satisfying $\gl(\pos)\subseteq G$.
  In this case, one obtains that there exists some injective function $\eta_{N,G}:\irr\stoch(T)\to\irr G$ for which
  \begin{align*}
    \cc[\Delta_{r,s}(G)]\hilb_F 
    \simeq
    \bigoplus_{\tau\in\irr\stoch(T)}\tau\otimes\eta_{N,G}(\tau).
  \end{align*}
  This duality was studied for the case $G=\gl(\pos)$ and $N=0$ in~\cite{ghgl}.
  Moreover, other interesting choices of $G$ could be the real Clifford group (in the quibt, i.e.\ $d=2$, case), and the CSS-ness preserving subgroup of the real Clifford group studied in~\cite{rebits}. 
  
  The difference to our case is that
  \begin{align}
    \label{eq:inclusion}
    \cc[\Delta_{r,s}(G)]\hilb_F \subseteq
    \cc[\Delta_{r,s}(\cliff)]\hilb_F,
  \end{align}
  where the inclusion can be strict for $G\neq\cliff$.
  While we have not attempted to prove that this inclusion is in fact strict for any of the choices for $G$ introduced above, we certainly expect this to be the case. 
  This is because the commutant of $\Delta_{r,s}(G)$ will typically strictly contain $\com{r,s}$ for sufficiently high values of $t$.
  
  Whenever the inclusion~\eqref{eq:inclusion} is actually an equality for all $N\in\grass\cup\grass^0$, the proof methods used in this paper may be straightforwardly modified to fully decompose $\Delta_{r,s}(G)$.
  The decomposition will have the same form as the decomposition in our main theorem, Thm.~\ref{thm:main}.
  If, however, the inclusion is strict for some $N$, the proof methods in this paper only provide a partial decomposition of $\Delta_{r,s}(G)$.
\end{remark}

\begin{lemma}
  \label{lem:independent of N}
  Let $N$ and $\eta_N$ be as in Thm.~\ref{thm:eta duality}.
  Consider some $N'$ satisfying $\dim N'=\dim N$ and $\ones_t\in N'\iff\ones_t \in N$. 
  Then, $\stoch(T)^N\simeq\stoch(T)^{N'}$ are conjugate in $\stoch(T)$ and, for any $\tau\in\stoch(T_N)$, $\eta_N(\tau)\simeq\eta_{N'}(\tau)$. 
\end{lemma}
\begin{proof}
  By Lem.~\ref{lem:stoch single orbit} there exists an $O\in\stoch(T)$ such that $ON=N'$.
  Then a short calculation shows that $O\stoch(T)^NO^{-1}=\stoch(T)^{N'}$.
  Let $P(N,\tau)$ be the projector onto the $\tau\otimes\eta_N(\tau)$ component in $C_N$.
  Then the isomorphism is afforded by $R(O)$.
  This is because, by the first claim, $R(O)P(N,\tau)R^\dagger(O)=P(N',\tau)$ and so,
  \begin{align*}
    R(O)P(N,\tau) \Delta_{r,s}(U) P(N,\tau) R^\dagger(O)
    &=
    P(N',\tau)\Delta_{r,s}(U)P(N',\tau),
    \quad
    \forall\ U\in\cliff,\\
    R(O)P(N,\tau) R(O') P(N,\tau) R^\dagger(O)
    &=
    P(N',\tau) R(OO'O^{-1}) P(N',\tau),
    \quad
    \forall\ O'\in\stoch(T)^N.
  \end{align*}
\end{proof}

\subsection{The main theorem}
\label{sec:main theorem}

In this section we show our main result: the decomposition of $\Delta_{r,s}$.
For this, it is instructive to first consider the case where $r-s=0\mod D$ (so that the all-ones vector is isotropic, $q_{r,s}(\ones_t)=0$).%
\footnote{Recall that $D=d$ if $d$ is odd, and if $d=2$ then $D=4$.} 
In this case, the code space $\cones$ corresponding to the space $N=\langle\ones_t\rangle$ is a $\stoch(T)\times\cliff$ subrepresentation of $\hilb_{n,t}$.
In other words, $\pones$ is in the commutant of both $\stoch(T)$ and $\cliff$.
Moreover, this code space is an $\com{r,s}$-subrepresentation: 
Because the elements of the stabilizer group of $\cones$ are tensor power Pauli operators, i.e.\ $W^{\otimes(r,s)}$ with $W\in\pauli$, it holds that $\pones\in\cc[\Delta_{r,s}]$ commutes with any tensor power CSS code projector $P(N)$.
The starting point of our main theorem is thus the elementary decomposition
\begin{align}
  \label{eq:elementary}
  \hilb_{n,t}=\cones\oplus\cones^\perp.
\end{align}
as a  representation of the algebra  $\com{r,s}\Delta_{r,s}$ generated by $\com{r,s}$ and $\Delta_{r,s}$.
For notational uniformity, we let $\cones:=\{0\}$ whenever $r-s\neq0\mod D$---this way Eq.~\eqref{eq:elementary} holds also in this case.

The two terms in Eq.~\eqref{eq:elementary}  behave rather differently as Clifford representations.
For example, $\cones$ is precisely the maximal $\cliff$-subrepresentation of $\Delta_{r,s}$ which has the Pauli group in its kernel.
This way, $\Delta_{r,s}|_{\cones}$ realizes a homomorphism $\cliff\to\cliff/\pauli\simeq\sp(V)$.
We will thus decompose each of the two terms in Eq.~\eqref{eq:elementary} separately.
The main result in this paper is Thm.~\ref{thm:main}.

Recall that the group $\stoch(T)^N:=\{O\in\stoch(T)\div ON=N\}$ acts on the code space $C_N$ through a representation $R_N$ (which simply corresponds to $R|_{C_N}$). 
This representation satisfies, furthermore, that $\stoch(T)^N/\ker(R_N)\simeq\stoch(T_N)$, $T_N=N^\perp/N$.
More generally -- because of this homomorphism at the level of groups -- arbitrary representations of $\stoch(T_N)$, say $\rho$, may be seen as representations of $\stoch(T)^N$ by extending them trivially through
\begin{align*}
  \stoch(T)^N\to\stoch(T_N)\to\rho(\stoch(T_N)).
\end{align*}
This gives a canonical injection $\irr\stoch(T_N)\subset\irr\stoch(T)^N$.
In particular, $\stoch(T)^N$-subrepresentations of $C_N$ are identified in this way with $\stoch(T_N)$ representations.

\begin{theorem}
  \label{thm:main}
  Let $t\leq n$. 
  Consider a subset $\{N_i\}_i\subset\grass\cup\grass^0$ that contains exactly one element per  $\stoch(T)$ orbit.
	In particular, for each non-empty orbit $\grass_m$ (resp. $\grass_m^0$) there is exactly one $i_m\in\nn$ such that $N_{i_m}\in\grass_m$ (resp. one $i_m^0\in\nn$ such that $N_{i_m^0}\in\grass_m^0$).
  Let $T_i:=N_i^\perp/N_i$.
  Then, there exists an injective function 
  \begin{align*}
    \eta:\bigcup_i \irr\stoch(T_i)\to\irr\cliff,
  \end{align*}
  fir which
  \begin{align*}
    \Delta_{r,s}
    \simeq
    \bigoplus_i
    \bigoplus_{\substack{\tau\in\\ \irr\stoch(T_i)} }
    \ind{\stoch(T)^{N_i}}{\stoch(T)}(\tau)\otimes\eta(\tau).
  \end{align*}
	For each $\tau\in\irr\stoch(T_i)$ it holds that $\dim\eta(\tau)\geq\dim\tau$ and $\rk\ \eta(\tau)=\dim T_i$.
	Moreover $\pauli\subset\ker\eta(\tau)$ for all $\tau\in\irr\stoch(T_i)$ if  and  only  if $T_i\in\grass^0$.
\end{theorem}

Consider the operators
\begin{align*}
  C_m' := \sum_{N\in\grass_m}    P(N),\qquad
  D_m := \sum_{N\in\grass_m^0}  P(N)
\end{align*}
(the use of a ``prime'' symbol in $C_m'$ will be justified soon).
These commute with both the stochastic orthogonal group \emph{and} the Clifford group.
Thus, the spaces
\begin{align*}
  \cm{m}':=\range C_m', \qquad
  \dm{m}:=\range D_m
\end{align*}
are $\stoch(T)\times\cliff$-subrepresentations.
We may further see that
\begin{align*}
  \cm{m}'=\vspan\{C_N\div N\in\grass_m\},\qquad
  \dm{m}=\vspan\{C_N\div N\in\grass_m^0\}.
\end{align*}

Recall that $\pones$ commutes with all $P(N)$ and thus with each operator $C_m'$ and $D_m$.
The projection acts trivially on $\dm{m}$, while it acts on $C_m'$ as 
\begin{align}
  \label{eq:pones commutes}
  \pones C_m' = C_m'\pones = D_{m+1}.
\end{align}
This follows because, by $\ones_t\in N^\perp$ (stochasticity), $\pones P(N) = P(\langle N,\ones_t\rangle)$.
Thus $C_m' = (\ii-\pones)C_m' + D_{m+1}$ and, at the level of spaces,
\begin{align*}
  \cm{m}' \simeq (\cones^\perp\cap\cm{m}')\oplus\dm{m+1}
\end{align*}
is a $\stoch(T)\times\cliff$-decomposition.
Since our goal is to decompose $\cones$ and $\cones^\perp$ separately, it makes sense to consider
\begin{align*}
  C_m := (\ii-\pones)C_m',\qquad
  \cm{m} := \range C_m = \cm{m}'\cap\cones^\perp.
\end{align*}
In particular, $\dm{1}=\cones$ and $\cm{1}=\cones^\perp$.
Throughout, we let $P(\cm{m}')$, $P(\cm{m})$ and $P(\dm{m})$ be the projectors onto the  corresponding spaces.

While our goal is to decompose $\hilb_{n,t}$ as an $\stoch(T)\times\cliff$ representation, it is useful to consider first the action of the larger algebra $\com{r,s}\Delta_{r,s}$ (as the remarks around Eq.~\eqref{eq:elementary} suggest).

\begin{lemma}
  \label{lem:mod out ideals}
  It holds that
  \begin{enumerate}
    \item
    the algebra $\com{r,s}\Delta_{r,s}$ acts on $\cm{m}$ and $\dm{m}$,
    \item
    the action of $\com{r,s}$ on $\langle\dm{m},\cm{m}\rangle^\perp$ contains the ideal $\com{r,s}^m$ in its kernel,
    \item
    if $r-s=0\mod d$, the action on $\dm{m}^\perp$ contains the ideal $\com{r,s}^{m,0}$ in its kernel.
  \end{enumerate}
\end{lemma}
\begin{proof}
  Consider any $N\in\grass\cup\grass^0$.
  Then $\cliff$ acts on $C_N$ and thus on $\cm{m}$ and $\dm{m}$ (by Prop.~\ref{prop:gnw}).
  Moreover, Eq.~\eqref{eq:o on code} shows that $\stoch(T)$ also acts on these spaces. 
  Now, notice that $P(N)$ preserves $\cm{m}'$ by Prop.~\ref{prop:semi group}, Point 2.
  Then, $P(\cm{m})=P(\cm{m}')\pones$ commutes with $P(N)$.
  This proves the  first point.

  The second point follows from the fact that for each $R(O)P(N)\in\semigroup_{r,s}^m$, it holds that $C_N\subseteq\langle\dm{m},\cm{m}\rangle$. 

  The third point follows similarly: if $R(O)P(N)\in\semigroup_{r,s}^{m,0}$, it holds that $C_N\subseteq\dm{m}$.
\end{proof}

\begin{remark}
  \label{rem:eigenspaces}
  As noted above, the spaces $\cm{m}$ and $\dm{m}$ are representations of the full algebra  $\com{r,s}\Delta_{r,s}$, i.e.\ $P(\cm{m})$ and $P(\dm{m})$ commute with both $\com{r,s}$ and $\Delta_{r,s}$.
  The operators $C_m$ and $D_m$, however, do not share this property:
  In fact $C_m$ and  $D_m$ commute only with the action of $\stoch(T)\times\cliff$.
  
  This hints at  the possibility of studying the  eigenspaces of these operators to obtain explicit descriptions of the $\stoch(T)\times\cliff$-subrepresentations of $\hilb_{n,t}$.
  Looking at the spectrum  of  these types of operator has, moreover, already been useful when attempting to build approximations to the ‘‘Clifford twirl’’ operation~\cite{heinrichthesis},
  \begin{align*}
    \endom(\hilb_{n,t})\ni
    A\mapsto\sum_{U\in\cliff} U^{\otimes  t} A (U^\dagger)^{\otimes  t}.
  \end{align*}
  This avenue we leave for future work.
\end{remark}

By the discussion above, the following inclusions hold
\begin{align*}
  \cm{m+1}\subseteq\cm{m}, \quad
  \dm{m+1}\subseteq\dm{m}.
\end{align*}
In this way, we may define $\spinful{m}:=\cm{m}\cap\cm{m+1}^\perp$ and $\spinless{m}:=\dm{m}\cap\dm{m+1}^\perp$ so that, by Lem.~\ref{lem:mod out ideals}, the following equations are decompositions as $\com{r,s}\Delta_{r,s}$ representations:
\begin{align}
  \label{eq:H=K+L}
  \cones^\perp=\bigoplus_{m}\spinful{m}, \qquad
  \cones=\bigoplus_{m}\spinless{m}.
\end{align}

The goal is thus to decompose the spaces $\spinful{m}$ and $\spinless{m}$ as $\stoch(T)\times\cliff$ representations.
We do so by conisdering their intersection with CSS codes.
At several points, there will be statements that hold true both for $\spinful{m}$ and $\spinless{m}$.
Whenever this is the case, we use the placeholder $\egal{m} = \spinful{m}$ or $\spinless{m}$.
Moreover, $\grass(\egal{m})=\grass_m$ if $\egal{m}=\spinful{m}$, and $\grass(\egal{m})=\grass_m^0$ if $\egal{m}=\spinless{m}$.

\begin{lemma}
  \label{lem:span of intersections}
  Each space $\egal{m}$ 
	satisfies $\egal{m}  = \vspan\{ C_N\cap\egal{m} \div N\in\grass(\egal{m})\}$.
\end{lemma}
\begin{proof}
	Consider an arbitrary element $\ket{\Psi}\in\egal{m}$, which we may expand as
  \begin{align*}
    \ket{\Psi}
    =
    \sum_{N\in\grass(\egal{m})}\ket{\Psi_N},
    \qquad
    \ket{\Psi_N}\in C_N.
  \end{align*}
  Then,
  \begin{align*}
    \ket{\Psi}=P(\egal{m})\ket{\Psi}
    =
    \sum_{N\in\grass(\egal{m})}P(\egal{m})\ket{\Psi_N},
  \end{align*}
  where $P(\egal{m})\ket{\Psi_N}\in C_N\cap\egal{m}$ since
  \begin{align*}
    P(N)P(\egal{m})\ket{\Psi_N}
    =
    P(\egal{m})P(N)\ket{\Psi_N}
    =
    P(\egal{m})\ket{\Psi_N}.
  \end{align*}
  This proves that $\ket{\Psi}$ may be expanded in terms of elements in $\{\egal{m}\cap C_N \div N\in\grass(\egal{m})\}$.
\end{proof}

The intersection spaces $C_N\cap\egal{m}$ are $\stoch(T)^N\times\cliff$ subrepresentations.
Notice that the action of $O\in\stoch(T)\setminus\stoch(T)^N$ non-trivially permutes these intersection spaces while keeping $\egal{m}$ invariant. 
Thus $C_N\cap\egal{m}$ is not a $\stoch(T)$-subrepresentation.

\begin{lemma}
  \label{lem:subcode rank-deficient}
  Assume that $N\in\grass(\egal{m})$ and $t-2m\leq n$. 
  Then, 
  \begin{align*}
    C_N\cap\Delta_{r,s}^{(t-2m)}
    =
    C_N\cap\egal{m}.
  \end{align*}
\end{lemma}

\begin{remark}
	\label{rem:rank-deficient}
	The concept behind this lemma is easiest to understand in the case where $r-s\neq0\mod D$.
	In this case 
	\begin{align*}
		\egal{m}=\spinful{m} 
		=
		\vspan\{ C_N \div \dim N = m \} \cap \vspan\{ C_N \div \dim N = m+1 \}^\perp,
	\end{align*}
	and,
	\begin{align*}
		C_N\cap\egal{m}
		=
		C_N \cap \vspan\{ C_{N'} \div \dim N' = m+1 \}^\perp.
	\end{align*}
	Thus, Lem~\ref{lem:subcode rank-deficient} states that the ‘‘maximal rank subspace’’ $C_N\cap\Delta_{r,s}^{(t-2m)}$ of $C_N$ is equal to the subspace of $C_N$ which is orthogonal to all $C_{N'}$ with $\dim N'>\dim N$.
	Equivalently, the aforementioned lemma states that all ‘‘rank-deficient’’ subrepresentations of $C_N$ (that is, subrepresentations with rank \emph{strictly smaller} than $t-2m$) live in the span of lower-rank codes, i.e.\ in $C_N\cap\cm{m+1}'$.
\end{remark}

\begin{proof}
  By Lem.~\ref{lem:light cone}, the commutant of $\cliff$ in $\endom(C_N\cap\egal{m})$ is contained in $\cc[\semigroup_{r,s}^N]$.
  Because $\dm{m}\subset\cones$, Lem.~\ref{lem:mod out ideals} implies that any element of $\com{r,s}^{m+1}$ has $\spinful{m}$ in its kernel and, analogously, the same is true for $\com{r,s}^{m+1,0}$ and $\spinless{m}$.
	In symbols:
  \begin{align}
    \label{eq:layer kernel}
		\com{r,s}^{m+1}|_{\spinful{m}}=0,
		\qquad
		\com{r,s}^{m+1,0}|_{\spinless{m}}=0.
  \end{align}
  This way, for any $R(O)P(N')\in\semigroup_{r,s}^N$,
  \begin{align*}
    R(O)P(N')|_{C_N\cap\egal{m}}
    =
    \begin{cases}
      R(O)|_{C_N\cap\egal{m}},\qquad & N'=N,\\
      0,\qquad & N'\supsetneq N.
    \end{cases}
  \end{align*}
  Thus, the commutant of $\cliff$ in $\endom(C_N\cap\egal{m})$ is generated by the subgroup of $\stoch(T)$ for which $R(O)\cdot(C_N\cap\egal{m})=C_N\cap\egal{m}$.
  Because $R(O)\egal{m}=\egal{m}$ for all $O\in\stoch(T)$, it is sufficient to require $R(O) C_N= C_N$, which happens if and only if $O\in\stoch(T)^N$.
  By Lem.~\ref{lem:kernel on code action} the action of $\stoch(T)^N$ on $C_N$ realizes the homomorphism $\stoch(T)^N\to\stoch(T_N)$.
  
  Then, the commutant of $\cliff$ in $\endom(C_N\cap\spinful{m})$ is generated by $R(\stoch(T)^N)|_{C_N}$ and thus there exists some injective function $\t\eta:\irr\stoch(T_N)\to\irr(\cliff)$ for which
  \begin{align*}
    C_N\cap\egal{m}
    \simeq
    \bigoplus_{\tau\in\irr\stoch(T_N)}\tau\otimes\t\eta(\tau),
  \end{align*}
  where, by the same argument as in the proof of Thm.~\ref{thm:eta duality}, the sum ranges over all $\irr\stoch(T_N)\subseteq\irr\stoch(T)^N$ since $t-2m\leq n$.
  
  Every $F\in\hom(\pos\to N^\perp)$ for which $\rank FM_{r,s}F^T=t-2m$, is orthogonal to any subcode $C_{N'}\subsetneq C_N$, and therefore
  \begin{align}
    \label{eq:from comment}
    C_N\cap\Delta_{r,s}^{(t-2m)}
    \subseteq
    C_N\cap\egal{m},
  \end{align}
  which implies that $\eta_N(\tau)\subseteq\t\eta(\tau)$ for each $\tau$ (where $\eta_N$ is as in Thm.~\ref{thm:eta duality}).
  However, $\t\eta(\tau)$ is irreducible so that $\t\eta(\tau)=\eta_N(\tau)$ and 
  \begin{align*}
    C_N\cap\Delta_{r,s}^{(t-2m)}
    \simeq
    C_N\cap\egal{m},
  \end{align*}
  and so the result follows.
\end{proof}

The previous lemma implies that, by Thm.~\ref{thm:eta duality}, each intersection space decomposes, as a $\stoch(T)^N\times\cliff$ representation, as
\begin{align}
	\label{eq:cn egal}
  C_{N}\cap\egal{m} \simeq \bigoplus_{\tau\in\stoch(T_N)}\tau\otimes\eta(\tau).
\end{align}
Here, by Lem.~\ref{lem:independent of N}, $\eta$ is dependent only on whether $\egal{m}=\spinful{m}$ or $\egal{m}=\spinless{m}$, not on the specific subspace $N(\egal{m})$ chosen. 

Lem.~\ref{lem:lin indep} shows that $\egal{m}$ is  simply a direct sum of  the spaces $C_N\cap\egal{m}$.
This allows us to directly use Eq.~\eqref{eq:cn egal} to decompose $\egal{m}$.
Taken together, this lemma and Thm.~\ref{thm:rank deficient} are analogous to~\cite[Lem.~3.3]{rank-deficient}.
The  idea of the proof of Lem.~\ref{lem:lin indep} is the  following:
For any Clifford irrep $\rho\subset C_N\cap\egal{m}$, we find a vector $\ket{\Psi}\in\rho$ which satisfies
\begin{align*}
	\ket{\Psi}\notin\vspan\{C_{N'}\cap\egal{m} \div N'\in\grass(\egal{m}),\ N'\neq N\,\}
	=: \egal{m,(N)}.
\end{align*}
It follows that $\rho\cap\egal{m,(N)}=\{0\}$ trivially intersect.
This is because $\egal{m,(N)}$ is Clifford invariant and the Clifford orbit of $\ket{\Psi}$,
\begin{align*}
	\{ \Delta_{r,s}(U)\ket{\Psi} \}_{U\in \cliff},
\end{align*}
spans $\rho$ (since it is irreducible).
These results then imply the direct sum decomposition.

\begin{lemma}
  \label{lem:lin indep}
  Let $m$ be such that $t-m\leq n$.
  Then, for any $N\in\grass(\egal{m})$, the space $C_N\cap\egal{m}$ intersects trivially with $\egal{m,(N)}$.
	In other words, we have the direct sum decomposition
  \begin{equation}
    \label{eq:lin indep}
      \egal{m} = \bigoplus_{N\in\grass(\egal{m})}(C_N\cap\egal{m}).
  \end{equation}
\end{lemma}

\begin{remark}
  \label{rem:orthogonality}
  Notice that the terms in the direct sums in Eq.~\eqref{eq:lin indep} are \emph{not} in general orthogonal to each other.
\end{remark}

\begin{proof}
  Throughout the proof, we let 
  \begin{align*}
    \supp C_N := \bigcup_{\psi\in C_N} \supp\psi.
  \end{align*}
  Consider one of the terms in the right-hand side of~\eqref{eq:lin indep}.
  By Lem.~\ref{lem:subcode rank-deficient}, $C_N\cap\egal{m}$ is the span of all irreps with maximal rank in $C_N$, i.e.,
  \begin{align*}
    C_N\cap\egal{m} 
    =
    \vspan\{ 
      \Delta_{r,s}(U)\Psi_B \div U\in\cliff,\ 
      \Psi_B\in C_N\text{ wght. vec. with } \rank B = t-2m
    \}.
  \end{align*}
  We claim that all weight vectors $\Psi_B$ with rank $t-2m$ in $C_N$ are linear combinations of $\ket{[F]_N}$ where $\range F = N^\perp$.
  To see this, write
  \begin{align*}
    \Psi_B = \sum_{F'\in\hom(\pos\to N^\perp)} c_{F'}\ket{[F']_N}.
  \end{align*}
  Now, for any $F'$ with $c_{F'}\neq0$, according to Lem.~\ref{lem:weights tensor power},
  \begin{align*}
    \rank\Psi_B = t-2m = \rank\beta_{r,s}|_{\range F'} = \rank\beta_{r,s}|_{N^\perp}.
  \end{align*}
  This implies that $\langle\range F', N\rangle=N^\perp$.
  This way, because $t-m\leq n$, there exists some $F\in[F']_N$ for which $\range F = N^\perp$.
  Furthermore, for any two distinct coset states $\ket{[F]_N}\neq\ket{[F']_N}$ their supports are disjoint, so no cancellations happen when combining different coset states. 
  In particular, there at least one $F\in\supp\Psi_B$ for which $\range F=N^\perp$.
  
  The corresponding standard basis vector $\ket{F}$ is orthogonal to all other codes $C_{N'}$, where $N'\in\grass(\egal{m})$ and $N'\neq N$.
  To prove this we argue by contradiction.
  If $F \in \supp(C_N)\cap \supp(C_{N'})$ then $N^\perp=\range F \subseteq (N')^\perp$.
  But this can not hold because $N^\perp$ and $(N')^\perp$ are of the same dimension and distinct.
  
  This way,
  \begin{align*}
    \{\Psi_B\in C_N\text{ wght. vec. with } \rank B = t-2m\}
    \cap
    \vspan\{ C_{N'}\div N'\in\grass(\egal{m}), \, N'\neq N\}
    =
    \emptyset.
  \end{align*}
  Consider an  arbitrary irreducible $\cliff$-subrepresentation $\rho\subseteq\egal{m}\cap C_N$ with a maximal rank weight vector, say, $\ket{\Psi_B}$. 
  For any $\ket{\psi}\in\rho$, the set $\Delta_{r,s}(\cliff)\ket{\psi}$ spans $\rho$.
  Therefore, for any $\grass(\egal{m})\ni N'\neq N$, it holds that  if
  \begin{align}
    \label{eq:rho intersect}
    \rho\cap
    \vspan\{ C_{N'}\div N'\in\grass(\egal{m}), \, N'\neq N\}
  \end{align} 
  is non-empty, then it is equal to  $\rho$.
  However, this would imply that
  \begin{align*}
    \ket{\Psi_B}\in\vspan\{ C_{N'}\div N'\in\grass(\egal{m}), \, N'\neq N\},
  \end{align*}
  a contradiction.
  Thus, the  intersection in Eq.~\eqref{eq:rho intersect} is empty and  the claim follows.
\end{proof}

By Lems.~\ref{lem:subcode rank-deficient} and~\ref{lem:span of intersections}, 
\begin{align*}
  \egal{m} = \vspan\{ C_N\cap\Delta_{r,s}^{t-2m} \div N \in \grass(\egal{m})\} 
  \subseteq
  \Delta_{r,s}^{t-2m}.
\end{align*}
Since
\begin{align*}
  \hilb_{n,t}
  =
  \begin{cases}
    \bigoplus_{m=0}^{m(T)}\spinful{m} \quad &\text{if} \quad  r-s\neq0\mod D,\\
    \bigoplus_{m=0}^{m(T)}\spinful{m}\oplus\spinless{m+1}
    \quad &\text{if} \quad  r-s=\mod D,\\
  \end{cases}
\end{align*}
this allows us to conclude that, for all $m\geq \frac{t-n}{2}$,
\begin{align}
  \label{eq:rank code connection}
  \Delta_{r,s}^{(t-2m)}=
  \begin{cases}
    \spinful{m}, \quad &\text{if} \quad r-s\neq0\mod D \text{ or } m=0,\\
    \spinful{m}\oplus\spinless{m}, \quad &\text{if} \quad  r-s=0\mod D,\ m>0.
  \end{cases}
\end{align}
We now look at the representation spaces on the right-hand side of Eq.~\eqref{eq:rank code connection}. 

\begin{theorem}
  \label{thm:rank deficient}
  Assume that $m,r,s$ are such that $\grass(\egal{m})$ is non-empty and that $t-2m < n$.
  Let $N\in\grass(\egal{m})$.
  Then, there exists an injective map $\eta:\irr\stoch(T_N)\to\irr\cliff$ such that, as a representation of $\stoch(T)\times\cliff$,
  \begin{align}
    \label{eq:km decomposition}
    \egal{m}
    \simeq 
    \bigoplus_{\tau\in\irr\stoch(T_N)} 
    \ind{\stoch(T)^N}{\stoch(T)}(\tau)\otimes\eta(\tau),
  \end{align}
  where $\dim\eta(\tau)\geq\dim\tau$ for all $\tau$.
\end{theorem}

\begin{remark}
  \label{rem:on thm rank deficient}
  If $\grass(\egal{m})=\grass_m^0$ in Thm~\ref{thm:rank deficient}, the assumption of the theorem implies that $r-s=0\mod D$ and $m\leq m(T)+1$.
  In this case, $\egal{m}\subseteq\cones$ and $\range(\eta)\subseteq\irr\sp(V)\subseteq\irr\cliff$.
\end{remark}

\begin{proof}
  By Thm.~\ref{thm:eta duality}, there is a $\stoch(T)^N\times\cliff$-subrepresentation $\mcal{R}_N^\tau\subset C_N\cap\egal{m}$ for which 
  \begin{align*}
    \mcal{R}_N^\tau
    \simeq
    \tau\otimes\eta_N(\tau),
    \qquad
    \tau\in\irr\stoch(T_N).
  \end{align*}
  (Notice that $\mcal{R}_N^\tau$ is \emph{not necessarily} $\stoch(T)$-invariant.)
  Acting with $\stoch(T)$ on $\mcal{R}_N^\tau$, we get some $\stoch(T)\times\cliff$-representation $\hilb_{n,t}\supseteq\mcal{R}^{\tau}\simeq I(\tau)\otimes\eta_N(\tau)$, where $I(\tau)$ is some representation of $\stoch(T)$.
  We will show that $I(\tau)\simeq\ind{\stoch(T)^N}{\stoch(T)}(\tau)$.
  
  We may identify the spaces in the orbit $\{ON \div O\in\stoch(T)\}=\grass(\egal{m})$ (see Lem.~\ref{lem:stoch single orbit}) with right cosets $\stoch(T)/\stoch(T)^N$.
  Define a set of representatives $\{O_i\}$, one for each such coset, and let $\mcal{R}_i^\tau=R(O_i)\mcal{R}_N^\tau$.
  By Lem.~\ref{lem:lin indep}, the spaces $\{\mcal{R}_i^\tau\}$ are linearly independent,
  i.e.\ $\mcal{R}^{\tau}=\oplus_i\mcal{R}_i^{\tau}$.
  We now compute the action of $\stoch(T)$ on $\mcal{R}^{\tau}$.
  It is sufficient to compute it on vectors of the form
  \begin{align*}
    R(O_i)(\psi\otimes\phi), \quad
    \psi\otimes\phi\in \mcal{R}_N^\tau,
  \end{align*}
  where, for notational simplicity, we have kept the isomorphism
  $\mcal{R}_N^\tau\simeq\tau\otimes\eta_N(\tau)$
  implicit.
  For any $O\in\stoch(T)$ there is a permutation $\pi:i\to\pi(i)$ and an $O'\in\stoch(T)^N$ such that $OO_i = O_{\pi(i)}O'$.
  This implies that
  \begin{align*}
    R(O)R(O_i)(\psi\otimes\phi)
    =
    R(O_{\pi(i)})\big((\tau(O')\psi)\otimes\phi\big),
  \end{align*}
  so $I(\tau)\simeq\ind{\stoch(T)^N}{\stoch(T)}(\tau)$ as claimed.
\end{proof}

\begin{proof}[Proof of the main theorem, Thm.~\ref{thm:main}]
  Thm.~\ref{thm:rank deficient} and eq.~\eqref{eq:H=K+L} show that the claimed equation is a valid decomposition of $\hilb_{n,t}$ as a $\stoch(T)\times\cliff$ representation.
  
  We now show that $\eta$ is injective.
  By Thm.~\ref{thm:rank deficient}, $\eta|_{\irr\stoch(T_i)}$ is injective for each $i$.
	Moreover, let $N_1$ and $N_2$ be isotropic stochastic subspaces of $T$ and let $\tau_1\in\irr\stoch(T_{N_1})$ and $\tau_2\in\irr\stoch(T_{N_2})$. 
	If $m_1 := \dim N_1 \ neq  \dim N_2 =: m_2$, then by Thm.~\ref{thm:rank deficient}, 
  \begin{align*}
    m_1=\rank\eta(\tau_1)\neq\rank\eta(\tau_2)=m_2,
  \end{align*}
  and hence $\eta(\tau_1)\not\simeq\eta(\tau_2)$.
	On the other hand, if $N_1\in\grass_m$ and $N_2\in\grass_m^0$, then $\pauli\subseteq\ker\eta(\tau_2)$, while, on  the  contrary, $\pauli\not\subseteq\ker\eta(\tau_1)$.
\end{proof}

We can rederive a classical result from the invariant theory of the Clifford group~\cite{nebe_invariants,nebe-book,runge96} as a corollary of our results.

\begin{corollary}
  \label{cor:invariants}
  The representation $\Delta_{r,s}$ contains a $\cliff$-trivial subrepresentation if and  only if either of the two following conditions hold:
  \begin{enumerate}
    \item $d$ is odd, $r-s=0\mod d$ (and thus $t=2t'$  is  even), and  $s=t'\mod 2$,
    \item $d=2$, and 
    \begin{align}
      r-s=0\mod 8. \label{eq:mod 8}
    \end{align}
    Here, Eq.~\eqref{eq:mod 8} implies that $t=2t'$ is even.
  \end{enumerate}
  In both cases considered above, the trivial component is equal to $\spinless{t'}$.
\end{corollary} 
\begin{proof}
  The representation $\Delta_{r,s}$ only has a trivial component (say, $\rho$)  if $r-s=0\mod D$, in which case $\rho\subset\cones$.
  Furthermore $\rk(\rho)=0$ so that $\rho\subseteq\Delta_{r,s}^{(0)}$.
  In particular, necessarily it holds that $t=2t'$, and in this case
  \begin{align*}
    \Delta_{r,s}^{(0)} = 
    \spinless{t'}.
  \end{align*}
  Here we used eq.~\eqref{eq:rank code connection} together with the fact that a stochastic isotropic $N$ must contain $\ones_t$ to be maximal.
  
  Now, $\Delta_{r,s}^{(0)}$ is non-zero if and only if $\grass_{t'}^0$ to be non-empty.  
  If $d$ is odd, this happens exactly when $T\simeq\hh^{\oplus t'}$, or equivalently, when $\discr(\beta_{r,s})=(-1)^s=(-1)^{t'}$.
  In this case, $\Delta_{r,s}^{(0)}=\rho$ by Lem.~\ref{lem:rank 0}.
  
  If $d=2$, $\Delta_{r,s}^{(0)}$ may contain non-trivial components according to Lem.~\ref{lem:rank 0}.
  In particular, $(\omega_8\ii)^{\otimes(r-s)}=1$ if and only if~\eqref{eq:mod 8} holds.
  This way, $\rho$ is non-trivial only if the latter equation holds, in which case $\rho=\Delta_{r,s}^{(0)}$.
  We now claim that $\Delta_{r,s}^{(0)}$ is non-zero whenever~\eqref{eq:mod 8} holds.
  
  To prove this, we show that $\grass_{t'}^0$ is not empty.
  By Lem~\ref{lem:rs equivalence}, $\Delta_{r,s}\simeq\Delta_{t',t'}$ and by Prop.~\ref{prop:equivalence qrs} $q_{r,s}\sim  q_{t',t'}$.
  We realize $q_{t',t'}$ within $T$ in the same way as we defined $q_{r,s}$ on this space, simply taking $r=s=t'$.
  Then, the space
  \begin{align*}
    \vspan\{e_i+e_{i+t'}\div i=1,\dots,t'\}
  \end{align*}
  is isotropic and of maximal dimension $t'$.
  Indeed, 
  \begin{align*}
    q_{t',t'}(e_i+e_{i+t'}) = 
    q_{t',t'}(e_i) + q_{t',t'}(e_{i+t'}) + 2\beta_{t',t'}(e_i,e_{i+t'})
    =
    1-1+0=0,
  \end{align*}
  where $\beta_{t',t'}$ is the generalized polarization of $q_{t',t'}$.
  As before, $\beta_{t',t'}$ is simply the dot product on $T$, and thus $\beta_{t',t'}(e_i+e_{i+t'},e_j+e_{j+t'})=0$.
  
  But then, the equivalent form $q_{r,s}$ also has a $t'$-dimensional isotropic subspace, i.e., $\grass_{t'}^0$ is not empty.
\end{proof}

We finally point out how the spaces corresponding to all representations with a fixed rank, $\langle\spinful{m},\spinless{m}\rangle$, are singled out rather canonically by the action of $\com{r,s}$ on them.

\begin{lemma}
  \label{lem:ideal kernel}
  The space $\spinful{m}\oplus\spinless{m}$ is the maximal $\com{r,s}$-subrepresentation of $\hilb_{n,t}$ which has $\com{r,s}^{m+1}$ in its kernel but not $\com{r,s}^m$.
\end{lemma}
\begin{proof}
  Consider a distinct pair of subspaces $\mcal{S}_i=\spinful{m_i}\oplus\spinless{m_i}$, with $i=1,2$ and $m_1\neq m_2$.
  Each of these spaces is a $\cliff$ representation and, moreover, they share no irrep in common.
  That is, if $\chi_i$ is the character of $\mcal{S}_i$, then the character inner product vanishes, $\langle\chi_1,\chi_2\rangle_{\cliff}=0$.
  To see this, notice that by eq.~\eqref{eq:rank code connection}, every irrep in $\mcal{S}_i$ has rank $m_i$.

  Thus, by Schur's lemma, each space $\spinful{m}$, $\spinless{m}$ is preserved by $\com{r,s}$.
  In particular one may block diagonalize any $A\in\com{r,s}$ as
  \begin{align*}
    A = \bigoplus_m A_m,
  \end{align*}
  where $A_m$ acts on $\spinful{m}\oplus\spinless{m}$.
  By Lem.~\ref{lem:mod out ideals}, $\spinful{m}\oplus\spinless{m}$ mods out $\com{r,s}^{m+1}$. 
  On the other hand, $C_m\in\com{r,s}^{m}$ acts positively on $\spinful{m}\oplus\spinless{m}$ and is thus not modded out by any $\com{r,s}$-subrepresentation of this space.
  
  Analogously, any $\com{r,s}$-irrep in some $\spinful{m'}\oplus\spinless{m'}$ with $m'\neq m$ either mods out $\com{r,s}^m$ (if $m'<m$), or does not mod out $\com{r,s}^{m+1}$ (if $m<m'$).
\end{proof}


\subsection{Stabilizer tensor powers}
\label{sec:stab}

In~\cite{gnw} it was shown that the space spanned by stabilizer tensor powers is the trivial $\stoch(T)$-subrepresentation of $\hilb_{n,t}$:
\begin{align*}
  \cc\stabs_{n,t}
  :=
  \vspan\left\{
    \ket{s}^{\otimes t} \div \ket{s}\in\stabs_n
  \right\}
  =
  \hilb_{n,t}^{\stoch(T)}.
\end{align*}
Independently,~\cite{rank-deficient} decomposes the trivial $\orth(T)$-subrepresentation of the same Hilbert space (for $d$ odd) as
\begin{align}
  \label{eq:trivial orth rep}
  \hilb_{n,t}^{\orth(T)}
  \simeq
  \bigoplus_r \eta(\id_{\stoch(T_r)}).
\end{align}
Here we show, with an analogous calculation, that the space $\cc\stabs_{n,t}$ decomposes in a  similar fashion to Eq.~\eqref{eq:trivial orth rep}.

Consider the decomposition of $\hilb_{n,t}$ into $\stoch(T)$-isotypes,
\begin{align}
  \label{eq:stoch isotypes}
  \hilb_{n,t}
  \simeq
  \bigoplus_{\tau\in\irr\stoch(T)}\tau\otimes\Theta(\tau),
\end{align} 
where $\Theta(\tau)$ is a (possibly reducible) $\cliff$-representation.
We are interested in decomposing $\Theta(\id_{\stoch(T)})$.

Let $N\in\grass\cup\grass^0$ and $\tau\in\irr\stoch(T_N)$. 
Comparing~\eqref{eq:stoch isotypes} to Thm.~\ref{thm:main}, we see that the multiplicity of $\eta(\tau)$ in $\Theta(\id_{\stoch(T)})$ is equal to the multiplicity of $\id_{\stoch(T)}$ in $\ind{\stoch(T)^{N}}{\stoch(T)}(\tau)$, ie.
\begin{align*}
  \langle\Theta(\id_{\stoch(T)}),\ \eta(\tau)\rangle_\cliff
  =
  \langle\id_{\stoch(T)},\ \ind{\stoch(T)^{N}}{\stoch(T)}(\tau)\rangle_{\stoch(T)}.
\end{align*}
By Frobenius reciprocity,
\begin{align*}
  \left\langle
  \id_{\stoch(T)},\ \ind{\stoch(T)^{N}}{\stoch(T)}(\tau)
  \right\rangle_{\stoch(T)}
  =
  \left\langle
  \res{\stoch(T)^{N}}(\id_{\stoch(T)}),\ \tau
  \right\rangle_{\stoch(T)^{N}}
  =
  \left\langle 
  \id_{\stoch(T)^{N}},\tau
  \right\rangle_{\stoch(T)^{N}}
  =
  \delta_{\tau,\id_{\stoch(T_N)}}.
\end{align*}
This proves:

\begin{lemma}
  \label{lem:stab tensor powers}
	Let $t\leq n$ and
	consider, as in Thm.~\ref{thm:main}, a subset $\{N_i\}_i\subset\grass\cup\grass^0$ that contains exactly one element per $\stoch(T)$ orbit.	 
	Finally, let $T_i:=N_i^\perp/N_i$. 
	Then, it holds that
  \begin{align}
    \label{eq:stab tensor powers}
    \cc\stabs_{n,t}
    \simeq
    \bigoplus_{i}
    \eta(\id_{\stoch(T_i)}).
  \end{align}
\end{lemma}

Notice that the rank $t-2m$ component of this space decomposes as
\begin{align}
	\label{eq:stab t-2m}
  \egal{m}\cap\cc\stabs_{n,t}&\simeq\eta(\id_{\stoch(T_i)}),
  \qquad N_i\in\grass(\egal{m}).
\end{align}

By the decomposition~\eqref{eq:stab tensor powers}, the span of stabilizer tensor powers is a multiplicity-free Clifford representation.
It follows that the compression of the algebra $\com{r,s}$ to $\cc\stabs_{n,t}$ is Abelian.
Moreover, by Eq.~\eqref{eq:stab t-2m}, this compression  generated by the action of $\{P(\spinful{m}),\ P(\spinless{m})\}_m$ on that space.
In the following we provide a more explicit basis for this compression.

Lem.~\ref{lem:compression to stabs} provides an alternative -- arguably more explicit -- description of the compression of $\com{r,s}$ to $\cc\stabs_{n,t}$.

\begin{lemma}
  \label{lem:compression to stabs}
  The operators $C_m$, $D_m$ (defined in Sec.~\ref{sec:main theorem}) act on $\cc\stabs_{n,t}$.
  Moreover, the compression of $\com{r,s}$ to $\cc\stabs_{n,t}$, 
  \begin{align*}
    P(\cc\stabs_{n,t})\com{r,s}P(\cc\stabs_{n,t}),
  \end{align*}
  where $P(\cc\stabs_{n,t})$ projects onto $\cc\stabs_{n,t}$, has the following basis
  \begin{align*}
    \{C_m P(\cc\stabs_{n,t}), D_{m+1} P(\cc\stabs_{n,t})\}_{m=0}^{m(T)}.
  \end{align*} 
\end{lemma}
\begin{proof}
  The first claim simply follows from the fact that $C_m$ and $D_m$ commute with $\stoch(T)$, since
  \begin{align*}
    P(\cc\stabs_{n,t}) = \frac{1}{|\stoch(T)|}\sum_{O\in\stoch(T)}R(O).
  \end{align*}
  To prove the second claim, consider an arbitrary element $P(N)R(O)\in\semigroup_{r,s}$ with $N\in\grass(\egal{m})$.
  Then,
  \begin{align*}
    \sum_{\substack{O_1,\ O_2\\ \in \stoch(T)}} R(O_1)P(N)R(O)R(O_2)
    &=
    \sum_{\substack{O_1,\ O_2\\ \in \stoch(T)}} P(O_1N)R(O_1OO_2)\\
    &=
    \sum_{\substack{O_1,\ O_2\\ \in \stoch(T)}} P(O_1N)R(O_2)\\
    &\propto
    \sum_{N'\in\grass(\egal{m})} P(N')P(\cc\stabs_{n,t})\\
    &\propto
    \begin{cases}
      C_m'P(\cc\stabs_{n,t}), \qquad &N\in \grass_m,\\
      D_mP(\cc\stabs_{n,t}), \qquad  &N\in \grass_m^0.
    \end{cases}.
  \end{align*}
  Now, using $C_m'=C_m+D_{m+1}$, we obtain the result.
\end{proof}

\begin{corollary}
  \label{cor:cm dm stab}
  It holds that
  \begin{align*}
    \left[
    C_m, C_{m'}
    \right]P(\cc\stabs_{n,t})
    =
    \left[
    D_m, D_{m'}
    \right]P(\cc\stabs_{n,t})
    =
    \left[
    C_m, D_{m'}
    \right]
    P(\cc\stabs_{n,t})
    =0
  \end{align*}
  for all $m$ and $m'$.
\end{corollary}

In particular, 
\begin{align*}
	C_mP(\cc\stabs_{n,t})\propto P(\cm{m})P(\cc\stabs_{n,t}),
	\quad\text{and}\quad
	D_mP(\cc\stabs_{n,t})\propto P(\dm{m})P(\cc\stabs_{n,t}).
\end{align*}
This way, there exists a pair of real numbers $a_m,b_m\in\rr$ for which
\begin{align*}
	(a_mC_m-b_mC_{m+1})P(\cc\stabs_{n,t}) = P(\spinful{m})P(\cc\stabs_{n,t}),
\end{align*}
and, similarly, there  exist $c_m,d_m\in \rr$ satisfying 
\begin{align*}
	(c_mD_m-d_mD_{m+1})P(\cc\stabs_{n,t}) = P(\spinless{m})P(\cc\stabs_{n,t}).
\end{align*}


\subsection{Exact dualities for low tensor powers}
\label{sec:exact}

For some combinations of $r,s,d$, the resulting representation $\Delta_{r,s}$ gives rise to an exact correspondence between $\stoch(T)$ and $\cliff$.
This happens exactly when there are no isotropic stochastic subspaces of $T$.
For qubits, this is exactly the case if $r=t=3$. 
Here one obtains a duality between $\cliff$ and $\stoch(\zz_2^3)\simeq S_3$, and this proves that $\cliff$ is a unitary 3-design.

If $d>2$, on the other hand, Ref.~\cite[Thm.~3]{nwstabnet} provides a list of tensor power representations which give rise to this exact duality.
Here we include a short proof that this list is complete.

\begin{lemma}
  \label{lem:exact duality}
  Let $n\geq2$, $d$ be odd and $r+s\geq 2$.
  Then the commutant of $\Delta_{r,s}$ is spanned by $R(O)$ where $O\in\stoch(T)$ if and only if $d,r,s$ satisfy $r+s\leq 3$, $rs=0$, and if $r+s=3$ then $\ell(3)=-\ell(-1)$.
\end{lemma} 
\begin{proof}
  A necessary and sufficient condition for $R$ to span the commutant of $\Delta_{r,s}$ is $|\grass_m|=|\grass_m^0|=0$ for all $m$.
  This happens if and only if both $\ones_t$ and $\ones_t^\perp$ are anisotropic.
  \begin{align*}
    T = \langle\ones_t\rangle \oplus \ones_t^\perp,
  \end{align*}
  So by the Chevalley-Warning theorem $\dim\ones_t^\perp=t-1\leq2$.
  
  Anisotropicity of $\ones_t$ is equivalent to $r-s\neq0\mod d$.
  If $t=2$, this implies $s\in\{0,2\}$ and hence $rs=0$.
  Anisotropicity of $\ones_t^\perp$ is equivalent to the following: there exist $a,b\neq0$ for which
  \begin{align}
    \label{eq:isometry}
    \beta_{r,s}|_{\ones_t^\perp} \sim 
    \begin{cases}
      a\beta_{1,0},\quad &t=2,\\
      \diag(1,b),\quad &t=3, 
    \end{cases}
  \end{align}
  where $\ell(b)\neq\ell(-1)$ (which is equivalent to the condition that $\ones_t^\perp$ is not a hyperbolic plane).
  By multiplicativity of the discriminant we have that
  \begin{align*}
    \discr(\beta_{r,s})
    &=\ell((-1)^s)\\
    &=\discr(\langle\ones_t\rangle)\discr(\ones_t^\perp)\\
    &=
    \begin{cases}
      \ell(r-s)\ell(a),\quad &t=2,\\
      \ell(r-s)\ell(b),\quad &t=3.
    \end{cases}
  \end{align*}
  
  If $t=2$, there always exists an $a$ satisfying the conditions above.
  We conclude that if $r+s=2$ and $rs=0$, then $|\grass_m|,\ |\grass_m^0|=0$ for all $m$.
  
  If $t=3$, in contrast, $b$ is subject to the following \emph{two} conditions derived above:
  \begin{align*}
    \ell(b) = \ell(r-s)\ell((-1)^s),
    \qquad
    \ell(b)=-\ell(-1).
  \end{align*}
  A solution to these equations exists if and only if
  \begin{align*}
    -1=\ell\left((-1)^{s+1}(r-s)\right).
  \end{align*}
  If $s=1,2$ this equation implies $1=-1$ and does not hold.
  If $s=0,3$ it holds if and only if $\ell(3)=-\ell(-1)$.
\end{proof}

As a direct consequence of this, we have that, for any $r,s,d$ as in Lem~\ref{lem:exact duality}, 
\begin{align*}
  \Delta_{r,s}
  \simeq\bigoplus_{\tau\in\irr\stoch(T)}\tau\otimes\theta(\tau),
\end{align*}
for some injective function $\theta:\irr\stoch(T)\to\irr\cliff$.

\begin{remark}
  \label{rem:2 mod 3}
  The statement~\cite[Thm.~3]{nwstabnet} deals with the case $r=t=3$, where the condition $d=2\mod 3$ is obtained.
  This condition is equivalent to our condition $\ell(3)=-\ell(-1)$.
  To see this, recall that our condition is equivalent to $\ones_3^\perp$ being anisotropic, which is equivalent to the equations
  \begin{align*}
    1+x_1^2+x_2^2=0=1+x_1+x_2,
  \end{align*}
  having no solution over $\zz_d$.
  A short calculation shows that these equations have a solution if and only if there exists an $x$ for which $1+x+x^2=0$ (in which case, the solution is $x_1=x=x_2^{-1}$).
  Finally, as pointed out in~\cite{nwstabnet}, this polynomial is reducible (and hence contains a root over $\zz_d$) if and only if $d\neq2\mod3$.
\end{remark}


\section{Real Clifford action on $\cones$}
\label{sec:rcliff}

Consider the action of $\rcliff$ on $\cones$ when $d=2$ and $r-s=0\mod 4$. 
Since $\pauli\in\ker(\cones)$, this action realizes the homomorphism $\rcliff\to\rcliff/\pauli$ where the orthogonal group $\orth(V)\subset\sp(V)$ preserves the form $\kappa(v)=v_z\cdot v_x$.
For simplicity, we call this representation $\Delta$.

Recall the coset basis of $\cones$, 
\begin{align*}
  \ket{[F]_{\ones_t}} = 
  2^{-n/2}
  \big(\ket{f_1}+\ket{f_1+\ones_t}\big)\otimes
  \cdots
  \otimes\big(\ket{f_n}+\ket{f_n+\ones_t}\big),
\end{align*}
where $f_i\in\ones_t^\perp$.

Now, let $T'$ be a subspace of $\ones_t^\perp$ for which $\ones_t^\perp=\langle\ones_t\rangle\oplus T'$ so that
\begin{align*}
  T' \simeq \ones_t^\perp/\ones_t.
\end{align*}
By Prop.~\ref{prop:orthogonal to symplectic},  $\beta':=\beta|_{T'}$ is a symplectic product.
Let $\t\sp(T')$ be the embedding of the symplectic group of $T'$ into $\gl(T)$ which acts trivially on $(T')^\perp$.
That is $S\in\t\sp(T')$ satisfies
\begin{align*}
  S|_{(T')^\perp}=\ii_{(T')^\perp}, 
\end{align*}
and,
\begin{align*}
  \beta'(Su,Sv)=\beta'(u,v), \qquad \forall\ u,v\in T'.
\end{align*}
We may define a representation $\t\Delta$ of $\t\sp(T')$ on $\cones$ by
\begin{align*}
  \t\Delta(S)\ket{[F]_{\langle\ones_t\rangle}} = \ket{[SF]_{\langle\ones_t\rangle}}.
\end{align*}

\begin{lemma}
  \label{lem:sp-o commute}
  The representations $\Delta$ and $\t\Delta$ commute with each other.
\end{lemma}
\begin{proof}
  A straightforward but bulky calculation shows that, for any $S\in\t\sp(T')$ and for any basis element $\ket{[F]_{\langle\ones_t\rangle}}$, the following holds:
  \begin{align*}
    \Delta(\h_i)\t\Delta(S)\ket{[F]_{\langle\ones_t\rangle}}
    &=
    \t\Delta(S)\Delta(\h_i)\ket{[F]_{\langle\ones_t\rangle}}\\
    \Delta(X_i)\t\Delta(S)\ket{[F]_{\langle\ones_t\rangle}}
    &=
    \t\Delta(S)\Delta(X_i)\ket{[F]_{\langle\ones_t\rangle}}\\
    \Delta(\cnot_{ij})\t\Delta(S)\ket{[F]_{\langle\ones_t\rangle}}
    &=
    \t\Delta(S)\Delta(\cnot_{ij})\ket{[F]_{\langle\ones_t\rangle}},
  \end{align*}
  where $i$ is arbitrary and $j\neq i$.
  This implies the claim since these group elements generate $\rcliff$.
\end{proof}

We now consider a second basis, dual to this first one, which we call the \emph{Weyl basis}.
The starting point is the identification
\begin{align*}
  \hilb_{n,t} \simeq \hilb_{2n,t/2}\simeq \endom(\hilb_n)^{\otimes t/2},
\end{align*}
obtained first by grouping factors, and then using the inverse vectorization map.
Then, letting $\ket{W(a)}:=2^{-n/2}\vect{W(a)}$ with $a\in V$, $W(a)$ a \emph{real} Wey operator, and $t':=\frac t2 -1$, the basis is given by
\begin{align*}
  \{\ket{\Psi_A}:=\ket{W(a_1)}\otimes\cdots\otimes\ket{W(a_{t'})}\otimes\ket{W(a_1+\dots+a_{t'}}\}
\end{align*}
where $A$ is a $t'\times 2n$ matrix with rows $a_i$.

\begin{lemma}
  \label{lem:weyl basis}
  The set $\{\Psi_A\}$ introduced above is an orthonormal basis for $\cones$.
\end{lemma}
\begin{proof}
  Orthonormality follows from $\tr W^T(a)W(b)=2^{n}\delta_{a,b}$.
  Furthermore, this set contains one element for each matix in $\zz_2^{t'\times 2n}$, which gives
  \begin{align*}
    |\{\Psi_A\}| = 2^{n(t-2)}=\dim\cones.
  \end{align*}
  
  It is therefore sufficient to prove that
  \begin{align*}
    W^{\otimes(r,s)}(v)\ket{\Psi_A}=W^{\otimes t}(v)=\ket{\Psi_A},
  \end{align*}
  for all $v\in V$, where $W(v)$ is a \emph{real} Weyl operator (this is because $\center(\pauli)$ is modded out because $r-s=0\mod 4$.
  \begin{align*}
    W^{\otimes 2}(v)\vect{W(a_i)} 
    =
    \vect{W(v)W(a_i)W^T(v)}
    =
    (-1)^{[v,a_i]+\kappa(v)}\vect{W(a_i)}.
  \end{align*}
  This way
  \begin{align*}
    W^{\otimes t}(v) \ket{\Psi_A} = 
    (-1)^{\sum_i[v,a_i] + [v,\sum_i a_i]}\ket{\Psi_A}=\ket{\Psi_A}
  \end{align*}
\end{proof}

Following a similar argumentation as in~\cite[App.~B]{gracefully}, we can see that any $O\in O(V)$ acts by permuting the Weyl basis,
\begin{align*}
  \Delta(O)\ket{\Psi_A} = \ket{\Psi_{AO^T}}.
\end{align*}

\begin{lemma}
  \label{lem:weyl transform}
  Let $u',v'\in\zz_2^{t'}$ and
  \begin{align*}
    u := (u', \ones_{t'}\cdot u),
    \qquad
    v := (v', \ones_{t'}\cdot v)
    \qquad \in \zz_2^{t/2}.
  \end{align*}
  Then,
  \begin{align*}
    2^{-t/4}
    \sum_{w\in \zz_2^{t/2}}
    (-1)^{u\cdot w}\vect{W((w,v))}
    =
    \frac{1}{\sqrt{2}}\Big(
    \ket{(v, v+u)^T}+\ket{(v, v+u)^T+\ones_t}
    \Big)
  \end{align*}
\end{lemma}

Let $u_1',v_2',\dots,u_n',v_n'$ be the columns of $A$, 
\begin{align*}
  u_i := (u_i', \ones_{t'}\cdot u_i),
  \qquad
  v_i := (v_i', \ones_{t'}\cdot v_i),
\end{align*}
and let $A_Z$ (resp. $A_X$) be the $(t/2)\times n$ matrix with columns $u_i$ (resp. $v_i$).
Then, as a corollary of the result above,
\begin{align*}
  2^{-tn/4}
  \sum_{M\in \zz_2^{(t/2)\times n}}
  (-1)^{\tr(M^T A_Z)}\ket{\Psi_{(M,A_X)}}
  =
  \ket{\left[\begin{pmatrix} A_X\\ A_X+A_Z \end{pmatrix}\right]_{\langle\ones_t\rangle}}
\end{align*}

We now calculate the dimension of the commutant of $\Delta$, which is equal to the number of orbits of $O(V)$ on $V^{\times(t')}$.

Consider some $t'$-tuple $\mbf{v}$ of $V$ vectors, let the orbit containing this point be $\orb(\mbf{v})$.
Associated to $\mbf{v}$ are a set of index sets $I\subset\{1,\dots,t'\}$ for which $\{v_i\}_{i\in I}$ is linearly independent.
Ordering these subsets lexicographically, we let $I(\mbf{v})$ be the maximal such subset.
Further, let $M(\mbf{v})$ be the $(t'-|I(\mbf{v})|)\times|I(\mbf{v})|$ matrix such that,
\begin{align*}
  v_j = \sum_{i\in I(\mbf{v})} M(\mbf{v})_{ji}v_i, \qquad \forall j\notin I(\mbf{v}).
\end{align*}

\begin{lemma}
  \label{lem:orbit classification}
  {Can probably be simplified with Takagi's orbit invariants.}
  A point $\mbf{u}$ is in the orbit $\orb(\mbf{v})$ if and only if the following conditions hold:
  \begin{enumerate}
    \item
    $I(\mbf{u})=I(\mbf{v})$,
    \item
    $[u_i,u_j] = [v_i,v_j]$, for all $i<j\in I(\mbf{v})$,
    \item
    $\kappa(u_i)=\kappa(v_i)$, for all $i\in I(\mbf{v})$,
    \item
    $M(\mbf{v})=M(\mbf{u})$.
  \end{enumerate}
\end{lemma}
\begin{proof}
  The \emph{only if} direction follows simply from the facts that $O\in\orth(V)$ preserves $[\darg,\darg]$ and $\kappa(\darg)$, and that any linear relation $\mbf{v}\cdot a=0$, where $a\in\zz_2^{t'}$, implies $O\mbf{v}\cdot a=0$.
  
  Conversely, let $O$ be an element of $O(V)$ for which $Ov_i=u_i$ for all $i\in I:=I(\mbf{v})$ ---such an $O$ exists by the Cahit-Arf theorem.
  Let $M:=M(\mbf{v})$.
  Then, for each $j\notin I$, it holds that
  \begin{align*}
    u_j = \sum_{i\in I} M_{ji}u_i = \sum_{i\in I} M_{ji}Ov_i=Ov_j,
  \end{align*}
  and thus $\mbf{u}=O\mbf{v}$.
\end{proof}

In the regime where $t'\ll n$, the vast majority of orbits will contain $t'$ linearly independent vectors.
In each of these orbits $M(\mbf{v})=0$ and $I(\mbf{v})=\{1,\dots,t'\}$, so this class of orbits is labeled by the numbers $[v_i,v_j]$ and $\kappa(v_i)$.
Since each of these numbers may be chosen independently, there are $2^{t'^2}$ such orbits.
Up to subleading order corrections, this coincides with $|\sp(T')|$.
This leads to the intuition that $\sp(T')$ captures most of the structure of the commutant of $\Delta$.
In analogy to Lem.~\ref{lem:subcode rank-deficient}, one would expect that a subspace $\mcal{L}\subseteq\cones$ gives rise to an exact duality between $\rcliff$ and $\sp(T')$.

\begin{conjecture}
  \label{conj:rcliff duality}
  For any fixed $t$ and sufficiently large $n$ the following holds.
  There exists a subspace $\mcal{L}\subset\cones$ with  
  \begin{align*}
    \frac{\dim\cones-\dim\mcal{L}}
         {\dim\cones}
    =
    o(\exp(-n)),
  \end{align*}
  and an injective function $\theta:\irr\sp(T')\to\irr\orth(V)\subset\irr\rcliff$ such that, as a $\sp(T')\times\rcliff$ representation,
  \begin{align*}
    \mcal{L}\simeq \bigoplus_{\tau\in\irr\sp(T')}\tau\otimes\theta(\tau).
  \end{align*}
\end{conjecture}


\section{Conjugating black box Cliffords}
\label{sec:black box}

Suppose one is given $t$ uses of a black box Clifford unitary $U$.
How large does $t$ have to be in order to implement $\bar U$?
The simplest case to analyse here is when the implementation is \emph{parallel}, which is to say when there exist isometries $V_1$, $V_2$ for which $\bar U = V_2 U^{\otimes t} V_1$.
This question is equivalent to asking what is the minimal $t$ for which $\Delta_{0,1}\subset\Delta_{t,0}$, i.e.\ that the conjugate representation of $\cliff$ is a subrepresentation of the $t$-th tensor power representation.

\begin{lemma}
  \label{lem:conjugate rep}
  The minimal $t$ for which $\Delta_{0,1}$ is a subrepresentation of $\Delta_{t,0}$ is
  \begin{enumerate}
    \item $t=7$ if $d=2$,
    \item $t=2d-1$ if $d=1\mod4$,
    \item $t=4d-1$ if $d=3\mod4$.
  \end{enumerate}
\end{lemma}
\begin{proof}
  It is clear that $t>1$ is necessary, and so any subrepresentation $\rho$ of $\Delta_{t,0}$ isomorphic to $\Delta_{0,1}$ will be rank-deficient and by Thm.~\ref{thm:rank deficient} is in the span of all codes $C_N$ with $t-2\dim N=1$.
  This can only happen if $t$ is odd.
  Moreover, $\pauli \not\subseteq\ker\Delta_{0,1}$ so that 
  \begin{align*}
    \rho\subseteq\spinful{m},\quad m=\frac{t-1}{2}.
  \end{align*}
  If $d$ is odd and $t>2$, then by the Chevalley-Warning theorem $\grass_m$ is non-empty.
  If $d=2$ and $t=7$, then $\grass_3$ is non-empty by the following example:
  \begin{align*}
    N=\langle
    (1111000),\
    (0011110),\
    (1010101)
    \rangle,
  \end{align*}
  where vectors are written in the orthonormal basis of $q_{t,0}$.
  
  Now, because for every $N,N'\in\grass_m$, there is an $O\in\stoch(T)$ for which
  \begin{align*}
    R(O)P(N) R^\dagger(O) = P(N'),
  \end{align*}
  it follows that these are isomoprhic as Clifford representations and we can assume without loss of generality that $\rho\subseteq C_N$ for some code $N$.
  Now, this $N$ is such that $\dim T_N=1$, so that $T_N =\langle[\ones_t]_N\rangle$ and $\stoch(T_N)=\{\ii\}$.
  By~\ref{thm:eta duality}, $C_N$ is irreducible and so $C_N\simeq\Delta_{0,1}$.
  
  We can use Lem.~\ref{lem:rs equivalence}, to re-express
  \begin{align*}
    \Delta_{t,0}\simeq \Delta_{r,s},
  \end{align*} 
  for some $r$ and $s$ that are subject to the conditions of that lemma.
  By Lem.~\ref{lem:restriction to css}, $C_N\simeq \Delta_{r-m,s-m}$ must be isomorphic to $\Delta_{0,1}$, and thus we must be able to choose $r$ and $s$ such that $r-m=0$, $s-m=1$.
  Thus, $t$ must be such that
  \begin{align*}
    \Delta_{t,0}\simeq \Delta_{m,m+1}.
  \end{align*} 
  Here we argue by cases:
  If $d=1\mod4$ we require $t\mod d = m-(m+1)=-1$, the smallest odd $t$ for which this equation holds is $t=2d-1$.
  If $d=3\mod4$ we require furthermore that $s=\frac{t+1}{2}$ is even.
  The smallest $t$ for which these two conditions hold is $t=4d-1$.
  If $d=2$, we require instead that $t\mod 8 = -1$, in which case we can take $t=7$.
\end{proof}

By encoding $\hilb\mapsto C_N$, where $C_N$ is as in the proof of Lem.~\ref{lem:conjugate rep}, we obtain an implementation of $\bar U$.
Namely, if $U_N$ is the encoding isometry, $U_N^\dagger U^{\otimes t} U_N=\bar U$.
Using this result, one may use the teleportation trick from~\cite{marcotulio} to probabilistically implement $U^\dagger$ in a heralded fashion. 

The protocol for conjugating Cliffords above is considerably simpler than the protocols studied in~\cite{marcotulio,complexconjug} which conjugate \emph{arbitrary unitaries}.
Two properties contrast these two cases.
First, the isometry $U_N$ is a Clifford operation, while the corresponding isometry from~\cite{marcotulio} could require a high $T$-gate count.
Second, in Ref.~\cite{marcotulio} it is shown that in order to implement $\bar U$, at least $t=d^n-1$ black box uses are necessary (even if one is content with a heralded implementation with positive probability).
On the other hand, our protocol requires only on the order of $d$ black box uses to implement $\bar U$ deterministically.

\section{Acknowledgements}

We thank Oscar Garcia, Shamgar Gurevich, Markus Heinrich, Sepehr Nezami, and Michael Walter for very insightful discussions.
We also thank Markus Heinrich and Pascal Bassler for bringing the problem of black box Clifford conjugation to our attention.
This work has
been supported by Germany's Excellence Strategy – Cluster of Excellence Matter and Light for Quantum Computing (ML4Q)
EXC 2004/1 - 390534769
and the German Research Council (DFG) via contract GR4334/2-2.

\section{Author declarations}
\subsection{Conflict of interest}
The authors have no conflicts to disclose.


\appendix

\section{Deferred proofs}
\label{app:proofs}

\subsection{Proofs from Sec.~\ref{sec:prelim}}
\label{app:proofs prelim}

\begin{proof}[Proof of Prop.~\ref{prop:sym alt tensors}.]
  Let $\beta\in\bil(K)$ and $\phi$ be its tensor representation.
  Then, $\beta\in\sym(K)$ is equivalent to
  \begin{align*}
    \phi(u\otimes v) = \beta(u,v) = \beta(v,u) = \phi(\pi(u\otimes v)),
    \quad
    \forall\ u,v \in K.
  \end{align*}
  This way, $\phi = \phi\circ\pi$.
  Furthermore, $\beta\in\alt(K)$ is equivalent to
  \begin{align*}
    \phi(u\otimes u) =0, \quad
    \forall\ u\in K.
  \end{align*}
  Because $(K\otimes K)^\pi=\vspan\{u\otimes u \div u\in K\}$, the equation above is equivalent to $\phi|_{(K\otimes K)^\pi}=0$.
\end{proof}

\begin{proof}[Proof of Prop.~\ref{prop:refinable bilinear}.]
  Let $\phi_\beta$ be the tensor representation of $\beta$, then
  \begin{align*}
    \phi_\beta(u\otimes v)
    =
    \beta(u,v)
    &=
    q(u+v)-q(u)-q(v)\\
    &=
    \phi((u+v)^{\otimes 2}-u^{\otimes 2}-v^{\otimes 2})\\
    &=
    \phi(u\otimes v + v\otimes u).
  \end{align*}
\end{proof}

We define the \emph{generalized polarisation map} $\t\Xi:\t\q(K)\to\sym(K)$ to be the additive map $\t\Xi(q)=\beta$, where $q$ is a generalized refinement of $\beta$. 

\begin{proposition}
  \label{prop:generalized xi}
  Let $d=2$. Then 
  \begin{align*}
    \ker\t\Xi=2K^* :=\{2f \div f\in K^*\},
    \qquad
    \range\t\Xi = \sym(K).
  \end{align*}
\end{proposition}
\begin{proof}
  If $\t\Xi(q)=0$ for some $q\in\t\q(K)$, then for all $u,v\in K$ it holds that
  \begin{align}
    \label{eq:actually linear}
    q(u+v)=q(u)+q(v).
  \end{align}
  Using $v=u$ we see that $q(u)=2f(u)$ for some $f:K\to\zz_2$. 
  But by~\eqref{eq:actually linear}, $f\in K^*$. 
  
  Now we constructively show that every $\beta\in\sym(K)$ has a generalized quadratic refinement $q\in\t\q(K)$.
	Throughout the rest of the proof we will use $\{\{\dots\}\}$ to denote a “$\zz_2$ pocket inside of a $\zz_4$ environment,'' that is if $a$ is a $\zz_2$-valued expression, then $\{\{a\}\}=1\in\zz_4$ if $a=1\in\zz_2$ and $\{\{a\}\}=0$ otherwise.
  We furthermore use $s_u:= \supp(u)$ for any $u\in K$ with respect to the standard basis $\{e_i\}$.
	   
  We claim that $q$ defined by
  \begin{align*}
    q(u) = 
    \sum_{i\in s_u} \{\{\beta(e_i,e_i)\}\}
    + 
    \sum_{\substack{i<j\\ \in s_u}} 2\{\{\beta(e_i,e_j)\}\},
  \end{align*}
  is a generalized refinement of $\beta$.
	Consider $u,v\in K$ and let $I_1,\ I_2,\ I_3$ be disjoint subsets of $\{1,\dots,\dim K\}$ such that
	\begin{align*}
		s_u = I_1\cup I_2, \qquad s_v = I_2\cup I_3.
	\end{align*}
	Moreover, for   $a,b\in\{1,2,3\}$, let
	\begin{align*}
		[a;b]:= 
		\sum_{i\in I_a}
		\sum_{\substack{j\in I_b\\ j\neq i}}
		\beta(e_i,e_j) \mod 2.
	\end{align*}
  Then,
  \begin{align*}
    q(u+v)  
    &= 
    \sum_{i\in I_1,I_3}\{\{\beta(e_i,e_i)\}\} +
    2\Big\{\Big\{ [1;1]+[1;3]+[3;3]\Big\}\Big\},\\
    q(u)
    &=
    \sum_{i\in I_1,I_2}\{\{\beta(e_i,e_i)\}\} +
    2\Big\{\Big\{ [1;1]+[1;2]+[2;2]\Big\}\Big\},\\
    q(v)
    &=
    \sum_{i\in I_2,I_3}\{\{\beta(e_i,e_i)\}\} +
    2\Big\{\Big\{ [2;2]+[2;3]+[3;3]\Big\}\Big\},
  \end{align*}
  and so
  \begin{align*}
    q(u+v)-q(u)-q(v)
    &=
    \sum_{i\in I_2}2\{\{\beta(e_i,e_i)\}\} +
    2\Big\{\Big\{ [1;2]+[2;3]+[1;3]\Big\}\Big\}\\
    &=
    2\Big\{\Big\{
      \sum_{i\in I_2}\beta(e_i,e_i)
      +[1;2]+[2;3]+[1;3]
    \Big\}\Big\}\\
    &=
    2\{\{\beta(u,v)\}\},
  \end{align*}
  where the last line follows from the fact that $\beta$ is symmetric.
\end{proof}

\begin{proof}[Proof of Prop.~\ref{prop:form restriction}]
  If $d$ is odd,  
  \begin{align*}
    \discr(\beta_N)=(-1)^m\discr(\beta_{r,s}) = (-1)^{s-m}.
  \end{align*}
  This way $\beta_{r-m,s-m}\simeq\beta_N$ since they have the same rank and discriminant.
  
  Now turn to the case $d=2$.
  It suffices to prove the claim when $m=1$. 
  The form $\beta_N$ is of odd type since $\ones_t\notin N$ and so, for at least some $u\in T_N$,
  \begin{align*}
    \beta_N(u,u)=\beta_N(\ones_t,u)\neq0.
  \end{align*}
  This way $q_N\simeq q_{r',s'}$ where $r'+s'=t-2$.
  Now, we can write the isometry $T\simeq T_0\oplus T_0^\perp$, where $T_0$ is the subspace of $N^\perp$ isometric to $T_N$.
  By~\eqref{eq:garf qrs} and the additivity of the generalized Arf invariant,
  \begin{align*}
    \garf(q_N) + \garf\left(q_{r,s}|_{T_0^\perp}\right) = 
    r'-s' + \garf\left(q_{r,s}|_{T_0^\perp}\right) = 
    \garf(q_{r,s})
    =
    r-s\mod 8.
  \end{align*}
  
  We end the proof by showing that $\garf(q_{r,s}|_{T_0^\perp}) = 0$.
  Let $N=\langle a \rangle$, where $a\notin\langle\ones_t\rangle$. 
  It is sufficient to show that there exists a vector $b\in T$ such that
  \begin{align*}
    q_{r,s}(b)=0,\qquad \beta_{r,s}(a,b)=a\cdot b=1.
  \end{align*}
  Indeed, these equations imply that $T_0^\perp = \langle a,b\rangle$ and so
  \begin{align*}
    q_{r,s}|_{T_0^\perp}\sim 2q_{\hh}^0.
  \end{align*}
  
  We prove this in three cases. 
  Let $\alpha_r$ be $|\supp(a)\cap\{1,\dots,r\}|$ and $\alpha_s = |\supp(a)| - \alpha_r$.
  Then the cases are: 
  \emph{i)} either $s=0$ or $r=0$, 
  \emph{ii)} $r,s\geq1$ and either $\alpha_r=0$ or $\alpha_s=0$,
  \emph{iii)} $r,s,\alpha_r,\alpha_s\geq1$.
  
  \emph{Case i)} Without loss of generality take $s=0$ and write
  \begin{align}
    \label{eq:a wlog}
    a = e_1+e_2+\cdots+e_{\alpha_r-1}+e_{\alpha_r},
  \end{align}
  where $4\leq \alpha_r <r$, and the right-hand inequality follows from $a\notin\langle\ones_t\rangle$.
  Then $b=e_1+e_2+e_3+e_r$.
  
  \emph{Case ii)} Similarly, take $\alpha_s=0$ and $a$ as in~\eqref{eq:a wlog}.
  Then $b=e_1+e_{r+1}$.
  
  \emph{Case iii)} Finally, without loss of generality consider 
  \begin{align}
    \label{eq:a wlog 2}
    a = e_1+\cdots+e_{\alpha_r} + e_{r+1} + \cdots e_{r+\alpha_s},
  \end{align}
  where at least one of the two inequalities $\alpha_s<s,$ $\alpha_r<r$ hold.
  Without loss of generality we can assume the first inequality holds and take $b=e_1+e_t$.
\end{proof}

\begin{proof}[Proof of Prop.~\ref{prop:even symmetric}.]
  The two enumerated claims follow from the classification of non-degenerate symmetric forms (see, e.g.,~\cite[Prop.~A.1.4]{klausthesis}). 
  
  Now, consider a form $\beta'$ with the same rank and type as $\beta$, and let $g\in\gl(K)$ be such that $\rad(\beta')=g\rad(\beta)$.
  Then, $(g\beta)|_{K/\rad(\beta')}$ is a non-degenerate form with the same type as $\beta_0'=\beta'|_{K/\rad(\beta')}$.
  Thus, by~\cite[Prop.~A.1.4]{klausthesis}, there exists some 
  \begin{align*}
    h\in\gl^{\rad(\beta')}(K) 
    := \{ g'\in\gl(K) \div g'\rad(\beta')=\rad(\beta')\},
  \end{align*}
  for which 
  \begin{align*}
    (hg\beta)|_{K/\rad(\beta')}=\beta_0'.
  \end{align*}
  This way $hg\beta=\beta'$ and the two forms are equivalent.
\end{proof}

\subsection{Proofs from Sec.~\ref{sec:tensor powers}}
\label{app:proofs tensor}

\deltars*

\begin{proof}
  In all the cases of the lemma, $q_{r,s}\sim q_{r',s'}$.
  Let $g\in\gl(T)$ be some transformation for which $q_{r',s'}(g\darg)=q_{r,s}(\darg)$.
  If $d$ is odd, the condition $r'-s'=r-s\mod d$ allows us to choose $g$ such that $g\ones_t = \ones_t$.
  If $d=2$, on the other hand, $\beta_{r,s}=\beta_{r',s'}$ and thus $g$ is an isometry of $\beta_{r,s}$, that is 
  \begin{align*}
    \beta_{r,s}(g\darg,g\darg)=\beta_{r,s}(\darg,\darg)
  \end{align*}
  Because $\beta_{r,s}(u,u)=\beta_{r,s}(\ones_t,u)$, the equation above implies that $g\ones_t=\ones_t$.
  
  Then, the isomorphism $U$ is can be expressed in the standard basis by $U:\ket{F}\mapsto\ket{gF}$, where $F\in\hom(\pos\to T)$.
  To verify this claim, we act on generators.
  
  First, we can see that $\cadd^{\otimes(r,s)}=\cadd^{\otimes t}=\cadd^{\otimes(r',s')}$ and compute that
  \begin{align*} 
    U\cadd^{\otimes t}U^\dagger=\cadd^{\otimes t}.
  \end{align*}
  
  Since all the other generators are single-qudit, we set $n=1$ for the rest of the proof.
  This way,
  \begin{align*}
    U\h^{\otimes(r,s)}U^\dagger 
    =
    \sum_{u,v\in T}
    (-1)^{\beta_{r,s}(g^{-1}u,g^{-1}v)}\ketbra{u}{v}
    =
    \h^{\otimes(r',s')},
  \end{align*}
  where $\ii_2$ is the two dimensional identity.
  
  If $d=2$,
  \begin{align*}
    \p^{\otimes(r,s)} = \sum_{u\in T} \tau^{q_{r,s}(u)} 
    \ketbra{u},
  \end{align*}
  and so
  \begin{align*}
    U\p^{\otimes(r,s)} U^\dagger
    = 
    \sum_{u\in T} \tau^{q_{r,s}(g^{-1}u)} 
    \ketbra{u}
    = 
    \sum_{u\in T} \tau^{q_{r',s'}(u)} 
    \ketbra{u}
    =
    \p^{\otimes(r',s')}.
  \end{align*}
  This proves point 3.
  
  If $d$ is odd, then using 
  \begin{align*}
    \p^{\otimes(r,s)} 
    = 
    \sum_{u\in T} \omega^{2^{-1}q_{r,s}(u)+2^{-1}\beta_{r,s}(\ones_t,u)} 
    \ketbra{u},
  \end{align*}
  we conclude
  \begin{align*}
    U\p^{\otimes(r,s)} U^\dagger 
    = 
    \sum_{u\in T} \omega^{2^{-1}q_{r,s}(g^{-}u)+2^{-1}\beta_{r,s}(g^{-1}\ones_t,g^{-1}u)} 
    \ketbra{u}
    =
    \p^{\otimes(r',s')}.
  \end{align*}
  
  Finally, in the odd $d$ case we must check the isomorphism on the Pauli $Z$ operator too:
  \begin{align*}
    U Z^{\otimes(r,s)} U^\dagger
    =
    \sum_{u\in T}
    \omega^{\beta_{r,s}(\ones_t,u)} U\ketbra{u}U^\dagger
    =
    \sum_{u\in T}
    \omega^{\beta_{r,s}(\ones_t,g^{-1}u)}\ketbra{u}
    =
    \sum_{u\in T}
    \omega^{\beta_{r',s'}(\ones_t,u)}\ketbra{u}
    =
    Z^{\otimes(r',s')},
  \end{align*}
  where we used $g^{-1}\ones_t=\ones_t$.
  This concludes the proof of Points 1. and 2.
\end{proof}

\codereps*

\begin{proof}
  We construct an explicit isomorphism. 
  Prop.~\ref{prop:form restriction} implies that $q_N\sim q_{r-m,s-m}$ and hence $\beta_N\sim\beta_{r-m,s-m}$.
  Let the corresponding isometry be $\nu: T_N\to\zz_d^{t-2m}$.
  Now, for each coset $[F]_N$, where $F\in\hom(\pos\to N^\perp)$, there corresponds a $F_0\in\hom(\pos\to \zz_d^{t-2m})$ such that $\nu[Fx]_N= F_0x$ for all $x\in\pos$.
  Furthermore, it is clear that 
  \begin{align*}
    q_N([\ones_t]_N)=
    q_{r,s}(\ones_t)=
    r-s\mod D = 
    q_{r-m,s-m}(\ones_{t-2m}),
  \end{align*}
  and since $\ones_t\in N^\perp\setminus N$, we can choose $\nu[\ones_t]_N = \ones_{t-2m}$.
  Then the isomorphism is given by 
  \begin{align*}
    \iota:\ket{[F]_N} \mapsto \ket{\nu F_0}\in\hilb_{n,t-2m}.
  \end{align*}
  To show this, we evaluate the action of generators.
  
  For the generators $\h$, $\p$ and $X$ we will take $n=1$, since the code spaces $C_N$ are $n$-th tensor powers.
  In this case, 
  \begin{align*}
    \iota\h^{\otimes(r,s)}|_{C_N}\iota^\dagger
    &=
    \iota\left(
    d^{-t/2}\sum_{u,v\in T}\omega^{\beta_{r,s}(u,v)}P(N)\ketbra{u}{v}P(N) 
    \right)\iota^\dagger\\
    &=
    d^{-m-t/2}\sum_{[u]_N,[v]_N\in T_N} 
    \left(\sum_{u'\in[u]_N}\sum_{v'\in[v]_N}
    \omega^{\beta_{r,s}(u',v')}
    \right)
    \iota\ketbra{[u]_N}{[v]_N}\iota^\dagger\\
    &=
    d^{m-t/2}\sum_{[u]_N,[v]_N\in T_N} 
    \omega^{\beta_N([u]_N,[v]_N)}\iota\ketbra{[u]_N}{[v]_N}\iota^\dagger\\
    &=
    d^{m-t/2}
    \sum_{a,b\in\zz_d^{t-2m}}
    \omega^{\beta_{r-m,s-m}(a,b)}
    \ketbra{a}{b}
  \end{align*}
  where the third line follows from $\beta_{r,s}$ being well defined on $T_N$ and the last follows from the identification $a:=\nu[u]_N$, $b:=\nu[v]_N$.
  
  Similarly,
  \begin{align*}
    \iota\p^{\otimes(r,s)}|_{C_N}\iota^\dagger
    =
    \sum_{[v]_N\in T_N}\tau^{q_N([v]_N)}\iota\ketbra{[v]_N}\iota^\dagger
    =
    \sum_{a\in\zz_d^{t-2m}}\tau^{q_{r-m,s-m}(a)}\ketbra{a}.
  \end{align*}
  
  For the Pauli $X$ case, $X^{\otimes(r,s)}=X^{\otimes t}$, and thus for all $v\in N^\perp$,
  \begin{align*}
    \iota X^{\otimes t}\ket{[v]_N}
    =
    \iota\ket{[v]_N+[\ones_t]_N}
    =
    \ket{\nu[v]_N + \ones_{t-2m}}
    =
    X^{\otimes(t-2m)}\ket{\nu[v]_N}
    =
    X^{\otimes(t-2m)}\iota\ket{[v]_N}.
  \end{align*}
  
  Finally, to compute the action of $\cadd^{\otimes(r,s)}=\cadd^{\otimes t}$, take $n=2$ and act on $\ket{[v_1]_N}\ket{[v_2]_N}$,
  \begin{align*}
    \cadd^{\otimes t}\ket{[v_1]_N}\ket{[v_2]_N}
    &=
    d^{-m}\sum_{u_1,u_2\in N} \cadd^{\otimes t}\ket{v_1+u_1}\ket{v_2+u_2}\\
    &=
    d^{-m}\sum_{u_1,u_2\in N} \ket{v_1+u_1}\ket{v_1+v_2+u_1+u_2}\\
    &=
    \ket{[v_1]_N}\ket{[v_1+v_2]_N}.
  \end{align*}
  Then, using $a_i := \nu[v_i]_N$,
  \begin{align*}
    \iota\cadd^{\otimes t}\ket{[v_1]_N}\ket{[v_2]_N}
    =
    \ket{a_1}\ket{a_1+a_1}
    =
    \cadd^{\otimes(t-2m)}\iota\ket{[v_1]_N}\ket{[v_2]_N}.
  \end{align*}
\end{proof}

\quditsubcode*

\begin{proof}
  Consider $\Delta_{r,s}|_{\cones}$. 
  Clearly $\pauli$ is contained in the kernel of this representation and so
  \begin{align*}
    \Delta_{r,s}(W\mu(S))|_{\cones} = \mu^{\otimes(r,s)}(S)|_{\cones}.
  \end{align*}
  But then~\cite[Lem.~2.7,~Cor.~2.3]{rank-deficient} imply that this representation is of the claimed form.
  Finally, this representation is a faithful representation of $\sp(V)$ and we get $\ker(\cones)=\pauli$.
\end{proof}

\subsection{Proofs from Sec.~\ref{sec:rank}}
\label{app:proofs rank}

\diagonals*

\begin{proof}
  Consider the homomorphism $\varphi:\cliff\to\cliff/\pauli\simeq\sp(V)$.
  Diagonal Cliffords preserve all $Z$-type Pauli matrices so that
  \begin{align}
    \label{eq:diag to borel}
    \varphi(\dcliff)
    =
    \left\{
      \begin{pmatrix}
        \ii & S\\ 0 & \ii
      \end{pmatrix}
      \DIV
      S\in\sym(\pos)
    \right\} =: \borel,
  \end{align}
  where we have used the basis $\{e_1,\dots,e_n,f_1,\dots,f_n\}$ for which $W(e_i)=Z_i$.
  We may further specify
  \begin{align*}
    \varphi(\p_i)       &= \ii + e_ie_i^T\\
    \varphi(\cphase_{ij}) &= \ii + e_ie_j^T + e_je_i^T,
  \end{align*}
  where $\cphase_{ij}:=\h_i\h_j\cadd_{ij}\h_i^\dagger \h_j^\dagger$.
  Thus, $\langle \varphi(\p_i), \varphi(\cphase_{ij})\rangle_{ij} = \borel$, and by~\eqref{eq:diag to borel}, 
  \begin{align}
    \label{eq:dcliff generators}
    \langle \p_i, \cphase_{ij}, Z_i\rangle_{ij} = \dcliff.
  \end{align}
  Since each of these generators is of the form~\eqref{eq:diag quad}, the first claim follows.
  The second claim follows from the fact that $2\pos^*\subset\t\q(\pos)$ when $d=2$.
\end{proof}

\rankzero*

\begin{proof}
  Let $\rho$ have rank zero, and consider the group
  \begin{align*}
    G :=
    \begin{cases}
      \ker\left(\res{\rcliff}\rho\right),\qquad &d=2,\\
      \pauli\ker\left(\rho\right),\qquad &d>2.
    \end{cases}
  \end{align*}
  In the qubit case, $\rk(\rho)=0$ implies $\rpauli\subseteq\ker(\res{\rcliff}\rho)$, where $\rpauli$ is the subgroup of real multi-qubit Pauli matrices. 
  Notice
  \begin{align*}
    G\normaleq \cliff; \quad G\subseteq\rcliff,\quad \text{if } d=2.
  \end{align*}
  Then, 
  \begin{align*}
    \t G :=
    \begin{cases}
      G/\rpauli,\qquad d=2,\\
      G/\pauli,\qquad d>2,
    \end{cases}
  \end{align*}
  satisfies $\t G \normaleq \orth(V)$ in the qubit case, and $\t G\normaleq \sp(V)$ when $d>2$.
  Furthermore, $\t G$ is non-trivial: in the qubit case it contains $\rpauli\rdcliff/\rpauli\simeq Q(\pos)$, whereas in the $d>2$ case it contains the subgroup 
  \begin{align*}
    \borel := \left\{\begin{pmatrix}\ii & A\\ 0 & \ii\end{pmatrix} \div A \in \sym_{n\times n} \right\} \subseteq\sp(V).
  \end{align*}
  
  If $d=2$, $\t G = \orth(V)$ because $\orth(V)$ is simple for $n\geq3$~\cite[Sec.~1.4]{CarterSimple}, and thus $G=\rcliff\subseteq\ker\rho$.
  This implies that the subgroup $H=\ker(\rho)$ satisfies
  \begin{align*}
    \rcliff \subseteq H \normaleq \cliff.
  \end{align*}
  Consider the group $\langle i\ii,\ H\rangle=\{H,\ iH\}\subseteq\cliff$.
  Then, $\orth(V)\subseteq \t H:= \langle i\ii,\ H\rangle/\pauli \normaleq \sp(V)$, but since $\sp(V)$ is simple for $n\geq3$~\cite[Sec.~1.3]{CarterSimple} we have that $\t H = \sp(V)$. 
  
  Now, we show that in fact $i\ii\in H$. 
  Because $\t H =\sp(V)$, there is some $W\in\rpauli$ and a phase $\alpha$ for which $\alpha W\p_1\in H$. 
  Using $\rpauli\subset H$,
  \begin{align*}
    1
    =
    \rho(X_1)
    =
    \rho(\alpha W\p_1 X_1 \p_1^\dagger W^\dagger \alpha^*)
    =
    \rho(\pm iWX_1Z_1W^\dagger)
    =
    \rho(i\ii)\rho(\pm X_1Z_1)
    =
    \rho(i\ii).
  \end{align*}
  This way $\pauli\subset H$, and thus $\{H,\omega_8 H\}=\cliff$ where $\omega_8$ is a primitive eigth root of unity.
  Thus, an arbitarary Clifford has the form $g=\omega_8^a h$, where $h\in H$ and $a\in\{0,1\}$, and $\rho(g)=\rho(\omega_8^a\ii)$.
  This equation implies two things:
  \begin{enumerate}
    \item 
    $\rho$ is Abelian and thus one-dimensional,
    \item
    if $\rho'$ has rank zero as well and $\rho'(\omega_8^a\ii)=\rho(\omega_8^a\ii)$, then $\rho'(g)=\rho(g)$ for all $g\in\cliff$.
  \end{enumerate}
  Moreover $\res{\center(\cliff)}\rho$ has as kernel equal to either $\center(\cliff)$ or $\langle i\ii\rangle$.
  Because of this, $\res{\center(\cliff)}\rho$ is a $\zz_2$ representation and thus $\pm1$ valued.
  
  If $d>2$, on the other hand, $\t G = \sp(V)$ because $\borel\subseteq\ker\rho$ and $\sp(V)$ is generated by $\borel$ conjugates.
  This way $G=\cliff$ and $\sp(V)\subseteq \ker(\rho)$.
  For any $S\in\sp(V)$ let $v\in V$ be such that $Sv\neq v$.
  Then,
  \begin{align*}
    W^\dagger(v)\mu(S)W(v)
    =
    W^{\dagger}(v)W(Sv)\mu(S)
    =
    W(Sv-v)\mu(S) \in \ker\rho,
  \end{align*}
  so that $W(Sv-v)\in\ker\rho$ and $\pauli\subset\ker(\rho)$.
  This implies, finally, that $\ker(\rho)=\cliff$.
\end{proof}

\subsection{Proofs from Sec.~\ref{sec:classification}}
\label{app:proofs class}

\hfregular*

\begin{proof}
  The ‘‘if’’ direction of the first statement is clear.
  ‘‘Only if:’’ 
  
  We first show that $\ker F=\ker J$.
  Consider any $x\in\ker J$, then for all $y\in\pos$,
  \begin{align*}
    q_N(Fy)=q_{r,s}(Jy) = q_N(J(x+y)) = q_N(Fx+Fy) = q_N(Fx)+q_N(Fy)+2\beta_N(Fx,Fy).
  \end{align*}
  Using $q_N(Fx)=q_N(Jx)=0$, we obtain that $\beta_N(Fx,Fy)=0$.
  Since $F$ is surjective onto $T_N$ and $\beta_N$ is non-degenerate, it follows that $x\in\ker F$.
  This way, $\ker J\subseteq \ker F$.
  However, $F$ is surjective, so $\range J\subseteq \range F$, and $\dim\ker J \geq \dim\ker F$.
  
  Because $F$ and $J$ are surjective, their rows are linearly independent and thus there exists some $g\in\gl(\pos)$ for which $Fg = J$.
  By $q^F=q^J$, $g$ is a $q^F$-isometry:
  \begin{align*}
    q^F(g\darg)=q^F(\darg).
  \end{align*}
  Because $\ker F=\ker J$, we may choose $\pos_J=\pos_F$ and let $i_J$ be the inverse of $J|_{\pos_F}$.
  Then,
  \begin{align*}
    \ii_{T_N} = J i_J = Fg i_J,
  \end{align*}
  so $g i_J = i_F$.
  Thus $g:\ker F\to\ker F$ and $g:\pos_F \to \pos_F$.
  Furthermore, we may choose $g|_{\ker  F}=\ii_{\ker F}$,  so
  \begin{align*}
    g
    =
    \begin{pmatrix}
      \ii_{\ker F} & \\
       & g_0
    \end{pmatrix},
  \end{align*}
  where $g_0:=g|_{\pos_F}$.
  
  Because
  \begin{align*}
    q_N(v) = q^F(i_F(v))= q^F(g_0i_F(v)) = q_N(F i_F i_F^{-1}g_0 i^F(v)) = q_N(i_F^{-1}g_0 i^F(v)),
  \end{align*}
  it follows that $O_g:=i^{-1}_F g_0 i_F \in \orth(T_N)$ is a $q_N$-isometry.
  Moreover, for any $x\in F^{-1}[\ones_t]_N=J^{-1}[\ones_t]_N$, Eq.~\eqref{eq:fof} implies 
  \begin{align*}
    [\ones_t]_N = Jx = O_g Fx = O_g[\ones_t]_N,
  \end{align*}
  so that $O_g\in\stoch(T_N)$.
  This implies  $J=O_g F$, for $O_g\in\stoch(T_N)$, as claimed.

  We now prove the second statement.
  Consider an arbitrary computational basis state $\ket{OF}\in\hilb_F$, where $O\in\stoch(T_N)$.
  We have, for $L, \  R  \in \stoch(T_N)$,
  \begin{align*}
    L OF R_F^{-1} = L O R^{-1} F.
  \end{align*}
  Because $F$ is surjective, the $\stoch(T_N)$ action on the orbit $\{ OF \div O\in\stoch(T_N)\}$ has no fixed points.
  Thus, the isomorphism to the regular representation is given by
  \begin{align*}
    \ket{OF}\mapsto\ket{O}\in\cc[\stoch(T_N)].
  \end{align*}
\end{proof}

\section{Matrix representations for generalized quadratic forms}
\label{app:gen quad}

Matrix representations are useful when thinking about quadratic forms, particularly so in the thorny $d=2$ case.
While we have not used matrix representations of generalized quadratic forms in this case, we include this short note on them in order to provide some intuition to this slightly uncommon object.

We will use $\zz_4$-valued matrices to represent generalized quadratic forms.
For this, we will inject $\zz_2^k\to\zz_4^k$ by interpreting the entry of a vector modulo 4 (rather than modulo 2).
For any vector $u\in\zz_2^k$, we denote its image under this injection by $\t u$.

An $M_q\in\zz_4^{k\times k}$ is a matrix representation of $q\in\t\q(\zz_2^k)$ if
\begin{align*}
  \t u^T M_q \t u = q(u),
  \qquad
  \forall\ u\in\zz_2^k.
\end{align*} 
As with regular quadratic forms, matrix representations of generalized quadratic forms are not unique.%
\footnote{
  In this appendix we have implicitly introduced a ‘‘preferred’’ basis by considering forms over the space $\zz_2^k$ rather than the simply a $k$-dimensional space over $\zz_2$.
  In the latter case, the matrix representation of a form also depends on the choice of basis with respect to which the representation is defined---as is the case for matrix representations of regular quadratic forms.
  This way we may say that a matrix representation of a generalized quadratic form is not uniquely determined by the basis used.
}

\begin{proposition}
  \label{prop:gen q rep}
  Let $M,M'\in\zz_4^{k\times k}$ be matrix representations of a generalized quadratic form $q\in\t\q(\zz_2^k)$.
  Then $M'=M+A$, where $A$ is an alternating $\zz_4$-matrix, i.e.
  \begin{align*}
    A+A^T=0,
    \qquad
    A_{ii}=0
    \quad\forall\ i.
  \end{align*}
\end{proposition}
\begin{proof}
  Let $e_i$ be the standard basis of $\zz_2^k$.
  
  Necessity:
  The diagonal vanishing property follows from
  \begin{align*}
    \t e_i^T (M-M') \t e_i 
    =
    q(e_i)-q(e_i)
    =
    0.
  \end{align*}
  Moreover, letting $e_{ij}:=e_i+e_j$ (and thus $\t e_{ij} = \t e_i + \t e_j$) and using the equation above
  \begin{align*}
    0
    =
    q(e_{ij})-q(e_{ij})
    =
    \t e_{ij}^T (M-M') \t e_{ij} 
    =
    (M-M')_{ij}+(M-M')_{ji}.
  \end{align*}
  Thus, $(M-M')+(M-M')^T=0$.
  
  Sufficiency:
  In a similar fashion as above, one may see that for alternating $\zz_4$-matrices $A$ it holds that $\t u^T A \t u = 0$ for all $u\in\zz_2^k$.
\end{proof}

We point out two basic facts:
First, the diagonal of a matrix representation $M_q$ of $q\in\t\q(\zz_2^k)$ is canonically associated to $q$---namely, $(M_q)_{ii}=q(e_i)$.
Second, we may always find an upper-triangular matrix representation and, moreover, this upper-triangular representation is unique.
A short calculation shows that this representation is given by
\begin{align*}
  (M_q)_{ij}
  =
  \begin{cases}
    q(e_i),         \quad &i=j,\\
    2\beta(e_i,e_j),\quad &i<j,\\
    0,              \quad &\text{otherwise.}
  \end{cases}
\end{align*}

We finish by working out some examples.
Consider a regular quadratic form $q\in\q(\zz_2^k)$.
Then, the upper triangual representation $M_{2q}$ of $2q\in\t\q(\zz_2^k)$ has matrix elements in $\{0,2\}$.
Moreover, 
\begin{align}
  \label{eq:m2=2m}
  M_{2q}=2M_q,
\end{align}
where $M_q$ is the (also unique) upper-triangular representation of $q$.
Warning! The equation above only holds for the upper-triangular representation.
For example, the matrix
\begin{align*}
  M = 
  \begin{pmatrix}
    0 & 2\\ 2&0
  \end{pmatrix}
\end{align*}
represents the \emph{trivial} generalized form $q=0$.

Now consider the form $q_{r,s}$ defined as in the main text.
Its upper-triangular representation is given by $\diag(\ones_r,-\ones_s)$.
In the simple case $r=s=1$, its set of matrix representations is given by
\begin{align*}
  \begin{pmatrix}
    1 & 0\\
    0 & -1
  \end{pmatrix},
  \quad
  \begin{pmatrix}
    1 & 1\\
    -1 & -1
  \end{pmatrix},
  \quad
  \begin{pmatrix}
    1 & -1\\
    1 & -1
  \end{pmatrix},
  \quad
  \begin{pmatrix}
    1 & 2\\
    2 & -1
  \end{pmatrix}.
\end{align*}

Now consider the form $q_t:=q_{t,0}$ for $t=2\mod4$, and take the restriction $q=q_t|_{\ones_t^\perp/\langle\ones_t\rangle}$.%
\footnote{
If it were the case that $t=0\mod4$ or $t=\pm1\mod4$, this case would be easy to analyze. In the former case, $q$ has radical $\langle\ones_t\rangle$ and is a direct sum of hyperbolic planes on $\ones_t^\perp/\langle\ones_t\rangle$.
In the latter case, it is either $q_{t-2,1}$ or $q_{t-1}$ respectively if $t=-1\mod4$ or $t=1\mod4$.
This way the case $t=2\mod4$ is particularly awkward to describe.
}
Letting $\beta_t$ be $q_t$'s polarization and $\beta=\beta_t|_{\ones_t^\perp/\langle\ones_t\rangle}$, we see that $\beta$ is a form of even type:
Indeed $\ones_t$ is the Wu class of $\beta_t$.

The space $\ones_t^\perp$ has as a basis the vectors
\begin{align*}
  b_i &= e_i+e_{i+1},\quad
  i=1,\dots,t-2,\\
  b_{t-1} &= \ones_t.
\end{align*}
With respect to this basis we identify $\ones_t^\perp\simeq\zz_2^{t-1}$.

Notice the following relations:
\begin{align*}
  q(b_i)&=2,\\
  \beta(b_i,b_j) &=
  \begin{cases}
    1, \quad & i=j\pm1,\ i,j<t-1,\\
    0, \quad & \text{otherwise},
  \end{cases} 
\end{align*}
Then,
\begin{align*}
  M_{\beta_t|_{\ones_t^\perp} }
  =
  \begin{pmatrix}
    0 & 1 &   &         &   &   &   & 0 \\
    1 & 0 & 1 &         &   &   &   & 0 \\
      & 1 & 0 & 1       &   &   &   & 0 \\
      &   &   & \ddots  &   &   &   & \vdots  \\
      &   &   &         & 1 & 0 & 1 & 0 \\
      &   &   &         & 0 & 1 & 0 & 0 \\
    0 & 0 & 0 & \cdots  & 0 & 0 & 0 & 0
  \end{pmatrix}
\end{align*}
has rank $t-2$ and the radical of $\beta_t|_{\ones_t^\perp}$ is spanned by $\ones_t=\sum_ib_{2i-1}$. 
Moreover, since $\ones_t$ is the Wu class of $\beta_t$, $\beta$ is a full rank alternating form.
This implies $q\sim2q'$ for a quadratic form $q'$ which is a sum of $t/2$ hyperbolic planes.
Indeed, a basis of hyperbolic pairs $\{p_i,q_i\}_{i=1}^{\frac t2 - 1}$ is given by
\begin{align*}
  p_i = \sum_{j\leq 2i-1} e_j + e_{2i},
  \quad
  q_i = \sum_{j\leq 2i-1} e_j + e_{2i+1}.
\end{align*}
We can compute $q(p_i)=q(q_i)=2i\mod4$, and clearly $q'(p_i)=q'(q_i)=0$.
Moreover $\beta(p_i,q_j)=\delta_{ij}$ and the $p_i$'s span a Lagrangian (same for the $q_i$'s).
This way, with respect to the hyperbolic basis, its upper-triangular representation is
\begin{align*}
  M_q
  =
  \begin{pmatrix}
    2 & 2 & \\
      & 2 & \\
      &   & 0 & 2 \\
      &   &   & 0 \\
      &   &   &   & 2 & 2 & \\
      &   &   &   &   & 2 & \\
      &   &   &   &   &   & \ddots \\
      &   &   &   &   &   &        & 2 & 2 & \\
      &   &   &   &   &   &        &   & 2 & \\
  \end{pmatrix}.
\end{align*}
This way,
\begin{align*}
  q'\sim (q_\hh^1)^{\oplus (t+2)/4}\oplus(q_\hh^0)^{\oplus (t-2)/4},
  \quad
  q=2q',
\end{align*}
and
\begin{align*}
  \arf(q')=\frac{t+2}{4}\mod2,\quad
  \garf(q)=4\arf(q')=t+2\mod8.
\end{align*}

As mentioned before, different choices of basis yield different matrix representations.
For example, with respect to the $b_i$ basis, the unique upper triangular representation of $q$ is
\begin{align*}
  M_q
  =
  \begin{pmatrix}
    2 & 2 &   &   &   & \\
      & 2 & 2 &   &   & \\
      &   & 2 & 2 &   & \\
      &   & \ddots& & & \\
      &   &   &   & 2 & 2 & 
  \end{pmatrix}.
\end{align*}

\section{The orthogonal stochastic group}
\label{app:orthogonal stochastic}

A natural question, and one which seems to not have been addressed in the literature, is \emph{what is the relationship between $\stoch(T)$ and the classical groups when $d=2$.}
We provide some details of this relationship in this appendix.
Throughout, we let $d=2$, $q=q_{t,0}$ and $\beta=\beta_{t,0}$.

The core of the problem is the following.
It is clear that that $\stoch(T)\subset\iso(\beta)$, where
\begin{align*}
  \iso(\beta) = \{ O\in\gl(T) \div \beta(O\darg,O\darg)=\beta(\darg,\darg)
\end{align*}
is the isometry group of $\beta$.%
\footnote{
  This inclusion can be strict. For example when $t=4$ the \emph{anti-identity} matrix $\ii_4+\ones_4\ones_4^T$ is in $\iso(\beta)$ but not in $\stoch(T)$.
}
When the characteristic is even, however, $\beta$ is of odd type and thus does not polarize any $t$-dimensional quadratic form.
In this way, this isometry group is not necessarily isomorphic to any $t$-dimensional orthogonal group.
We would like to find a relationship between $\iso(\beta)$ and \emph{some} standard classical group---this way would enable us to leverage the variety of results characterising the representation theory of finite classical groups.

For this end, consider the space $T'=T\oplus T$ equipped with a quadratic form $\kappa(u\oplus v)=u\cdot v$.
The subspaces $L:=T\oplus0$ and $R:=0\oplus T$ are Lagrangians and a short calculation shows that $\arf(\kappa)=0$ and that this form is non-degenerate.
We therefore have that
\begin{align*}
  \orth_\kappa := \{ O\in\gl(T') \div \kappa(O\darg)=\kappa(\darg)\} = \orth^+_{2t}
  \subset \sp(T'),
\end{align*}
where $\orth^+_{2t}$ is the standard orthogonal group with trivial Arf invariant.

Consider the maximal subgroup of $\orth_\kappa$ that preserves $L$, i.e.\ whose matrices are of the form
\begin{align*}
  \begin{pmatrix}
    A & B\\
      & C
  \end{pmatrix}.
\end{align*}
This is known as the parabolic subgroup associated to $L$.
Let us look at the subgroup $\levi_{L,R}(\kappa)\subset\orth_\kappa$ which preserves both $R$ \emph{and} $L$---this is known as the \emph{Levi subgroup} of the aforementioned parabolic group.
A short calculation shows
\begin{align*}
  \levi(\kappa)
  =
  \left\{
    \begin{pmatrix}
      O & \\
        & O^T
    \end{pmatrix}
    \; : \;
    OO^T=\ii
  \right\}.
\end{align*}
Thus, one may canonically identify $\levi_{L,R}(\kappa)\simeq\iso(\beta)$.

In summary, we have identified $\iso(\beta)$ as the Levi subgroup of $\orth_\kappa$ on $T\oplus T$.
We now further characterize the relationship between $\iso(\beta)$ and $\stoch(T)$.

Matrices in $\iso(\beta)$ are in a one-one correspondence with orthonormal bases of $T$, i.e.\ bases $b_i$ satisfying $\beta(b_i,b_j)=\delta_{ij}$.
Notice that the $b_i$ necessarily have an odd support: $1=\beta(b_i,b_i)=|\supp(b_i)|\mod2$.
Now, the key point is that if $q(b_i)=3\mod4$ then there is no $O\in\stoch(T)$ for which $Oe_i=b_i$.
It is thus natural to define the following partition of the set of orthonormal bases:
\begin{align*}
  \mfrak{B}_k := \{ \text{orthonorm. bases }\{b_i\} \div
  \text{there's exactly $k$ indices $i$ s.t. }
  q(b_i)=3
  \}.
\end{align*}
This way, $\iso(\beta)$ corresponds to $\cup_k\mfrak{B}_k$ while $\stoch(T)$ corresponds to $\mfrak{B}_0$.
For which $k$'s is $\mfrak{B}_k$ non-empty?
If $k=0\mod4$ then the anti-id. basis can be singled out as a representative of $\mfrak{B}_k$.
Thus, any $B\in\mfrak{B}_k$ may be written as
\begin{align*}
  (OA_k\pi) E,
\end{align*}
for some $\pi\in S_t$ and some $O\in\stoch(T)$, where $E:=\{e_i\}$.

\begin{proposition}
  \label{prop:k multiple of 4}
  It holds that the set $\mfrak{B}_k\neq\emptyset$ is not empty if and only if $k=0\mod4$.
\end{proposition}
\begin{proof}
	\emph{If:} 
	Let $b_i$ be $e_i+(\ones_k,\mbf{0}^{t-k})^T$ for $i\leq k$. 
	Then,  the basis  $\{b_i\}_{i=1}^k\cup\{e_i\}_{i>k}$ is an  element of $\mfrak{B}_k$.
	Indeed $\beta(b_i,e_j)=0$, $\beta(e_i,e_j)=\delta_{ij}$ and
	\begin{align*}
		\beta(b_i,b_j) = \beta(e_i,e_j) + 1  + 1 + k = \delta_{ij}.
	\end{align*}
	\emph{Only if:}
  Suppose for the sake of contradiction that there is a non-empty $\mfrak{B}_k$ with $k\mod4=l\in\{1,2,3\}$.
  Without loss of generality, we may assume that the corresponding basis $\{b_i\}$ is such that
  \begin{align*}
    q(b_i)=
    \begin{cases}
      3, \quad &\forall\ i\leq k,\\
      1, \quad &\forall\ i> k.
    \end{cases}
  \end{align*}
  
  Now consider the vectors
  \begin{align*}
    v_i = e_i+(\ones_{k-l},0^{t-k+l})^T
    =
    e_i+\sum_{j\leq k-l} e_j,
    \quad
    i=1,\dots,k-l,
  \end{align*}
  and
  \begin{align*}
    v_i = e_i, \quad i>k.
  \end{align*}
  By the generalized Witt's lemma~\cite{wittmod4}, there is an $O\in\stoch(T)$ for which
  \begin{align*}
    Ob_i=v_i \quad \forall\ i\leq k-l, \text{ or } i>k.
  \end{align*}
  
  Now, consider the vectors $Ob_{k-j}$, with $0\leq j\leq l-1$.
  Because $\beta(b_{k-j},b_i)=\beta(Ob_{k-j},Ob_i)=\beta(Ob_{k-j},e_i)=0$ for all $i>k$, it must hold that $Ob_{k-j}$ is not supported in the last $t-k$ entries.
  Similarly, because the vectors $Ob_i$ with $i\leq k-l$ restrict to a basis of the subspace supported on the first $k-l$ entries, $\beta(b_{k-j},b_i)=0$ for these values of $i$ implies that $Ob_{k-j}$ is not supported in the first $k-l$ entries.
  Thus, all the vectors $\{Ob_{k-j}\}_{j=0}^{l-1}$ are supported in the same $l$ entries.
  If $l<3$ this contradicts $q(Ob_{k-j})=q(b_{k-j})=3$.
  If $l=3$, $q(b_{k-j})=3$ implies that the three vectors are the same and thus not linearly independent.
\end{proof}

\begin{remark}
  One may much more elegantly and simply prove a relaxation of the result above---namely, that $\mfrak{B}_k$ is non-empty only if $k$ is even.
  For this, notice that any basis $\{b_i\}\in\mfrak{B}_k$ is such that $\sum_ib_i=\ones_t$.
  This is because the map $Me_i=b_i$ is a $\beta$-isometry, $M\in\iso(\beta)$, and thus preserves the Wu class of $\beta$.
  Then, by orthogonality
  \begin{align*}
    t\mod4=q(\ones_t)=\sum_i q(b_i) = 3k+t-k\mod4,
  \end{align*}
  and so $0 = 2k\mod4$ or, equivalently, $k=0\mod2$.
\end{remark}

This way, any isometry $M\in\iso(\beta)$ may be written, as above, as
\begin{align*}
  M=OA_k\pi
\end{align*}
for some $O\in\stoch(T)$, some $k$ with $k=0\mod4$, and some $\pi\in S_t$.
Thus, the double cosets $\{\stoch(T)A_k S_t\}_k$ give a covering of the group.
Moreover, these different cosets do not intersect since $OA_k\pi=O'A_{k'}\pi'$ implies that $O^{-1}O':\mfrak{B}_{k'}\to\mfrak{B}_k$ which, by $O\in\stoch(T)$, implies $k=k'$.

This way, we have the disjoint union
\begin{align}
  \label{eq:iso normal form}
  \iso(\beta) = \dot\bigcup_k \stoch(T)A_k S_t,
\end{align}
associated to the normal form just derived.
In particular, any representation $\rho$ of $\iso(\beta)$ is determined by $\res{\stoch(T)}\rho$ together with $\{\rho(A_k)\}_k$.
Moreover, any matrix $M\in\stoch(T)A_k\stoch(T)$ maps the standard basis to an element of $\mfrak{B}_k$, so that $\stoch(T)A_k S_t=\stoch(T)A_k\stoch(T)$.
This way,
\begin{align*}
  \stoch(T) \backslash \iso(\beta)/\stoch(T) = \mrm{G}_A:=\langle A_k\rangle_k.
\end{align*}
(The left-hand side equality is not an isomorphism of groups.)

A short calculation shows that the $A_k$ matrices commute with each other, and thus
\begin{align*}
  \mrm{G}_A:=\langle A_k\rangle_k \simeq \zz_2^{\lfloor t/4\rfloor}.
\end{align*}
To see the right-hand side notice that $A_k\notin\langle A_{k'}\rangle_{k'<k}$.
Thus, any element in $\irr\iso(\beta)$ is identified with a pair $(\nu,v)$ where $\nu\in\irr(\stoch(T))$ and $v\in\zz_2^{\lfloor t/4\rfloor}$.

\newpage
\section{Table of symbols used}

\begin{table}[h!]
\begin{tabular}{c|p{10cm}}
  Symbol & Meaning/Definition\\
  \hline
  \hline
	$D$	& $=d$ if $d$ is odd, $=4$ if $d=2$\\
  $T$         & $\zz_d^t$ with generalized quadratic form $q_{r,s}$\\
  $\discr(\darg)$ & Discriminant of bilinear form\\
  $\arf(\darg)$, $\garf(\darg)$ & Arf invariant, generalized Arf invariant of quadratic form\\
  $\Xi(\darg)$, $\t\Xi(\darg)$ & Polarization and generalized polarization of quadratic form\\
  $\grass_m$  & Set of isotropic stochastic subspaces of $T$ not containing $\ones_t$ with dimension $m$\\
  $\grass_m^0$& Set of isotropic stochastic subspaces of $T$ containing $\ones_t$ with dimension $m$\\
  $T_N$, $q_N$& $N^\perp/N$ with $N\subset T$ isotr. stoch., $q_{r,s}|_{T_N}$\\
  $C_N$ & Code space $\range P(N)$\\
	$\cones$ & Code space $\range P(\langle\ones_t\rangle)$ if $r-s=0\mod D$, otherwise $\{0\}$\\
  $\Delta_{r,s}^{(k)}$ & Rank $k$ component of $\Delta_{r,s}$\\
  $\Delta$    & Action of $\rcliff$ on $\cones$\\
  $\semigroup_{r,s}$ & %
  $ \{R(O)P(N)\div N\subset T\text{ stoch. isotr., } O\in\stoch(T)\}$\\
  $\semigroup_{r,s}^N$ & %
  Subset $ \{R(O)P(N')\div N\subseteq N'\subset T\text{ stoch. isotr., } O\in\stoch(T)\}\subset\com{r,s}$\\
  $\com{r,s}$ & $\vspan\{\semigroup_{r,s}\}$ commutant algebra of $\Delta_{r,s}$\\
  $\com{r,s}^m$ & $\vspan\{P(N)R(O) \div \dim N \geq m\}$\\
  $\com{r,s}^{m,0}$ & $\vspan\{P(N)R(O) \div \dim N \geq m,\ \ones_t\in N\}$\\
  $\cm{m}$  & $\vspan\{C_N\div N\in\grass_m\}\cap\cones^\perp$\\
  $\dm{m}$  & $\vspan\{C_N\div N\in\grass_m^0\}$\\
  $\spinful{m}$   & $\cm{m}\cap\cm{m+1}$\\
  $\spinless{m}$  & $\dm{m}\cap\dm{m+1}$\\
\end{tabular}
\end{table}


\bibliographystyle{unsrt} 
\bibliography{cliff-arxiv}

\end{document}